\documentclass[preprint,showpacs,preprintnumbers,amsmath,amssymb,nofootinbib]{revtex4-1}

\usepackage{graphicx}
\usepackage{epstopdf}

\usepackage{dcolumn}
\usepackage{bm}
\usepackage{color}

\usepackage{feynmf}
\usepackage{slashed}
\usepackage{enumerate}

\usepackage[scientific-notation=true]{siunitx} 

\usepackage[utf8]{inputenc}

\usepackage[colorlinks=true,urlcolor=blue,anchorcolor=blue,citecolor=blue,filecolor=blue,linkcolor=blue,menucolor=blue,linktocpage=true]{hyperref} 

\usepackage{appendix}
\usepackage{amsmath,amsthm,amssymb,mathrsfs}
\usepackage{bbold}
\usepackage{bbm}

\usepackage{multirow}
\usepackage{makecell}
\usepackage{verbatim}

\begin{document}

\title{General discussions on the SU(2) vector boson dark matter model with a single Higgs multiplet --- Lagrangian, discrete subgroups, and scalar classifications}

\author{Chun-Xue Yuan$^a$}
\author{Zhao Zhang$^a$}

\author{Chengfeng Cai$^b$}
\thanks{caichf3@mail.sysu.edu.cn}
\author{Yi-Lei Tang$^a$}
\thanks{tangylei@mail.sysu.edu.cn}

\author{Hong-Hao Zhang$^a$}
\thanks{zhh98@mail.sysu.edu.cn}

\affiliation{$^a$School of Physics, Sun Yat-Sen University, Guangzhou 510275, China}
\affiliation{$^b$School of Science, Sun Yat-Sen University, Shenzhen 518107, China}

\begin{abstract}
The vector boson dark matter particles which stem from some broken gauge symmetries usually require some unbroken symmetries to keep them stable. In previous literature, some simplest cases have been discussed, in which the unbroken symmetry is provided by a remnant subgroup of the gauge group. It would be interesting to ask whether all the possible remnant subgroups as well as all the possible coupling forms can be enumerated. Classifying all the Higgs components into different mass degenerate representations to simplify the diagonalization processes is also necessary. Rather than the ambitious target of providing a general solution to all kinds of gauge groups configured with all forms of the Higgs multiplets, in this paper, we concentrate on the case of the $\text{SU(2)}_{\text{D}}$ gauge group together with a single Higgs multiplet. We enumerate all possible discrete subgroups that can survive up to $n=21$, where $n$ is the dimension of the Higgs multiplet. We also provide the general algorithms to enumerate all possible renormalizable operators, to write the general forms of the vacuum expectation value (VEV) configurations, and to give the detailed results of all the mass degenerate irreducible representations embedded in the Higgs multiplet.
\end{abstract}



\maketitle
\tableofcontents

\section{Introduction}

The Dark matter (DM), which occupies about 26\% of the energy density budget of our current Universe, is still  mysterious. In the literature, spin-0 and spin-1/2 DM candidates have been sufficiently discussed, with the vector bosons sometimes regarded as pure mediators to carry the interactions [see Refs.~\cite{Bertone:2004pz, Bertone:2016nfn, Profumo:2017hqp, Bertone:2018krk, Arbey:2021gdg} for reviews in which typical DM models are enumerated therein]. Although the vector boson has the possibility to be a DM candidate, most of the corresponding discussions in the literature seem to focus on some specific models while losing some generalities.

The main procedure to discuss a particular vector boson DM model usually commences with a Lagrangian accommodating some vector field multiplets as well as the corresponding Higgs configurations. To keep the DM stable, some discrete or continuous symmetry might be imposed by hand at the beginning~\cite{Lebedev:2011iq, Abe:2012hb, Farzan:2012hh, Baek:2012se, Domingo:2013tna, Yu:2014pra, Chen:2014cbt, Gross:2015cwa, Duch:2015jta, DiFranzo:2015nli, Barman:2018esi, Azevedo:2018oxv, Poulin:2018kap, Saez:2018off, Barman:2019lvm, Mohamadnejad:2019vzg, Glaus:2019itb, Arcadi:2020jqf, Abe:2020mph, Delaunay:2020vdb, Salehian:2020asa, YaserAyazi:2022tbn, Frandsen:2022klh, Das:2024tfe}, be satisfied accidentally~\cite{Hambye:2008bq, Davoudiasl:2013jma, Boehm:2014bia, Gross:2015cwa, DiChiara:2015bua, Karam:2015jta, Khoze:2016zfi, Karam:2016rsz, Ko:2016fcd, Arcadi:2016kmk, Poulin:2018kap, Belyaev:2018xpf, Ramos:2021omo, Hu:2021pln, Baouche:2021wwa, Chowdhury:2021tnm, Cai:2021wmu, Borah:2021ftr, Belyaev:2022shr, Elahi:2022hgj, Zhang:2022wek, Frandsen:2022klh, Benitez-Irarrazabal:2024ich, Acevedo:2024ava, Tran:2023lzv}, or can be induced naturally after the gauge symmetry spontaneously breaks down~\cite{Chiang:2013kqa, Baek:2013dwa, Khoze:2014woa, Gross:2015cwa, Karam:2016rsz, Arcadi:2016kmk, Poulin:2018kap, Choi:2019zeb, Ghosh:2020ipy, Nomura:2020zlm, Zhang:2022wek}. After some Higgs components acquire the vacuum expectation value (VEV), one then investigates in the parameter space so that vector bosons acquire masses and the stable one(s) become the DM candidate. Phenomenologies are then evaluated for comparison with various experimental data, demonstrating the viability of such a model.

In the literature, it has been so long realized that gravity breaks the global symmetry while keeping the gauge symmetry intact~\cite{Giddings:1987cg, Abbott:1989jw, Gilbert:1989nq, Coleman:1989zu, Kallosh:1995hi, Witten:2017hdv}.
It then becomes more natural that the dark matter stability is kept by some subgroups remaining from some broken gauge groups, which will not be further demolished by the gravity effects.
For a gauge symmetry beyond the simplest U(1) case, a general discussion of the remaining group is difficult, and the diagonalizing processes of the mass matrix of all scalar components is another issue, since the analytical calculations become extremely formidable.

Aiming at solving these problems at least in some simple cases, we study the vector boson DM model stemming from a dark $\text{SU(2)}_\text{D}$ gauge group accompanied by a single Higgs multiplet as an example. In this paper, as an alternative to computing the phenomenologies based upon a particular Lagrangian configuration, we try to achieve the following goals:
\begin{itemize}
    \item Give a general algorithm to enumerate the general renormalizable Lagrangian.
    \item Enumerate all the possible subgroups as well as the corresponding VEV configurations which are capable to keep the DM stable.
    \item Classify all the scalar components into the irreducible representations of the subgroup, and show how they are formulated by the combinations of the original Higgs multiplet components.
    We also list all the squared mass matrices of the Higgs components expressed by the parameters appearing in the Lagrangian, while the effective potential minimal condition equations are also listed and utilized to eliminate the parameters appearing in the squared mass matrices.
\end{itemize}

These are the very beginnings required to analyze all the DM models before detailed phenomenological calculations are performed. Actually, Ref.~\cite{Etesi:1996urw} has partly resolved the first and second goals in the odd-dimensional Higgs multiplet cases, while Ref.~\cite{Hamada:2015bra} has partly listed the potential for an SM $\text{SU}(2)_\text{L}$ scalar multiplet with dimensions from 4 to 7 with some particular settings of the hypercharge. Reference ~\cite{King:2018fke} also achieved parts of our results, but with a quite different method for several specific cases without enumerating the couplings. Alternatively we work on a more familiar format of the SU(2) representations and follow a more intuitive process. In our paper, we adopt algorithms similar to those of Ref.~\cite{Etesi:1996urw} to find the subgroups and the corresponding VEV configurations. However, we develop a completely different method to accomplish the complete set of the renormalizable couplings, regardless of the intricate algorithms based on the graph theory in Ref.~\cite{Etesi:1996urw}. We show how the irreducible representations of the SU(2) group fragment into the representations of the discrete subgroups in detail, with the detailed coefficients presented, about which we have seen no general discussions available in the literature. In our paper, our algorithms can be generalized so that all three goals can be achieved for an arbitrary SU(2)$_\text{D}$ Higgs multiplet in principle; however, we only show detailed processes for several typical examples and enumerate the results from $j=2$ up to $j=10$ multiplets.

\section{The Langrangian}

As we have mentioned, we introduce a new $\text{SU(2)}_\text{D}$ gauge group associated with the gauge field $A_\text{D}^{a\mu}$, where $a=1,2,3$ is the $\text{SU(2)}_\text{D}$ index.  One $n$-dimensional Higgs multiplet $\Phi_{(n)}$ charged under this gauge group is also provided. Here $n=2 j+1$ is the dimension of the multiplet, where $j$ is the ``angular quantum number'' of the irreducible representation to which $\Phi_{(n)}$ corresponds.

For each $n$ or $j$, the most familiar basis for the components of $\Phi_{(n)}$ is the ``$J_\text{D}^3$ representation,'' in which the $\Phi_{(n)}$ components are arranged in accordance with ``$|j, m\rangle_\text{D}$,'' where $m$ indicates the $J_\text{D}^3$ eigenvalues. Here we just borrow the symbols from quantum mechanics, and ``$|j, m\rangle$'' means the eigenstates under the $\text{SU(2)}_\text{D}$ group, rather than the rotational group of the real space; therefore, additional ``D'' subscripts are imposed. $\hat{J}_\text{D}^{1,2,3}$ are the corresponding generators of the $\text{SU(2)}_\text{D}$ group. For a particular $j$, we write down the components of $\Phi_{(n)}$ as
\begin{eqnarray}\label{eq:Phi}
	\Phi_{(n)}=\xi_{(n)}\left(\Delta_j,\Delta_{j-1},\dots,\Delta_{-j+1},\Delta_{-j}\right)^\mathrm{T},
\end{eqnarray}
where ``$\Delta_m$'' here indicates the scalar component corresponding to the ``$|j, m\rangle_\text{D}$'' eigenstate. The normalization factor $\xi_{(n)}=1~(1/\sqrt{2})$ when $n$ is even (odd).

For even $n$, $j = \frac{n-1}{2}$ is a half-integer, and all $\Delta_{m}$'s are complex. On the other hand, when $n$ is odd, and $j = \frac{n-1}{2}$ therefore becomes an integer, then the $n$-dimensional irreducible representation can be real. In this paper, we would like a minimal configuration, so $\Phi_{(n)}$ is appointed be the real $\text{SU(2)}_\text{D}$ multiplet when $n$ is odd. It requires $\Phi_{(n)}$ to satisfy following constraint:
\begin{eqnarray}
    \Phi_{(n)} = T_{(n)} \Phi_{(n)}^*, \label{Self_Conjugate}
\end{eqnarray}
where 
\begin{eqnarray}
    T_{(n)} = [T_{(n)}]_{m_1, m_2} \label{T_OddN}
\end{eqnarray}
is an $n \times n$ matrix, with all $(T_{(n)})_{m, -m} = 1$ for any $m$, while all other elements vanish.

Then we are ready to write the general form of the renormalizable Lagrangian. The dark sector is given by
\begin{eqnarray}
    \mathcal{L} \supset -\frac{1}{4} F_{\text{D} \mu \nu}^a F_\text{D}^{a\mu \nu } + (D^{\mu} \Phi_{(n)})^{\dagger} D_{\mu} \Phi_{(n)} - V(\Phi_{(n)}^{\dagger}, \Phi_{(n)}, H^{\dagger}, H), \label{Lagrangian}
\end{eqnarray}
where $F_{\text{D} \mu \nu}^a = \partial_{\mu} A_{\text{D} \nu}^a - \partial_{\nu} A_{\text{D} \mu}^a + g_\text{D} \epsilon^{abc} A_{\text{D} \mu}^{b} A_{\text{D} \nu}^{c}$, and $D_{\mu} \Phi_{(n)} = \partial_{\mu} \Phi_{(n)} - i g_\text{D} \tau_{(n)}^a A_{\text{D} \mu}^{a} \Phi_{(n)}$, with $g_D$ being the $\text{SU(2)}_\text{D}$ gauge coupling constant, $\epsilon^{a b c}$ being the Levi-Civita symbol, and $\tau_{(n)}^a$ being the generator matrix of the $\text{SU(2)}_\text{D}$ group under the $n$-dimensional irreducible representation. 

$V(\Phi_{(n)}^{\dagger}, \Phi_{(n)}, H^{\dagger}, H)$ is the effective potential, and its general form can be formulated as
\begin{eqnarray}
    & & V(\Phi_{(n)}^{\dagger}, \Phi_{(n)}, H^{\dagger}, H) \nonumber \\
    &=& \frac{1}{2}[(\mu_{p q}^{2 (2, 0)} + \lambda_{H\Phi p q}^{(2, 0)} H^{\dagger} H) \Delta_{p} \Delta_{q} + (\mu_{p q}^{2 (1, 1)} + \lambda_{H\Phi p q}^{(1, 1)} H^{\dagger} H) \Delta_{p} \Delta_{q}^* + \text{h.c.}] \nonumber \\
    &+& (\kappa^{(3, 0)}_{p q r} \Delta_{p} \Delta_{q} \Delta_{r} + \kappa^{(2, 1)}_{q p r} \Delta_{p} \Delta_{q} \Delta_{r}^* + \text{h.c.}) \nonumber \\
    &+& (\lambda^{(4, 0)}_{p q r s} \Delta_{p} \Delta_{q} \Delta_{r} \Delta_{s} + \lambda^{(3, 1)}_{p q r s} \Delta_{p} \Delta_{q} \Delta_{r} \Delta_{s}^* + \lambda^{(2, 2)}_{p q r s} \Delta_{p} \Delta_{q} \Delta_{r}^* \Delta_{s}^* + \text{h.c.}), \label{EffectivePotential}
\end{eqnarray}
where $H$ is the standard model (SM) Higgs doublet, which is defined as
\begin{eqnarray}
    H = \left( \begin{array}{c}
    G^+ \\
    \frac{v_{\text{EW}} + \tilde{h} + i G^0}{\sqrt{2}} \end{array} \right),
\end{eqnarray}
$v_{\text{EW}} \approx 246$ GeV is the electroweak VEV, $\tilde{h}$ is the SM-like Higgs field, and $G^{\pm, 0}$ are the corresponding Goldstone bosons. $\mu^{2(2, 0)}_{p q}$, $\mu^{2(1, 1)}_{p q}$, $\lambda^{(2, 0)}_{H\Phi p q}$, $\lambda^{(1, 1)}_{H\Phi p q}$, $\kappa^{(3, 0)}_{p q r}$, $\kappa^{(2, 1)}_{p q r}$, $\lambda^{(4, 0)}_{p q r s}$, $\lambda^{(3, 1)}_{p q r s}$, $\lambda^{(2, 2)}_{p q r s}$ are the coupling constants with $p$, $q$, $r$, $s$ being integers running from $-j$ to $j$. The two superscripts within the brackets indicate the number of the original $\Delta$'s and the conjugated $\Delta^*$'s, respectively, combined to follow the couplings. For example, $\lambda^{(3, 1)}_{p q r s}$ is joined with the $\Delta_{p} \Delta_{q} \Delta_{r} \Delta_{s}^*$ term, which is composed of three original $\Delta$'s and one conjugated $\Delta^*$ to formulate the $(3,1)$ in the superscript.

Gauge invariance requires that each term in Eq.~\eqref{EffectivePotential} should be an $\text{SU(2)}_\text{D}$ singlet. Then our purpose is to find the general possible $\mu^2$, $\lambda_{H\Phi}$, $\kappa$, $\lambda$ assignments, which can be regarded as the combinations of the Clebsch–Gordan coefficients. Before closing this section, we should point out that, for even or odd $n$, some terms in the above general potential obviously vanish or have simpler formations. Firstly, when $n$ is an even number, $\mu^{2(2, 0)}_{p q}$ and $\lambda^{(2, 0)}_{H\Phi p q}$ finally disappear with our field configurations, since there is only one way to contract a two-order tensor $\Delta_p \Delta_q$ into an $\text{SU(2)}_\text{D}$ singlet---$\mu^{2(2, 0)}_{p q} \propto \lambda^{(2, 0)}_{p q} \propto T_{(n)}$. The definition of the $T_{(n)}$ for odd $n$ has been given around Eq.~\eqref{T_OddN}; however, here we have to define $T_{(n)}$ in the case of even $n$,
\begin{eqnarray}
    T_{(n)} = [T_{(n)}]_{m_1, m_2}, \label{T_EvenN}
\end{eqnarray}
where $[T_{(n)}]_{m, -m}=-[T_{(n)}]_{-m, m}=1$ for all $m > 0$, and all other elements are set zero. Notice that $T[T_{(n)}]_{m_1, m_2}$ is antisymmetric, therefore contracting with the symmetric $\Delta_p \Delta_q$ terms kills both the $\mu^{2(2, 0)}_{p q}$ and $\lambda^{(2, 0)}_{p q}$ terms. For the $\Delta_p \Delta_q \Delta_r$, $\Delta_p \Delta_q \Delta_r^{*}$ combinations, the direct product of three half-integer spins cannot constitute a singlet, implying that all the trilinear terms should also disappear.

On the other hand, when $n$ is an odd number, all the conjugated $\Delta_p^*$ terms can be transformed into the original $\Delta_p$ by utilizing the real field condition Eq.~\eqref{Self_Conjugate}. Therefore, we can choose to preserve  $\mu^{2(1, 1)}_{p q}$, $\lambda^{(1, 1)}_{H\Phi p q}$, $\kappa_{p q r}^{(3, 0)}$, $\lambda^{(4, 0)}_{p q r s}$ since all other contributions can be attributed to them. In this case, it is convenient to neglect the bracketed superscripts and to write them in the form $\mu^{2}_{p q}$, $\lambda_{H\Phi p q}$, $\kappa_{p q r}$, $\lambda_{p q r s}$.

It is easy to learn from the Littlewood-Richardson rule~\cite{Littlewood:1934}, that the only way to contract the bilinear term $\Delta_p \Delta_q^*$ into an $\text{SU(2)}_\text{D}$ singlet is given by
\begin{eqnarray}
    \Phi_{(n)}^{\dagger} \Phi_{(n)} = \xi_{(n)}^2 \Delta_m \Delta_m^*.
\end{eqnarray}
Therefore, Eq.~\eqref{EffectivePotential} can be simplified to be
\begin{eqnarray}
    & & V(\Phi_{(n)}^{\dagger}, \Phi_{(n)}, H^{\dagger}, H) \nonumber \\
    &=& \frac{1}{2}[(\mu^2 + \lambda_{H\Phi} H^{\dagger} H) \Delta_p \Delta_p^* + \text{h.c.}] + (\kappa^{(3, 0)}_{p q r} \Delta_{p} \Delta_{q} \Delta_{r} + \text{h.c.}) \nonumber \\
    &+& (\lambda^{(4, 0)}_{p q r s} \Delta_{p} \Delta_{q} \Delta_{r} \Delta_{s} + \lambda^{(3, 1)}_{p q r s} \Delta_{p} \Delta_{q} \Delta_{r} \Delta_{s}^* + \lambda^{(2, 2)}_{p q r s} \Delta_{p} \Delta_{q} \Delta_{r}^* \Delta_{s}^* + \text{h.c.}), \label{EffectivePotential_MassSimplified}
\end{eqnarray}
where $\mu^2$ and $\lambda_{H\Phi}$ are the common factors extracted from the $\mu^{2(1, 1)}_{p q} \Delta_{p} \Delta_{q}^*$ and the $\lambda^{(1, 1)}_{H\Phi p q} \Delta_{p} \Delta_{q}^*$ contractions.

\section{Enumeration of the General Coupling Combinations}

In this section, we aim at enumerating all the possible combinations of $\kappa_{p q r}^{(3, 0)}$, $\kappa_{p q r}^{(2, 1)}$, $\lambda^{(4, 0)}_{p q r s}$, $\lambda^{(3, 1)}_{p q r s}$, and $\lambda^{(2, 2)}_{p q r s}$ from $j=0$ until $j=10$. The Littlewood-Richardson rule can in principle help us to find all possible contractions upon all the $\Delta_p \Delta_q \Delta_r$, $\Delta_p \Delta_q \Delta_r \Delta_s$, $\Delta_p \Delta_q \Delta_r \Delta_s^{*}$, and $\Delta_p \Delta_q \Delta_r^{*} \Delta_s^{*}$ combinations to formulate the $\text{SU(2)}_\text{D}$ singlets, since such a process is fundamentally to find the trivial representation subspace for the direct product space of the $\Delta_s^{*}$'s, similar to the bilinear case. However, manipulating the multiple miscellaneous Young diagrams becomes a cumbersome task, 
and thus we take an alternative approach. In fact, the character table method can at least tell us the multiplicity, or the dimension of the trivial representation subspace. With the guide of this multiplicity as the bound of the number of independent combination formulas, we then try to enumerate all possible contractions by inserting various operators between $\Phi_{(n)}^{\dagger}$ and $\Phi_{(n)}$ until the number of independent contractions touches the bound.

Let us take the $\lambda_{p q r s}^{(2, 2)} \Delta_p \Delta_q \Delta_r^{*} \Delta_s^{*}$ term as an example for illustration. Here $\Delta_p \Delta_q \Delta_r^{*} \Delta_s^{*}$ is a 4-order tensor. When an element $g \in \text{SU(2)}_\text{D}$ acts upon it, the tensor transforms as
\begin{eqnarray}
    \Delta_p \Delta_q \Delta_r^{*} \Delta_s^{*} \rightarrow \Delta^{\prime}_p \Delta^{\prime}_q \Delta^{\prime *}_r \Delta^{\prime *}_s = V_{(n)}(g)_{p \tilde{p}} V_{(n)}(g)_{q \tilde{q}}  V_{(n)}(g)^{\dagger}_{\tilde{r} r}  V_{(n)}(g)^{\dagger}_{\tilde{s} s} \Delta_{\tilde{p}} \Delta_{\tilde{q}} \Delta_{\tilde{r}}^{*} \Delta_{\tilde{s}}^{*},
\end{eqnarray}
where $V_{(n)}(g)$ is the $n \times n$ representation matrix of $g$ under the $n$-dimensional irreducible representation of the $\text{SU(2)}_\text{D}$ group. It is easy to realize that $\Delta_p \Delta_q \Delta_r^{*} \Delta_s^{*}$ is symmetric by swapping the $\{p,q\}$ and $\{r,s\}$ indices separately, so we can equivalently study the decomposition of a tensor $T_{p q r s}$ which transforms under the action of $g \in \text{SU(2)}_\text{D}$ in the following way
\begin{eqnarray}
    T_{p q r s} \rightarrow T_{p q r s} = V_{(n)}(g)_{p \tilde{p}} V_{(n)}(g)_{q \tilde{q}}  V_{(n)}(g)^{\dagger}_{\tilde{r} r}  V_{(n)}(g)^{\dagger}_{\tilde{s} s} T_{\tilde{p} \tilde{q} \tilde{r} \tilde{s}},
\end{eqnarray}
while satisfying the symmetric condition
\begin{eqnarray}
    T_{p q r s} = T_{q p r s} = T_{p q s r} = T_{q p s r}.
\end{eqnarray}

In order to find the dimension of the trivial representation subspace, we need to compute the character table of the tensor space $T_{p q r s}$ with a group of the normalized orthogonal basis $B_{(n)}^{(2,2)}(p^{\prime}, q^{\prime}, r^{\prime}, s^{\prime})_{p q r s} \in \{ T_{p q r s} \}$ defined as
\begin{eqnarray}
    B_{(n)}^{(2,2)}(p^{\prime}, q^{\prime}, r^{\prime}, s^{\prime})_{p q r s} =  \frac{ \delta_{p^{\prime} p} \delta_{q^{\prime} q} \delta_{r^{\prime} r} \delta_{s^{\prime} s} + \delta_{p^{\prime} q} \delta_{q^{\prime} p} \delta_{r^{\prime} r} \delta_{s^{\prime} s} + \delta_{p^{\prime} p} \delta_{q^{\prime} q} \delta_{r^{\prime} s} \delta_{s^{\prime} r} + \delta_{p^{\prime} q} \delta_{q^{\prime} p} \delta_{r^{\prime} s} \delta_{s^{\prime} r} }{\sqrt{2 \times 2}}   ,
\end{eqnarray}
 where the $\frac{1}{\sqrt{2 \times 2}}$ factor is a normalized factor. The ``trace'' of the transformation under the action of $g \in \text{SU(2)}_\text{D}$ can then be computed by
\begin{eqnarray}
    \sum_{p^{\prime} \leq q^{\prime}, r^{\prime} \leq s^{\prime}} B_{(n)}^{(2,2)}(p^{\prime}, q^{\prime}, r^{\prime}, s^{\prime})_{p q r s}^* V_{(n)}(g)_{p \tilde{p}} V_{(n)}(g)_{q \tilde{q}}  V_{(n)}(g)^{\dagger}_{\tilde{r} r}  V_{(n)}(g)^{\dagger}_{\tilde{s} s} B_{(n)}^{(2,2)}(p^{\prime}, q^{\prime}, r^{\prime}, s^{\prime})_{\tilde{p} \tilde{q} \tilde{r} \tilde{s}}. \label{TraceFormula}
\end{eqnarray}
Since all rotation transformations with the same angle $\omega$ form the same equivalence class, the simplest way to compute the character of this equivalence class is to adopt
\begin{eqnarray}
    V_{(n)} (e^{i \omega J_z})_{m \tilde{m}} = \delta_{m \tilde{m}} e^{i m \omega}.
\end{eqnarray}
Taking this into Eq.~\eqref{TraceFormula}, we acquire
\begin{eqnarray}
    \chi_{(n)}^{(2,2)}(\omega) = \sum_{p \leq q, r \leq s} e^{i (p + q - r - s) \omega}, \label{CharacterFormula}
\end{eqnarray}
where $\chi_{(n)}^{(2,2)}(\omega)$ is the character of the equivalence class $\omega$ under the tensor representation $T_{p q r s}$.

Then, the multiplicity of the trivial representation is given by the continuous inner product of the character tables
\begin{eqnarray}
    N_{(n)}^{(2,2)} &=& \frac{1}{4 \pi^2} \int_{-\pi}^{\pi} d \phi \int_{0}^{\pi} d \theta \sin \theta  \int_{0}^{2 \pi} d \omega \sin^2\left( \frac{\omega}{2} \right) \chi_1^*(\omega) \chi_{(n)}^{(2,2)}(\omega) \nonumber \\
    &=& \frac{1}{\pi} \int_{0}^{2 \pi} d \omega \left( \frac{1-\cos \omega}{2} \right) \chi_{(n)}^{(2,2)} (\omega), \label{TrivialDimension}
\end{eqnarray}
where $\chi_1^*(\omega) \equiv 1$ is the character table of the trivial representation.  Substituting Eq.~\eqref{CharacterFormula} into Eq.~\eqref{TrivialDimension}, we have
\begin{eqnarray}
    N_{(n)}^{(2,2)} = N_{(n)0}^{(2,2)} - N_{(n)1}^{(2,2)},
\end{eqnarray}
where $N_{(n)i}^{(2,2)}$ is defined by the number of all possible $-j \leq p \leq q \leq j$ and $-j \leq r \leq s \leq j$ combinations satisfying $p+q-r-s =i$.

As an example, when $n=4$ so that $j=\frac{n-1}{2}= \frac{3}{2}$, enumerating all possible $p+q-r-s =1$ and $p+q-r-s =0$ when $-j \leq p \leq q \leq j$ and $-j \leq r \leq s \leq j$ gives
\begin{eqnarray}
  1 &=& -\frac{3}{2} - \frac{1}{2} + \frac{3}{2} + \frac{3}{2} = -\frac{3}{2} + \frac{1}{2} + \frac{3}{2} + \frac{1}{2} = -\frac{3}{2} + \frac{3}{2} + \frac{3}{2} - \frac{1}{2} = -\frac{3}{2} + \frac{3}{2} + \frac{1}{2} + \frac{1}{2} \nonumber \\
  &=& -\frac{1}{2} - \frac{1}{2} + \frac{3}{2} + \frac{1}{2} = -\frac{1}{2} + \frac{1}{2} + \frac{3}{2} - \frac{1}{2} = -\frac{1}{2} + \frac{1}{2} + \frac{1}{2} + \frac{1}{2} = -\frac{1}{2} + \frac{3}{2} + \frac{3}{2} - \frac{3}{2} \nonumber \\
  &=& -\frac{1}{2} + \frac{3}{2} + \frac{1}{2} - \frac{1}{2} = \frac{1}{2} + \frac{1}{2} + \frac{3}{2} - \frac{3}{2} = \frac{1}{2} + \frac{1}{2} + \frac{1}{2} - \frac{1}{2} = \frac{1}{2} + \frac{3}{2} + \frac{1}{2} - \frac{3}{2} \nonumber \\
  &=& \frac{1}{2} + \frac{3}{2} - \frac{1}{2} - \frac{1}{2} = \frac{3}{2} + \frac{3}{2} - \frac{1}{2} - \frac{3}{2}, \label{pqrs_1}\\
  0 &=& -\frac{3}{2} - \frac{3}{2} + \frac{3}{2} + \frac{3}{2} = -\frac{3}{2} - \frac{1}{2} + \frac{3}{2} + \frac{1}{2} = -\frac{3}{2} + \frac{1}{2} + \frac{3}{2} - \frac{1}{2} = -\frac{3}{2} + \frac{1}{2} + \frac{1}{2} + \frac{1}{2} \nonumber \\
  &=& -\frac{3}{2} + \frac{3}{2} + \frac{3}{2} - \frac{3}{2} = -\frac{3}{2} + \frac{3}{2} + \frac{1}{2} - \frac{1}{2} = -\frac{1}{2} - \frac{1}{2} + \frac{3}{2} - \frac{1}{2} = -\frac{1}{2} - \frac{1}{2} + \frac{1}{2} + \frac{1}{2} \nonumber \\
  &=& -\frac{1}{2} + \frac{1}{2} + \frac{3}{2} - \frac{3}{2} = -\frac{1}{2} + \frac{1}{2} + \frac{1}{2} - \frac{1}{2} = -\frac{1}{2} + \frac{3}{2} + \frac{1}{2} - \frac{3}{2} = -\frac{1}{2} + \frac{3}{2} - \frac{1}{2} - \frac{1}{2} \nonumber \\
  &=& \frac{1}{2} + \frac{1}{2} + \frac{1}{2} - \frac{3}{2} = \frac{1}{2} + \frac{1}{2} - \frac{1}{2} - \frac{1}{2} = \frac{1}{2} + \frac{3}{2} - \frac{1}{2} - \frac{3}{2} = \frac{3}{2} + \frac{3}{2} - \frac{3}{2} - \frac{3}{2}.\label{pqrs_0}
\end{eqnarray}
Equation ~\eqref{pqrs_1} gives $N_{(7)1}^{(2,2)}=14$, and Eq.~\eqref{pqrs_0} gives $N_{(4)1}^{(2,2)}=16$, so $N_{(4)}^{(2,2)}=2$. That is to say, there are only two independent combinations of $\Delta_p \Delta_q \Delta_r^* \Delta_s^*$ to form the $\text{SU(2)}_\text{D}$ singlets.  Then the task is to look into the following quartic polynomial chains:
\begin{eqnarray}
    & & \Phi_{(n)}^{\dagger} \Phi_{(n)} \Phi_{(n)}^{\dagger} \Phi_{(n)},~~\Phi_{(n)}^{\dagger} \tau_{(n)}^a \Phi_{(n)} \Phi_{(n)}^{\dagger} \tau_{(n)}^a \Phi_{(n)},~~\Phi_{(n)}^{\dagger} \tau_{(n)}^a \tau_{(n)}^b \Phi_{(n)} \Phi_{(n)}^{\dagger} \tau_{(n)}^a \tau_{(n)}^b \Phi_{(n)}, \label{QuarticPolynomials} \\
    & & \Phi_{(n)}^{\dagger} \tau_{(n)}^a \tau_{(n)}^b \tau_{(n)}^c \Phi_{(n)} \Phi_{(n)}^{\dagger} \tau_{(n)}^a \tau_{(n)}^b \tau_{(n)}^c \Phi_{(n)}, ~~\Phi_{(n)}^{\dagger} \tau_{(n)}^a \tau_{(n)}^b \tau_{(n)}^c \tau_{(n)}^d \Phi_{(n)} \Phi_{(n)}^{\dagger} \tau_{(n)}^a \tau_{(n)}^b \tau_{(n)}^c \tau_{(n)}^d \Phi_{(n)},~\dots \nonumber 
\end{eqnarray}
to select two linearly independent polynomials. The elements of the SU(2) $n$-dimension irreducible representation generator matrices $\tau^{a,b,c,\dots}_{(n)}$, with $a,b,c,\dots=1,2,3$, are illustrated in Eqs.~\eqref{JpJm} and Eqs.~\eqref{J123}. Selecting the linearly independent polynomials from Eq.~\eqref{QuarticPolynomials} can be easily accomplished by expanding one and checking the ranks of the coefficient matrices composed by different polynomial combinations. The simplest result of the independent quartic terms when $n=4$ is given by
\begin{eqnarray}
    \Phi_{(4)}^{\dagger} \Phi_{(4)} \Phi_{(4)}^{\dagger} \Phi_{(4)},~~\Phi_{(4)}^{\dagger} \tau_{(4)}^a \Phi_{(4)} \Phi_{(4)}^{\dagger} \tau_{(4)}^a \Phi_{(4)}.
\end{eqnarray}

For all the other $\Delta_p \Delta_q \Delta_r \Delta_s$, $\Delta_p \Delta_q \Delta_r \Delta_s^{*}$ combinations under any $n$-dimensional irreducible representation, one can modify and follow similar processes from Eqs.~\eqref{CharacterFormula} to \eqref{QuarticPolynomials} to write the character table of a tensor representation with a specific symmetry pattern of the indices, and then compute its inner product with the trivial representation character table. Finally, the multiplicity of the trivial representation $N_{(n)}^{(a,b)}=N_{(n)0}^{(a,b)}-N_{(n)1}^{(a,b)}$ is acquired where $(a,b)$ corresponds to the superscripts appearing in the coefficients in Eq.~\eqref{EffectivePotential_MassSimplified}. Enumeration of the fields with the generator matrices contracted inside, like Eq.~\eqref{QuarticPolynomials}, should also be listed. Defining the charge conjugated field
\begin{eqnarray}
    \Phi_{(n)}^C = T_{(n)} \Phi_{(n)}^*,
\end{eqnarray}
the following quartic polynomials
\begin{eqnarray}
    & & \Phi_{(n)}^{C \dagger} \Phi_{(n)} \Phi_{(n)}^{C \dagger} \Phi_{(n)},~~\Phi_{(n)}^{C \dagger} \tau_{(n)}^a \Phi_{(n)} \Phi_{(n)}^{C \dagger} \tau_{(n)}^a \Phi_{(n)},~~\Phi_{(n)}^{C \dagger} \tau_{(n)}^a \tau_{(n)}^b \Phi_{(n)} \Phi_{(n)}^{C \dagger} \tau_{(n)}^a \tau_{(n)}^b \Phi_{(n)}, \label{QuarticPolynomials_40} \\
    & & \Phi_{(n)}^{C \dagger} \tau_{(n)}^a \tau_{(n)}^b \tau_{(n)}^c \Phi_{(n)} \Phi_{(n)}^{C \dagger} \tau_{(n)}^a \tau_{(n)}^b \tau_{(n)}^c \Phi_{(n)}, ~~\Phi_{(n)}^{C \dagger} \tau_{(n)}^a \tau_{(n)}^b \tau_{(n)}^c \tau_{(n)}^d \Phi_{(n)} \Phi_{(n)}^{C \dagger} \tau_{(n)}^a \tau_{(n)}^b \tau_{(n)}^c \tau_{(n)}^d \Phi_{(n)},~\dots \nonumber 
\end{eqnarray}
and
\begin{eqnarray}
    & & \Phi_{(n)}^{\dagger} \Phi_{(n)} \Phi_{(n)}^{C \dagger} \Phi_{(n)},~~\Phi_{(n)}^{\dagger} \tau_{(n)}^a \Phi_{(n)} \Phi_{(n)}^{C \dagger} \tau_{(n)}^a \Phi_{(n)},~~\Phi_{(n)}^{\dagger} \tau_{(n)}^a \tau_{(n)}^b \Phi_{(n)} \Phi_{(n)}^{C \dagger} \tau_{(n)}^a \tau_{(n)}^b \Phi_{(n)}, \label{QuarticPolynomials_31} \\
    & & \Phi_{(n)}^{\dagger} \tau_{(n)}^a \tau_{(n)}^b \tau_{(n)}^c \Phi_{(n)} \Phi_{(n)}^{C \dagger} \tau_{(n)}^a \tau_{(n)}^b \tau_{(n)}^c \Phi_{(n)}, ~~\Phi_{(n)}^{\dagger} \tau_{(n)}^a \tau_{(n)}^b \tau_{(n)}^c \tau_{(n)}^d \Phi_{(n)} \Phi_{(n)}^{C \dagger} \tau_{(n)}^a \tau_{(n)}^b \tau_{(n)}^c \tau_{(n)}^d \Phi_{(n)},~\dots \nonumber 
\end{eqnarray}
are the possible singlet contractions of $\Delta_p \Delta_q \Delta_r \Delta_s$ and $\Delta_p \Delta_q \Delta_r \Delta_s^{*}$ respectively. Utilizing computer to expand each of the quartic contractions and selecting the anterior independent polynomials up to the numbers of $N_{(n)}^{(4,0)}$ and $N_{(n)}^{(3,1)}$, the complete set of independent $\Delta_p \Delta_q \Delta_r \Delta_s$ and $\Delta_p \Delta_q \Delta_r \Delta_s^{*}$ contractions are eventually found.

For odd-$n$ and even $j$, the multiplicity of the trivial representation of $\Delta_p \Delta_q \Delta_r$ is calculated to be one. Since there is no simple contraction of three $\Phi_{(n)}^{(\dagger)}$'s with the generator matrices inserted inside, we could only resort to the Clebsch–Gordan coefficients. Fortunately, it is easy, since the only way to combine three $n$-dimensional representations into a singlet is to contract two of them into a $n$-dimensional representation and then pair with the rest of the $n$-dimensional representation to form a singlet. Therefore, we have for the odd $n$-dimensional representations $\Phi_{(n)}$ with $n=2 j + 1$,
\begin{eqnarray}
    \kappa_{p q r}^{(3,0)} \Delta_p \Delta_q \Delta_r \propto \xi_{p q (-r)}^{jjj} \Delta_p \Delta_q \Delta_r,
\end{eqnarray}
where $\xi_{p q r}^{j_1 j_2 j_3}$ are the Clebsch–Gordan coefficients defined by
\begin{eqnarray}
    \xi_{p q (-r)}^{j_1 j_2 j_3} =  \langle j_3,m_3=r | (|j_1, m_1=p\rangle \otimes |j_2, m_2=q\rangle),
\end{eqnarray}
for any integer $-|j_1-j_2| \leq j_3 \leq j_1+j_2$.

As a summary, we list all the possible renormalizable independent contractions of the $\Phi_{(n)}^{(\dagger)}$ fields up to $n=21$ in Tables \ref{tab:odd_operators} and \ref{tab:even_operators} to close this section.
\begin{table}[h]
    \centering
    \setlength\tabcolsep{0.5em}
    \renewcommand{\arraystretch}{1}
    \scriptsize
    \begin{tabular}{|c|c|c|}
     \hline
     $j$ & $\Delta_p \Delta_q \Delta_r$ & $\Delta_p \Delta_q \Delta_r \Delta_s$ \\ \hline
     \multirow{2}{*}{$1$} & $0$ & $1$\\ \cline{2-3} 
     &  & $(\Phi^{\dag} \Phi)^2$\\ \hline
     \multirow{2}{*}{$2$} &  $1$ & $1$\\ \cline{2-3} 
     & $\xi_{(5)}^{p q r} \Delta_p \Delta_q \Delta_r$ &
     $ (\Phi^{\dag} \Phi)^2$ \\ \hline
     \multirow{2}{*}{$3$} & $0$ & $2$\\ \cline{2-3}
     &  & $ (\Phi^{\dag} \Phi)^2, (\Phi^{\dag} \tau^a \tau^b
     \Phi)^2$\\ \hline
     \multirow{2}{*}{$4$} & $1$ & $2$\\ \cline{2-3}
     & $\xi_{(9)}^{p q r} \Delta_p \Delta_q \Delta_r$ &
     $(\Phi^{\dag} \Phi)^2, (\Phi^{\dag} \tau^a \tau^b \Phi)^2$\\ \hline
     \multirow{2}{*}{$5$} & $0$ & $2$\\ \cline{2-3}
     &  & $(\Phi^{\dag} \Phi)^2, (\Phi^{\dag} \tau^a \tau^b
     \Phi)^2$\\ \hline
     \multirow{2}{*}{$6$} & $1$ & $3$\\ \cline{2-3}
     & $\xi_{(13)}^{p q r} \Delta_p \Delta_q \Delta_r$ & $(\Phi^{\dag} \Phi)^2, (\Phi^{\dag} \tau^a \tau^b \Phi)^2, (\Phi^{\dag} \tau^a \tau^b \tau^c \tau^d \Phi )^2$\\ \hline
     \multirow{2}{*}{$7$} & $0$ & $3$\\ \cline{2-3}
     &  & $(\Phi^{\dag} \Phi)^2, (\Phi^{\dag} \tau^a \tau^b \Phi)^2, (\Phi^{\dag} \tau^a \tau^b \tau^c \tau^d \Phi )^2$\\ \hline
     \multirow{2}{*}{$8$} & $1$ & $3$\\ \cline{2-3}
     & $\xi_{(17)}^{p q r} \Delta_p \Delta_q \Delta_r$ & $(\Phi^{\dag} \Phi)^2, (\Phi^{\dag} \tau^a \tau^b \Phi)^2, (\Phi^{\dag} \tau^a \tau^b \tau^c \tau^d \Phi )^2$\\ \hline
     \multirow{2}{*}{$9$} & $0$ & $4$\\ \cline{2-3}
     &  & $(\Phi^{\dag} \Phi)^2, (\Phi^{\dag} \tau^a \tau^b \Phi)^2, (\Phi^{\dag} \tau^a \tau^b \tau^c \tau^d \Phi )^2, (\Phi^{\dag} \tau^a \tau^b \tau^c \tau^d \tau^e \tau^f \Phi )^2$\\ \hline
     \multirow{2}{*}{$10$} & $1$ & $4$\\ \cline{2-3}
     & $\xi_{(21)}^{p q r} \Delta_p \Delta_q \Delta_r$ & $(\Phi^{\dag} \Phi)^2, (\Phi^{\dag} \tau^a \tau^b \Phi)^2, (\Phi^{\dag} \tau^a \tau^b \tau^c \tau^d \Phi )^2, (\Phi^{\dag} \tau^a \tau^b \tau^c \tau^d \tau^e \tau^f \Phi )^2$\\ \hline
\end{tabular}
    \caption{Independent and renormalizable operators for odd dimensions $n=3,5,7,\cdots,21$. The representation dimension subscripts $n=2j+1$ are omitted. The bilinear $\Delta_p \Delta_q$ forms only one gauge invariant operator $\Phi^\dagger \Phi$ in any $n$.}
    \label{tab:odd_operators}
\end{table}

\begin{table}[h]
    \centering
    \setlength\tabcolsep{0.5em}
    \renewcommand{\arraystretch}{1}
    \scriptsize
    \begin{tabular}{|c|c|c|c|}
     \hline
     $j$ & $\Delta_p \Delta_q \Delta_r \Delta_s$ & $\Delta_p \Delta_q \Delta_r \Delta_s^*$ & $\Delta_p \Delta_q \Delta_r^* \Delta_s^*$\\ \hline
     \multirow{2}{*}{$\frac{1}{2}$} & $0$ & $0$ & $1$\\ \cline{2-4} 
     & & & $(\Phi^{\dag} \Phi)^2$ \\ \hline
     \multirow{2}{*}{$\frac{3}{2}$} & $1$ & $1$ & $2$\\ \cline{2-4} 
     & $( \Phi^{C\dagger} \tau^a \Phi )^2$ &
     $\Phi^{C\dagger} \tau^a \Phi \Phi^{\dag} \tau^a \Phi$
     & $( \Phi^{\dag} \Phi)^2, (\Phi^{\dag} \tau^a \Phi )^2$ \\ \hline
     \multirow{2}{*}{$\frac{5}{2}$} & $1$ & $1$ & $3$\\ \cline{2-4}
     & $( \Phi^{C\dagger} \tau^a \Phi )^2$ &
     $\Phi^{C\dagger} \tau^a \Phi \Phi^{\dag} \tau^a \Phi$
     & $( \Phi^{\dag} \Phi)^2, (\Phi^{\dag} \tau^a \Phi )^2,(\Phi^{\dag} \tau^a \tau^b \Phi)^2$ \\ \hline
     \multirow{2}{*}{$\frac{7}{2}$} & $1$ & $1$ & $4$\\ \cline{2-4}
     & $( \Phi^{C\dagger} \tau^a \Phi )^2$ &
     $\Phi^{C\dagger} \tau^a \Phi \Phi^{\dag} \tau^a \Phi$ & \makecell[c]{$( \Phi^{\dag} \Phi)^2, (\Phi^{\dag} \tau^a \Phi )^2, (\Phi^{\dag} \tau^a \tau^b \Phi)^2, (\Phi^{\dag} \tau^a \tau^b \tau^c \Phi)^2$}\\ \hline
     \multirow{2}{*}{$\frac{9}{2}$} & $2$ & $2$ & $5$\\ \cline{2-4}
     & \makecell[c]{$( \Phi^{C\dagger} \tau^a \Phi )^2,$ \\
     $( \Phi^{C\dagger} \tau^a  \tau^b  \tau^c \Phi )^2$} & \makecell[c]{$\Phi^{C\dagger} \tau^a \Phi \Phi^{\dag} \tau^a \Phi,$\\
     $\Phi^{C\dagger} \tau^a \tau^b \tau^c \Phi \Phi^{\dag} \tau^a \tau^b \tau^c \Phi$}
     & \makecell[c]{$( \Phi^{\dag} \Phi)^2, (\Phi^{\dag} \tau^a \Phi )^2,(\Phi^{\dag} \tau^a \tau^b \Phi)^2, (\Phi^{\dag} \tau^a \tau^b \tau^c \Phi)^2,$\\
     $(\Phi^{\dag} \tau^a \tau^b \tau^c \tau^d \Phi)^2$} \\ \hline
     \multirow{2}{*}{$\frac{11}{2}$} & $2$ & $2$ & $6$\\ \cline{2-4}
     & \makecell[c]{$( \Phi^{C\dagger} \tau^a \Phi )^2,$ \\
     $( \Phi^{C\dagger} \tau^a  \tau^b  \tau^c \Phi )^2$} & \makecell[c]{$\Phi^{C\dagger} \tau^a \Phi \Phi^{\dag} \tau^a \Phi$,\\
     $\Phi^{C\dagger} \tau^a \tau^b \tau^c \Phi \Phi^{\dag} \tau^a \tau^b \tau^c \Phi$}
     & \makecell[c]{$( \Phi^{\dag} \Phi)^2, (\Phi^{\dag} \tau^a \Phi )^2, \Phi^{\dag} \tau^a \tau^b \Phi)^2, (\Phi^{\dag} \tau^a \tau^b \tau^c \Phi)^2,$\\
     $ (\Phi^{\dag} \tau^a \tau^b \tau^c \tau^d \Phi)^2, (\Phi^{\dag} \tau^a \tau^b \tau^c \tau^d \tau^e \Phi)^2$} \\ \hline
     \multirow{2}{*}{$\frac{13}{2}$} & $2$ & $2$ & $7$\\ \cline{2-4}
     & \makecell[c]{$( \Phi^{C\dagger} \tau^a \Phi )^2,$ \\
     $( \Phi^{C\dagger} \tau^a  \tau^b  \tau^c \Phi )^2$} & \makecell[c]{$\Phi^{C\dagger} \tau^a \Phi \Phi^{\dag} \tau^a \Phi$,\\
     $\Phi^{C\dagger} \tau^a \tau^b \tau^c \Phi \Phi^{\dag} \tau^a \tau^b \tau^c \Phi$}
     & \makecell[c]{$( \Phi^{\dag} \Phi)^2, (\Phi^{\dag} \tau^a \Phi )^2, (\Phi^{\dag} \tau^a \tau^b \Phi)^2, (\Phi^{\dag} \tau^a \tau^b \tau^c \Phi)^2,$\\
     $(\Phi^{\dag} \tau^a \tau^b \tau^c \tau^d \Phi)^2, (\Phi^{\dag} \tau^a \tau^b \tau^c \tau^d \tau^e \Phi)^2,$\\
     $(\Phi^{\dag} \tau^a \tau^b \tau^c \tau^d \tau^e \tau^f \Phi)^2$} \\ \hline
     \multirow{2}{*}{$\frac{15}{2}$} & $3$ & $3$ & $8$\\ \cline{2-4}
     & \makecell[c]{$( \Phi^{C\dagger} \tau^a \Phi )^2,$ \\
     $( \Phi^{C\dagger} \tau^a  \tau^b  \tau^c \Phi )^2,$\\
     $( \Phi^{C\dagger} \tau^a  \tau^b  \tau^c \tau^d \tau^e \Phi )^2$} & \makecell[c]{$\Phi^{C\dagger} \tau^a \Phi \Phi^{\dag} \tau^a \Phi$,\\
     $\Phi^{C\dagger} \tau^a \tau^b \tau^c \Phi \Phi^{\dag} \tau^a \tau^b \tau^c \Phi$\\
      $\Phi^{C\dagger} \tau^a \tau^b \tau^c \tau^d \tau^e \Phi \Phi^{\dag} \tau^a \tau^b \tau^c \tau^d \tau^e \Phi$}
     & \makecell[c]{$( \Phi^{\dag} \Phi)^2, (\Phi^{\dag} \tau^a \Phi )^2, (\Phi^{\dag} \tau^a \tau^b \Phi)^2, (\Phi^{\dag} \tau^a \tau^b \tau^c \Phi)^2,$\\
     $(\Phi^{\dag} \tau^a \tau^b \tau^c \tau^d \Phi)^2, (\Phi^{\dag} \tau^a \tau^b \tau^c \tau^d \tau^e \Phi)^2,$\\
     $(\Phi^{\dag} \tau^a \tau^b \tau^c \tau^d \tau^e \tau^f \Phi)^2,(\Phi^{\dag} \tau^a \tau^b \tau^c \tau^d \tau^e \tau^f \tau^g \Phi)^2$} \\ \hline
     \multirow{2}{*}{$\frac{17}{2}$} & $3$ & $3$ & $9$\\ \cline{2-4}
     & \makecell[c]{$( \Phi^{C\dagger} \tau^a \Phi )^2,$ \\
     $( \Phi^{C\dagger} \tau^a  \tau^b  \tau^c \Phi )^2,$\\
     $( \Phi^{C\dagger} \tau^a  \tau^b  \tau^c \tau^d \tau^e \Phi )^2$} & \makecell[c]{$\Phi^{C\dagger} \tau^a \Phi \Phi^{\dag} \tau^a \Phi$,\\
     $\Phi^{C\dagger} \tau^a \tau^b \tau^c \Phi \Phi^{\dag} \tau^a \tau^b \tau^c \Phi$\\
      $\Phi^{C\dagger} \tau^a \tau^b \tau^c \tau^d \tau^e \Phi \Phi^{\dag} \tau^a \tau^b \tau^c \tau^d \tau^e \Phi$}
     & \makecell[c]{$( \Phi^{\dag} \Phi)^2, (\Phi^{\dag} \tau^a \Phi )^2, (\Phi^{\dag} \tau^a \tau^b \Phi)^2, (\Phi^{\dag} \tau^a \tau^b \tau^c \Phi)^2,$\\
     $(\Phi^{\dag} \tau^a \tau^b \tau^c \tau^d \Phi)^2, (\Phi^{\dag} \tau^a \tau^b \tau^c \tau^d \tau^e \Phi)^2,$\\
     $(\Phi^{\dag} \tau^a \tau^b \tau^c \tau^d \tau^e \tau^f \Phi)^2, (\Phi^{\dag} \tau^a \tau^b \tau^c \tau^d \tau^e \tau^f \tau^g \Phi)^2,$\\
     $(\Phi^{\dag} \tau^a \tau^b \tau^c \tau^d \tau^e \tau^f \tau^g \tau^h \Phi)^2$} \\ \hline
     \multirow{2}{*}{$\frac{19}{2}$} & $3$ & $3$ & $10$\\ \cline{2-4}
     & \makecell[c]{$( \Phi^{C\dagger} \tau^a \Phi )^2,$ \\
     $( \Phi^{C\dagger} \tau^a  \tau^b  \tau^c \Phi )^2,$\\
     $( \Phi^{C\dagger} \tau^a  \tau^b  \tau^c \tau^d \tau^e \Phi )^2$} & \makecell[c]{$\Phi^{C\dagger} \tau^a \Phi \Phi^{\dag} \tau^a \Phi$,\\
     $\Phi^{C\dagger} \tau^a \tau^b \tau^c \Phi \Phi^{\dag} \tau^a \tau^b \tau^c \Phi$\\
      $\Phi^{C\dagger} \tau^a \tau^b \tau^c \tau^d \tau^e \Phi \Phi^{\dag} \tau^a \tau^b \tau^c \tau^d \tau^e \Phi$}
     & \makecell[c]{$( \Phi^{\dag} \Phi)^2, (\Phi^{\dag} \tau^a \Phi )^2, (\Phi^{\dag} \tau^a \tau^b \Phi)^2, (\Phi^{\dag} \tau^a \tau^b \tau^c \Phi)^2,$\\
     $(\Phi^{\dag} \tau^a \tau^b \tau^c \tau^d \Phi)^2, (\Phi^{\dag} \tau^a \tau^b \tau^c \tau^d \tau^e \Phi)^2,$\\
     $(\Phi^{\dag} \tau^a \tau^b \tau^c \tau^d \tau^e \tau^f \Phi)^2, (\Phi^{\dag} \tau^a \tau^b \tau^c \tau^d \tau^e \tau^f \tau^g \Phi)^2,$\\
     $(\Phi^{\dag} \tau^a \tau^b \tau^c \tau^d \tau^e \tau^f \tau^g \tau^h \Phi)^2, (\Phi^{\dag} \tau^a \tau^b \tau^c \tau^d \tau^e \tau^f \tau^g \tau^h \tau^i \Phi)^2$} \\ \hline
    \end{tabular}
    \caption{Independent and renormalizable operators for even dimensions $n=2,4,6,\cdots,20$. The representation dimension subscripts $n=2j+1$ are omitted. The bilinear $\Delta_p \Delta_q^*$ forms only one gauge invariant operator $\Phi^\dagger \Phi$ in any $n$. The trilinear $\Delta_p \Delta_q \Delta_r$ and $\Delta_p \Delta_q \Delta_r^*$ cannot form a singlet in an even dimension.}
    \label{tab:even_operators}
\end{table}




\section{Vacuum Patterns with a Remnant Discrete Symmetry} \label{VEVPatternSection}

For an $n$-dimension representation of the SU(2) group, some discrete symmetry group might remain so that the vacuum is invariant under it. The authors of Ref.~\cite{Etesi:1996urw} enumerated all possible such discrete groups for odd $n$-dimensional multiplets up to $n=59$ by computing the product of the character tables. What they have done is to try all known point groups in a 3-dimensional space to see whether some trivial representations arise within the $n$-dimensional scalar multiplets. However, the problem still remains for an even $n$-dimensional representation when $j$ is a half-integer, since the multiplet does not restore to its original form after a $2 \pi$-rotation along any axis. Another thing that we want to do beyond the Ref.~\cite{Etesi:1996urw} is to show a systematic algorithm for finding the detailed formats of the VEV-patterns invariant under the discrete groups. In this section, we try up to $n=20$ to accomplish two goals:
\begin{itemize}
    \item To write all possible VEV configurations which are invariant under the transformations of the remained discrete groups.
    \item To show that only the $Z_N$ group can remain when an even $n$-dimensional multiplet is adopted.
\end{itemize}

For an $n$-dimensional multiplet $\Phi_{(n)}$, let us assume it breaks the $\text{SU(2)}_\text{D}$ symmetry into a nontrivial discrete symmetry $G$ by a vacuum configuration $\vec{v}_{(n)}$. Recall that any nontrivial discrete symmetry should include at least one $Z_N$ subgroup, without loss of the generality, we can set the axis for the $Z_N \subset G$ group to align with the $z$-axis. The elements of $\vec{v}_{(n)}$ can be denoted as 
\begin{equation}
	\vec{v}=(a_{j},a_{j-1},...,a_{-j+1},a_{-j})^T\quad\text{or}\quad \vec{v}=\sum_{m=-j}^{m=j}a_m \vec{v}_m, \label{VEV_z}
\end{equation}
with $j=(n-1)/2$. $\vec{v}_m$ are the eigenvectors of $z$-axis rotation generator $J_3$, and their corresponding expansion coefficients are denoted as $a_m$. Since the $z$-axis and the rotation axis direction overlap, the representation matrix of the generator of the $Z_N$ subgroup can be written as
\begin{equation}
	U(\vec{z},\alpha)=\text{Diag}\{e^{ij\alpha},e^{i(j-1)\alpha},...,e^{i(-j+1)\alpha},e^{i(-j)\alpha} \},
\end{equation}
where $\alpha=\frac{4\pi}{N}$ when $n$ is even and $\alpha=\frac{2\pi}{N}$ when $n$ is odd. Then the corresponding rotation operation is given by
\begin{equation}
	U(\vec{z},\alpha)\vec{v}=\sum_{m=-j}^{m=j}e^{im\alpha}a_m\vec{v}_m=\vec{v}.
\end{equation}
Then the invariance of $\vec{v}$ requires $a_m$ to vanish for all $m$ satisfying $e^{im\alpha}\neq1$. Thereby, the vacuum configuration can be written as
\begin{equation}\label{am}
	\left\{\begin{aligned}
			&\vec{v} =\sum_{m/N\in \mathbb{Z}~\&~|m|\leq j}a_m\vec{v}_m, \quad \text{when $n$ is odd}
			\\&\vec{v} =\sum_{2m/N\in \mathbb{Z} ~\&~ |m| \leq j}a_m\vec{v}_m, \quad \text{when $n$ is even}
		\end{aligned}\right.
\end{equation}
That is to say,
\begin{eqnarray}
  \left\{\begin{aligned}
    &{a_m = 0, \quad \text{for } \frac{m}{N} \notin \mathbb{Z} ~\&~ |m|\leq j}\text{ when $n$ is odd } \\
    &{a_m = 0, \quad \text{for } \frac{2m}{N} \notin \mathbb{Z} ~\&~ |m|\leq j}\text{ when $n$ is even }
  \end{aligned}\right.  \label{aConditions}
\end{eqnarray}
In this paper, we would like to discuss the nonvanishing VEV configurations, and we also avoid the situation of a remnant U(1) subgroup, so there should be at least one nonzero $|m| \leq j$ that satisfies $m/N \in \mathbb{Z}$ or $2m/N \in \mathbb{Z}$ for the odd or even $n$ respectively. This forces $N \leq j$ or $N \leq 2 j$ for the odd or even $n$ respectively.

Suppose there are other elements in $G$ except the ones in $Z_N$, i.e., $G$-$Z_N \neq \varnothing$, we can pick up another $Z_{N'} \in G$ with the rotation axis aligned with another direction denoted by $\vec{n}$. Without loss of generality, we can rotate this axis into the $x$-$z$ plane. This allows us to set a normalized axis with a form as $\vec{n}=(\sin\theta,0,\cos\theta)$. At the $J_3$ representation, diagonalizing $\vec{n} \cdot \vec{J}$ gives a group of the normalized eigenvectors
\begin{eqnarray}
    V_{(n)}(\vec{n} \cdot \vec{J}) \vec{v'}_{m}(\theta) = m \vec{v'}_{m}(\theta),\label{OtherDirectionEigens}
\end{eqnarray}
where again, $m = -j, -j+1, \dots, j-1, j$, and $\vec{v'}_{m}$'s with all $\vec{v'}^{\dagger}_{m} \vec{v'}_{m'} = \delta_{m m'}$ are the corresponding eigenstates under the $\vec{n} \cdot \vec{J}$ representation.

We then try to expand $\vec{v}$ with the basis $\vec{v'}_{m}$,
\begin{equation}
 	\vec{v}=\sum_{m=-j}^{m=j}b_{m} \vec{v'}_{m}(\theta)
\end{equation}
where $b_m$ are the coefficients. It is then easy to acquire the relationship between $a_m$ and $b_m$ by
\begin{equation}
    \vec{v'}^{\dagger}_{m'} \vec{v} =\sum_{m=-j}^{m=j} a_m \vec{v'}^{\dagger}_{m'} \vec{v}_m = \sum_{m=-j}^{m=j} b_m \vec{v'}^{\dagger}_{m'} \vec{v'}_m = \sum_{m=-j}^{m=j} b_m \delta_{m m'} = b_{m'}.
\end{equation}
Defining
\begin{equation}\label{eq:vmm}
    v_{m^{\prime} m}(\theta) = \vec{v'}^{\dagger}_{m'}(\theta) \vec{v}_{m}
\end{equation}
we then have
\begin{equation}
    b_{m^{\prime} } = \sum_{m=-j}^{m=j} a_m v_{m^{\prime} m}(\theta). \label{abRelations}
\end{equation}
Again, similar to Eq.~\eqref{am}, the invariance of the VEV under the $Z_{N'}$ symmetry requires
\begin{equation}\label{eq:ZN'bm}
	\left\{\begin{aligned}
			&\vec{v} =\sum_{m/{N^{\prime}}\in \mathbb{Z} ~\&~ |m|\leq j}b_m\vec{v'}_m, \quad \text{when $n$ is odd}
			\\&\vec{v} =\sum_{2m/{N^{\prime}}\in \mathbb{Z} ~\&~ |m| \leq j}b_m\vec{v'}_m, \quad \text{when $n$ is even}
		\end{aligned}\right. .
\end{equation}
Therefore, we can obtain the following linear equations
\begin{eqnarray}\label{bm}
  \left\{\begin{aligned}
    &{\sum_{m^{\prime}=-j}^{m^{\prime}=j}a_{m^{\prime}} v_{mm^{\prime}}(\theta) =  b_m = 0, \quad\text{for } \frac{m}{N^{\prime}} \notin \mathbb{Z} ~\&~ |m|\leq j}\text{ when $n$ is odd} \\
    &{\sum_{m^{\prime}=-j}^{m^{\prime}=j}a_{m^{\prime}} v_{mm^{\prime}}(\theta) =  b_m = 0, \quad\text{for } \frac{2m}{N^{\prime}} \notin \mathbb{Z} ~\&~ |m|\leq j}\text{ when $n$ is even}
  \end{aligned}\right. 
\end{eqnarray}
where the undetermined parameters are $a_{m^{\prime}}$ and $\theta$. If the solution space of Eq.~\eqref{bm} is nontrivial, then there exist nontrivial VEV configurations that are invariant under both $Z_N$ and $Z_{N^\prime}$. In most of the cases, these are the nontrivial VEV configurations which are invariant under the entire $G$ as well.

We have tried to enumerate all the possible VEV configurations with some remaining discrete symmetry groups for $n$ from 4 to 21. We summarize the results as follows:
\begin{itemize}
	\item When $n$ is odd, we adopt the discrete groups listed in Ref.~\cite{Etesi:1996urw} and show them in Appendix \ref{SubgroupsforSO3} 
    up to $n=21$. In this case, for the nontrivial discrete groups $G=A_4/S_4/A_5$, two generators rotating along two unparalleled axes corresponding to the two $Z_{N/N^{\prime}}$ subgroups respectively are sufficient to generate all elements in $G$, so the solution space of Eq.~\eqref{bm} is exactly the general VEV configurations. We just list the results in Section \ref{SubspaceEnumeration}.

	\item When $n$ is even, we were not able to find any information in the literature about whether there can be some discrete group $G$ other than the simplest $Z_N$ remaining. In order to answer this question, we can study the equivalent problem whether two nontrivial $Z_N$ and $Z_{N^{\prime}}$ symmetries along two different axes coexist. This requires the nonzero solution to the Eq.~\eqref{bm} for at least one specific $0<\theta < \pi$ value. The existence of such a $\theta$ is equivalent to the vanishing determinants of all largest submatrices of the coefficient matrix corresponding to the equation group Eq.~\eqref{bm} for the given $\theta$. We have calculated these determinants analytically for various $n$, and find that no such $0<\theta < \pi$ exists up to $n=20$. Therefore, when $n$ is even, the only discrete symmetries that can remain after the spontaneous breaking of $\text{SU(2)}_\text{D}$ are the $Z_N$ groups for all possible $N \leq 2 j$. Correspondingly, the general forms of the VEV configurations are $\vec{v} =\sum\limits_{2m/N\in \mathbb{Z}}a_m\vec{v}_m$, where $m$ runs all through the half-integers with the absolute values $|m| \leq j$.
\end{itemize}
 

\section{Scalar Classifications} \label{ScalarClassificationSection}
After the Higgs multiplet acquires the VEV, the vector bosons become massive by ``eating'' the Goldstone degrees of freedom. The next task is to diagonalize all the mass matrices. In this paper, usually the mass matrix of the scalar fields becomes too large for us to straightforwardly observe a simple solution, although one might expect some systematic degeneracy among different scalar combinations, cataloged by various irreducible representations of the remnant discrete symmetry. The question then becomes: is it possible to simplify the calculation by working out the detailed formats of these irreducible representations under the discrete symmetry prior to the diagonalizing processes? In order to solve it, in this section, we suggest a general algorithm to decompose the scalar multiplet into the irreducible subspaces under the discrete symmetry.

According to the Schur lemmas, no mixing arises between fields within different irreducible representations. All the fields within one subspace of one irreducible representation are mass-degenerate, so that the mass matrix within it is proportional to the identity matrix. Besides, the fields within a subspace of one irreducible representation $t$ with a multiplicity $N_t$ might mix. However only the corresponding fields representing the same component of the irreducible representation $t$ share the crossing-term of the mass matrix, so the masses of these $N_t$ multiplets will mix as a whole, giving a $N_t \times N_t$ mass matrix.

The other substantial problem is to find the minimum of the potential. Generally there are quite a number of equations indicating the minimal conditions, and it is difficult to find out whether some nontrivial VEVs can be induced. In this paper, we follow a reversed algorithm. If a particular VEV configuration does exist, then the coupling constants might be acquired by solving the minimal condition equations regarding them as unknown parameters. The problem then lies on the number of the equations. Are there too many equations for a solution of the coupling constants? Fortunately, if the VEV configuration we have selected respects some remaining symmetry, the number of the independent equations of the minimal conditions significantly reduces. This can be apprehended by realizing that each equation of the minimal condition indicates the derivative of the effective potential along a particular direction in the field space at the VEV point, and the remaining symmetry that this VEV respects transforms this direction to some others, making the derivatives along various directions equivalent. In Appendix~\ref{app:MinimalConditions}, we offer a proof that the number of independent minimal conditions is equal to or less than the multiplicity of the identical representation of the discrete subgroup appearing within the field space.


For the half-integer $j$ situations when $n$ is even, as we have mentioned before, up to $n=20$ the only discrete group which can remain from the broken $\text{SU(2)}_\text{D}$ is the $Z_N$ group, and all $Z_N$ irreducible representations are one-dimensional. Therefore, after rotating the axis of the $Z_N$ generator aligning with the $z$-axis, it is fairly easy for us to observe that each element of $\Phi_{(n)}$ becomes one irreducible subspace of $G$. When $j$ is an integer, so that $n$ is odd, discrete groups beyond $Z_N$ might appear, and at least two generators with the rotational axis unparallel to each other arise. In this section, we concentrate on this situation.

The character table algorithm is again utilized. Suppose $\text{SU(2)}_\text{D}$ is broken with a discrete $G$ remaining, the multiplicity of a irreducible representation $t$ in $\Phi_{(n)}$ is calculated by
\begin{equation}\label{chi}
	n_{t, \Phi_{(n)}}=\frac{1}{|G|}\sum_{g \in G} \bar{\chi}_{t}(g) \chi_{\Phi_{(n)}}(g),
\end{equation}
where $\chi_{t}(g)$ is the trace of the matrix of element $g \in G$ in the irreducible representation $t$ and $\chi_{\Phi_{(n)}}(g)$ is the trace of the matrix in the $\Phi_{(n)}$ space. The character tables of the irreducible representations of the discrete groups appearing in this paper will be listed in Appendix \ref{charactertableandgenerators}. To compute $\chi_{\Phi_{(n)}}(g)$, if we regard the rotation direction of $g$ as the $z$-axis, we have
\begin{eqnarray}
    V_{(n)}(g) = \text{Diag}\{e^{i j \varphi_g}, e^{i (j-1) \varphi_g}, \dots, e^{i (-j+1) \varphi_g}, e^{i (-j) \varphi_g}\},
\end{eqnarray}
where $\varphi_g$ is the rotation angle of $g$. Then,
\begin{eqnarray}
    \chi_{\Phi_{(n)}}(g) = e^{i j \varphi_g}+e^{i (j-1) \varphi_g}+\dots+e^{i (-j+1) \varphi_g}+e^{i (-j) \varphi_g} = \frac{\sin[(j+1/2)\varphi_g]}{\sin(\varphi_g/2)}.
\end{eqnarray}
For the irreducible representation $t$, if $n_{t, \Phi_{(n)}}\neq 0$, then we suggest the following processes to give a detailed format of the subspace(s) under the $J_3$ representation.

Consider a projection operator $P_t$ which projects $\Phi_{(n)}$ into a subspace which corresponds to the irreducible representation $t$ of the discrete group $G$,
\begin{equation}\label{Pnm}
	P_t	\Phi_{(n)}=	\Phi_t,
\end{equation}
where $\Phi_t$ is a $\mathrm{Dim}(t)$-dimensional vector and $\mathrm{Dim}(t)$ is the dimension of the irreducible representation $t$. Therefore, $P_t$ is a $\mathrm{Dim}(t) \times n$ matrix. We would like the inner product of the $\Phi_{(n)}$ space be inherited by $\Phi_t$, so we have
\begin{eqnarray}
    P_t P_t^{\dagger} = I_{\mathrm{Dim}(t) \times \mathrm{Dim}(t)}. \label{Normalized}
\end{eqnarray}
In this case, the columns of $P_t^{\dagger}$ can be regarded as the normalized orthogonal basis within the $\Phi_{(n)}$ space, which are projected by the projector operator $P_t$  into the simplest $(\vec{t}_{1}, \vec{t}_{2}, ..., \vec{t}_{\text{Dim}(t)})$ basis in the $\Phi_t$ space where $(\vec{t}_{i})_j = \delta_{i j}$.

We choose two (or more) generators of $S$ and $T$ from $G$. Their rotational axes should not be parallel. Their corresponding matrices under the $\Phi_{(n)}$ representation are formally written as
\begin{eqnarray}
    V_{(n)}(S)=e^{i V_{(n)}(\vec{J} \cdot \vec{n}_S) \phi_S}, ~~~V_{(n)}(T)=e^{i V_{(n)}(\vec{J} \cdot \vec{n}_T) \phi_T}, \label{S_T_Form}
\end{eqnarray}
where $\vec{n}_{S,T}$ are the normalized directions of the rotational axes of $S$ and $T$ respectively, while $\phi_{S,T}$ are the corresponding rotational angles. The detailed forms of Eq.~\eqref{S_T_Form} can be simply calculated by plugging in $V_{(n)}(\vec{J})$ according to Eqs.~\eqref{JpJm} and \eqref{J123}.

Since the projection operation keeps the representation structures, for both $S$ and $T$, we have 
\begin{equation}\label{PSP}
	P_t	V_{(n)}(S) \Phi_{(n)}=	S_t \Phi_t = S_t P_t \Phi_{(n)}, ~~P_t	V_{(n)}(T) \Phi_{(n)}=	T_t P_t	\Phi_{(n)}.
\end{equation}
%
Note that Eq.~\eqref{PSP} holds for any form of $\Phi_{(n)}$, so that
 \begin{equation}\label{PTP}
 	P_t V_{(n)}(S) = S_t P_t, ~~P_t V_{(n)}(T) = T_t P_t. 
 \end{equation}


In principle, solving Eqs.~\eqref{Normalized} and \eqref{S_T_Form} regarding the elements of $P_t$ as unknown parameters can give detailed results of the subspace of the irreducible representation $t$. However, Eq.~\eqref{Normalized} is a quadratic equation, making the analytic solutions too complicated to be acquired. In order to resolve this problem, we can again select $\vec{n}_S$ along the $z$-direction, so that $V_{(n)}(S)$ in Eq.~\eqref{S_T_Form} becomes diagonal. If we again select a group of basis in the $t$-representation to let $S_t$ become diagonal, we therefore immediately realize that $P_t$ can only connect the corresponding basis to the same eigenvalues within both $V_{(n)}(S)$ and $S_t$, rendering most of the $P_t$ elements zero. Usually, after this step, one can find that now all the row vectors of $P_t$ automatically become orthogonal with each other, and so that all the homogeneous functions induced by Eq.~\eqref{Normalized} can be neglected.

Then, in principle, $P_t V_{(n)}(T) = T_t P_t$ provides $\mathrm{Dim}(t) \times n$ homogeneous equations. One can then find the solution space of the remaining nonzero $P_t$ elements and finally normalize each of the $P_t$ row vectors to fit the non-homogeneous equations in Eq.~\eqref{Normalized}, and eventually, the detailed format of the irreducible representation $t$ subspace denoted in Eq.~\eqref{Pnm} is accomplished.

In this paper, we provide an algorithm which is a little bit different to further diminish the number of equations. If we diagonalize $T_t$ and $V_{(n)}(T)$ at the same time,
\begin{eqnarray}
    T^{\prime}_{t} = Q^{\dagger} T_{t} Q, ~~T_{(n)}^{\prime} = R^{\dagger} V_{(n)}(T) R \label{DiagonalizeT}
\end{eqnarray}
satisfying $Q^{\dagger} Q = Q Q^{\dagger} = I_{\mathrm{dim}(t) \times \mathrm{dim}(t)}$ and $R^{\dagger} R = R R^{\dagger} = I_{n \times n}$, and $T^{\prime}_{t}$, $T_{(n)}^{\prime}$ are the diagonalized matrices. $P_t V_{(n)}(T) = T_t P_t$ in Eq.~\eqref{PTP} then can be written with
\begin{eqnarray}
    P_t R T_{(n)}^{\prime} R^{\dagger} = Q T_t^{\prime} Q^{\dagger} P_t.
\end{eqnarray}
We define
\begin{eqnarray}
    P_t^{\prime} = Q^{\dagger} P_t R, \label{PtPrime}
\end{eqnarray}
and therefore,
\begin{eqnarray}\label{PprimetTprimen}
    P_t^{\prime} T_{(n)}^{\prime} = T_t^{\prime} P_t^{\prime}.
\end{eqnarray}
Similarly to $P_t$ in $P_t V_{(n)}(S) = S_t P_t$ in Eq.~\eqref{PTP}, $P_t^{\prime}$ can only connect the corresponding vectors with the same eigenvalues in $T_{(n)}^{\prime}$ and $T_t^{\prime}$, respectively, so some elements in $P_t^{\prime}$ are again constrained to be zero. These vanishing elements of $P_t^{\prime}$ in Eq.~\eqref{PtPrime} induce a group of homogeneous equations with the $P_t$ nonzero elements being unknown parameters. Finding the solution space of these equations again accomplishes the detailed format of $P_t$, as well as the  detailed format of the irreducible representation $t$ subspace denoted in Eq.~\eqref{Pnm}.




\section{Several Examples}
In the previous sections, we illustrate the general algorithm to find the VEVs and to separate the degenerate scalars in a $\Phi_{(n)}$ for any $n$ assignment. These descriptions seem too ``abstract'' for one to understand and follow. Therefore, in this section, we show some examples to clarify the algorithm.

In Sec. \ref{Dim4}, we will take $n=4$ as a typical example for illustrating the simplest case about the remnant $Z_N$ symmetry from the $\text{SU(2)}_\text{D}$ group breaking. In Sec. \ref{Dim9}, we will take $n=9$ as an example to show the readers how our general algorithm described in Secs. \ref{VEVPatternSection} and \ref{ScalarClassificationSection} works when $\text{SU(2)}_\text{D}$ breaks into an $A_4$ group. These general algorithms can by principle be generalized to any value of $n$; however, we found that for some larger $n$, sometimes the simple analytical results are not easy to acquire, since our general algorithm usually gives the results written in cumbersome tangled recursive radical signs, which are extremely difficult to untangle. If one only wants some numerical simulation processes, e.g., scanning the parameter space by applying some numerical packages like \texttt{micrOMEGAs}~\cite{Belanger:2001fz} or \texttt{MadDM}~\cite{Backovic:2013dpa}, these results in the form of tangled recursive radical signs can be easily transformed into their decimal evaluations up to our desired precision before they are input into these packages as parameters. However, in this paper, we focus on the more simplified ``analytical'' results. We thereby develop some special but simpler algorithms, other than those in Secs. \ref{VEVPatternSection} and \ref{ScalarClassificationSection}, to find the analytical results in some cases. One example is shown in Sub. \ref{Dim13}.

\subsection{4-d scalar of \text{SU(2)}} \label{Dim4}
Consider the dark sector $\text{SU(2)}_\text{D}$ symmetry to be completely broken by a scalar quartet. 
The eigenvectors of $\vec{n}\cdot \vec{J}$ with eigenvalues 3/2, 1/2, --1/2, --3/2 are given in Appendix \ref{sec:eigenvector}.
Assuming a vacuum invariant under a $Z_3$ symmetry and setting the rotation axis to be the $z$-axis, the vacuum configuration in the basis of the $J_3$ eigenvector should be
\begin{equation}
    \vec{v}=a_{3/2}\vec{v}_{3/2}+a_{-3/2}\vec{v}_{-3/2},
\end{equation}
where $m$ in $a_m,v_m$ is teh eigenvalue of the basis.

If there exists another nontrivial $Z_{3}$ rotation, Eq.\eqref{eq:ZN'bm} yields
\begin{equation}
    \vec{v}=b_{3/2}\vec{v}'_{3/2} + b_{-3/2}\vec{v}'_{-3/2},
\end{equation}
where $\vec{v}'_m$ are the eigenvectors of $(J_1\sin\theta+J_3\cos\theta)$. Then Eq.~\eqref{bm} gives
\begin{equation}
  \left\{\begin{aligned}
			& a_{3/2} \cot\left(\frac{\theta}{2}\right) + a_{-3/2} = 0
			\\& -a_{3/2}\tan\left(\frac{\theta}{2}\right) + a_{-3/2} = 0
		\end{aligned}\right. ,
\end{equation}
which have no nonzero solution so that no other $Z_3$ symmetry exists.

\subsection{9-d scalar of SU(2)} \label{Dim9}
 As is known from Appendix~\ref{SubgroupsforSO3}, the remnant symmetry of the vacuum in the SU(2) 9-dimensional representation can be $S_4$, $D_4$, etc.. Since the $D_4$ results are simple, for brevity we only show the processes for $S_4$, which has six twofold symmetry axes, four threefold symmetry axes, and three fourfold symmetry axes. One can pick up the following generators
\begin{eqnarray}
   S&=&e^{i J_3\frac{\pi}{2}}, \nonumber \\
    T&=&e^{i (J_1\sin\theta+J_3\cos\theta)\frac{2\pi}{3}}, \nonumber \\
    K&=&e^{iJ_1\pi}, \label{Generators9}
\end{eqnarray}
with one of the fourfold symmetry axes (or the axis of a $Z_4$-subgroup) along the $z$-direction, and with one threefold symmetry axis lying at the $x$-$z$ plane. Here $\theta=\arctan\sqrt{2}$. Again, under the 9-dimensional representation case, $V_{(9)}(S)$, $V_{(9)}(T)$, $V_{(9)}(K)$ are adopted alternatively by replacing $J_{1,2,3}$ in Eq.~\eqref{Generators9} with $V_{(9)}(J_{1,2,3})$. Their concrete form can be found in Appendix \ref{charactertableandgenerators}.

If the vacuum configuration remains invariant under the $V_{(9)}(S)$ transformations, according to Eq.~\eqref{aConditions} again, under the $J_3$ representation basis, the VEV is restricted in the form of
\begin{equation}
  \vec{v}=a_4\vec{v}_4+a_0\vec{v}_0+a_{-4}\vec{v}_{-4}.
\end{equation}
where $m$ in $a_m,\vec{v}_m$ are the eigenvalues of the basis, $a_0$ is real and $a_{-4}=a_4^*$. On the rotation axis corresponding to $T$, following the processes before Eq.~\eqref{eq:ZN'bm}, we therefore obtain the condition:
\begin{eqnarray}
    b_{\pm1}=b_{\pm2}=b_{\pm4}=0,
\end{eqnarray}
inducing the equations
\begin{equation}
	\vec{v'}^{\dagger}_k (\theta) \vec{v}=0, \quad (k=\pm1,\pm2,\pm4), \label{VDotVEquations}
\end{equation}
where $\vec{v^{\prime}}_k(\theta)$ are the eigenvectors of $V_{(9)}(J_1\sin\theta+J_3\cos\theta)$ satisfying  $V_{(9)}(T) \vec{v^{\prime}}_k(\theta) = e^{i k\frac{2\pi}{3}}\vec{v^{\prime}}_k(\theta)$. Then, by solving Eq.~\eqref{VDotVEquations} regarding $a_{0, \pm4}$ as the unknown parameters, we finally acquire
\begin{equation}
	\qquad a_{-4}=a_{4},\qquad a_0=-\sqrt{\frac{14}{5}}a_4.
\end{equation}
corresponding to the vacuum configuration
\begin{equation}
	\vec{v}=\left(a_4,0,0, 0,-\sqrt{\frac{14}{5}}a_4,0,0,0,a_4 \right)^\mathrm{T} \label{VEV9Dim}
\end{equation}



From Eqs.~\eqref{Generators9}, \eqref{JpJm}, and \eqref{J123}, we can write the matrix forms of $V_{(9)}(S,T,K)$. The character table algorithm tells us that the SU(2) 9-dimensional representation under the $S_4$ subgroup can be reduced \textbf{9}=\textbf{1}+\textbf{3}+\textbf{3'}+\textbf{2}. The character of each irreducible representation of the $S_4$ group is listed in Appendix \ref{charactertableandgenerators}. The $\textbf{1}$ subspace is equivalent to the VEV configuration Eq.~\eqref{VEV9Dim}. The other irreducible subspaces are then computed by following the processes described in Sec. \ref{ScalarClassificationSection}. In this paper, we only show the detailed process for finding the $\textbf{3}$ subspace as an example for brevity.

%
Following Eq.~\eqref{Pnm}, we consider a projection operator $P_{\textbf{3}}$ projecting the 9-d scalar $\Phi_{(9)}$ into the triplet representation $\textbf{3}$. We learn from Eq.~\eqref{S4_3d_SK} that $S_{\textbf{3}}$ takes three eigenvalues, $i$, $1$, and $-i$. According to Eq.~\eqref{Generators9}, we acquire
\begin{eqnarray}
    V_{(9)}(S) &=& e^{i V_{(9)}(J_3) \frac{\pi}{2}} = \text{Diag}\lbrace e^{i \frac{4 \pi}{2}}, e^{i \frac{3 \pi}{2}}, e^{i \frac{2 \pi}{2}}, e^{i \frac{\pi}{2}}, 1, e^{-i \frac{\pi}{2}}, e^{-i \frac{2 \pi}{2}}, e^{-i \frac{3 \pi}{2}}, e^{-i \frac{4 \pi}{2}}\rbrace \nonumber \\
    &=&\text{Diag}\lbrace 1, -i, -1, i, 1, -i, -1, i, 1 \rbrace.
\end{eqnarray}
Therefore, the projection operator $P_\textbf{3}$ should collect all the  $V_{(9)}(S)$ eigenvectors of the eigenvalues $i$, $1$, $-i$ into the corresponding $S_{\textbf{3}}$ eigenvectors by taking the form of
\begin{equation}
    P_{\textbf{3}}=\begin{pmatrix}
	0&0&0&p_1&0&0&0&p_2&0
	\\p_3&0&0&0&p_4&0&0&0&p_5
	\\0&p_6&0&0&0&p_7&0&0&0
	\end{pmatrix},
\end{equation}
where $p_k$ ($k=1,\dots,7$) are unknown parameters. Then Eq.~\eqref{PSP} becomes
\begin{equation}
	P_{\textbf{3}} V_{(9)}(S)\Phi_{(9)}=	S_{\textbf{3}}\Phi_{\textbf{3}},
\end{equation}
where $\Phi_{\textbf{3}}$ is the irreducible space of the $\textbf{3}$ representation.

Following Eq.~\eqref{DiagonalizeT}, we diagonalize $T_{\textbf{3}}$ in Eq.~\eqref{S4_3d_T} by
\begin{eqnarray}
    T^\prime_{\textbf{3}}=Q^\dagger T_{\textbf{3}}Q=\text{Diag}\{e^{-i\frac{2\pi}{3}},e^{i\frac{2\pi}{3}},1\}
\end{eqnarray}
where
\begin{eqnarray}
    Q=  \left( \begin{array}{ccc}
        \frac{\sqrt{3}-3}{6}  & -\frac{\sqrt{3}+3}{6}  & \frac{1}{\sqrt{3}} \\
        -\frac{1}{\sqrt{3}} & \frac{1}{\sqrt{3}} & \frac{1}{\sqrt{3}} \\
        \frac{\sqrt{3}+3}{6} & \frac{3-\sqrt{3}}{6}  & \frac{1}{\sqrt{3}} \\
    \end{array} \right).
\end{eqnarray}
Then, $V_{(9)}(T)$ in Eq.~\eqref{S4_n9_T} is also diagonalized by
\begin{eqnarray}
    T^\prime_{(9)} = R^\dagger V_{(9)}(T)R = \text{Diag}\{e^{i\frac{2\pi}{3}}, 1, e^{-i\frac{2\pi}{3}}, e^{i\frac{2\pi}{3}}, 1, e^{-i\frac{2\pi}{3}}, e^{i\frac{2\pi}{3}}, 1, e^{-i\frac{2\pi}{3}}\},
\end{eqnarray}
where $R$ can be computed by listing all the eigenvectors of $V_{(9)}(T) = e^{i \frac{2 \pi}{3} V_{(9)}(J_1 \sin\theta + J_3 \cos\theta)}$. The results are given by Eqs.~\eqref{JTheta_1}--\eqref{JTheta_4}. Arranging them within $R$ gives
\begin{equation}
    R=\{\vec{v}^\prime_{4},\vec{v}^\prime_{3},\vec{v}^\prime_{2},\vec{v}^\prime_{1},\vec{v}^\prime_{0},\vec{v}^\prime_{-1},\vec{v}^\prime_{-2},\vec{v}^\prime_{-3},\vec{v}^\prime_{-4}\}.
\end{equation}
Therefore, according to Eq.~\eqref{PprimetTprimen}, $P_{\textbf{3}}^{\prime}$ projecting $\Phi_{(9)}$ into the $\textbf{3}$ irreducible space from the $V_{(9)}(T) = e^{i \frac{2 \pi}{3} V_{(9)}(J_1 \sin\theta + J_3 \cos\theta)}$ representation satisfies
\begin{equation}
	P'_\textbf{3}	T'_{(9)}=T'_\textbf{3}P'_\textbf{3}. \label{PtPrime3}
\end{equation}
According to Eq.~\eqref{PtPrime}, the detailed formalism of $P_{\textbf{3}}^{\prime}$ depending on $p_{1,2,3,4,5,6,7}$ is computed by
\begin{eqnarray}
    P_{\textbf{3}}^{\prime} = Q^{\dagger} P_{\textbf{3}} R.
\end{eqnarray}
However, Eq.~\eqref{PtPrime} requires that $P_{\textbf{3}}$ only connects the corresponding eigenvectors with the same eigenvalues between $T_{\textbf{3}}^{\prime}$ and $T_{(9)}^{\prime}$, restricting its expression to be
\begin{equation}\label{Pprime3}
	P'_\textbf{3}=\left(
	\begin{array}{ccccccccc}
		0 & 0 & p'_1 & 0 & 0 & p'_2 & 0 & 0 & p'_3 \\
		p'_4 & 0 & 0 & p'_5 & 0 & 0 & p'_6 & 0 & 0 \\
		0 & p'_7 & 0 & 0 & p'_8 & 0 & 0 & p'_9 & 0 \\
	\end{array}
	\right),
\end{equation}
where $p'_k$($k=1,\dots,9$) are unknown parameters. The zero elements of $P_\textbf{3}^{\prime}$ in Eq.~\eqref{Pprime3} prompt the eighteen homogeneous equations
\begin{eqnarray}
    & & (P'_\textbf{3})_{(1,1)}=(P'_\textbf{3})_{(1,2)}=(P'_\textbf{3})_{(1,4)}=(P'_\textbf{3})_{(1,5)}=(P'_\textbf{3})_{(1,7)}=(P'_\textbf{3})_{(1,8)} \nonumber \\
    & = & (P'_\textbf{3})_{(2,2)}=(P'_\textbf{3})_{(2,3)}=(P'_\textbf{3})_{(2,5)}=(P'_\textbf{3})_{(2,6)}=(P'_\textbf{3})_{(2,8)}=(P'_\textbf{3})_{(2,9)} \nonumber \\
    & = & (P'_\textbf{3})_{(3,1)}=(P'_\textbf{3})_{(3,3)}=(P'_\textbf{3})_{(3,4)}=(P'_\textbf{3})_{(3,6)}=(P'_\textbf{3})_{(3,7)} =(P'_\textbf{3})_{(3,9)} = 0 
    \label{P3PrimeEquations}
\end{eqnarray}
depending on $p_k(k=1,\dots,7)$, where $(P'_\textbf{3})_{(i,j)}$ indicates the element of row $i$, column $j$ in  $P'_\textbf{3}$. Solving Eq.~\eqref{P3PrimeEquations} gives
\begin{equation}
	p_2= \frac{p_1}{\sqrt{7}},\quad p_3= \frac{2 p_1}{\sqrt{7}},\quad p_4= 0,\quad p_5= -\frac{2 p_1}{\sqrt{7}},\quad p_6= -\frac{p_1}{\sqrt{7}},\quad p_7= -p_1,
\end{equation}
with the only undetermined parameter $p_1$.

Finally, the requirement $P^\dagger_\textbf{3}P_\textbf{3}=I_{3\times3}$ normalizes $p_1$ by
\begin{equation}
	\frac{8}{7} p_1 p_1^*=1.
\end{equation}
Selecting $p_1= i \sqrt{\frac{7}{8}}$, the triplet subspace $\Phi_\textbf{3}$ is then given as 
\begin{equation}
    \Phi_\textbf{3} = \frac{1}{\sqrt{2}} \begin{pmatrix}
		i G^+\\G^0\\-i G^-
	\end{pmatrix}=
        \left(
	\begin{array}{c}
		\frac{i\sqrt{7}}{4}   \Delta _1+	\frac{i }{4}\Delta _{-3} \\
	-\text{Im$\Delta $}_4 \\
		-\frac{i \sqrt{7} }{4} \Delta _{-1}-\frac{i }{4} \Delta _3\\
	\end{array}
	\right) \label{S4_9d_GoldstoneS}
\end{equation}
Here we use the symbols $G^{0, \pm}$ to denote the elements in $\Phi_\textbf{3}$ due to the fact that these are actually three Goldstone bosons to be eaten by the longitudinal polarizations of the vector fields. This can also be verified by expanding the kinetic term $D^{\mu} \Phi_{(9)}^{\dagger} D_{\mu} \Phi_{(9)}$ in Eq.~\eqref{Lagrangian} relative to the vacuum configuration given by Eq.~\eqref{VEV9Dim},
\begin{eqnarray}
   \sqrt{2} \Phi_{(9)} = \left(a_4 + \Delta_4^{\prime},\Delta_3, \Delta_2, \Delta_1 , -\sqrt{\frac{14}{5}}a_4+ \Delta_0^{\prime}, \Delta_{-1}, \Delta_{-2}, \Delta_{-3}, a_4+\Delta'_{-4} \right)^\mathrm{T},
\end{eqnarray}
where
\begin{eqnarray}
    \Delta_4 &=& a_4+\Delta_4^{\prime}, \nonumber \\
    \Delta_0 &=& -\sqrt{\frac{14}{5}} a_4 + \Delta_0^{\prime}. 
\end{eqnarray}
Then, the bilinear terms connecting the partial derivatives of the three gauge fields with $\Delta_{1,2,3}$ give the formalism of the Goldstone degrees of freedom, which are exactly the same with Eq.~\eqref{S4_9d_GoldstoneS}.
The real Goldstone triplet formalism can also be written by defining 
\begin{equation}
	G_1=\frac{G^- - G^+}{\sqrt{2}i},\qquad
	G_2=-\frac{G^- + G^+}{\sqrt{2}},\qquad 
	G_3=G_0.
        \label{eq:Real_Glodstones}
\end{equation}
and therefore
\begin{equation}
	\begin{pmatrix}
		G_1
		\\G_2
		\\G_3
	\end{pmatrix}
	=\begin{pmatrix}
		-\frac{\sqrt{7}}{2}\text{Im}\Delta_1 + \frac{1}{2}\text{Im}\Delta_3 \\
		-\frac{\sqrt{7}}{2}\text{Re}\Delta_1 - \frac{1}{2}\text{Re}\Delta_3 \\
		-\sqrt{2}\text{Im}\Delta_4
	\end{pmatrix}.
\end{equation}

For the rest of the subspace classifications, we neglect the detailed processes and only show the final results in Sec. \ref{SubspaceEnumeration}.

\subsection{13-d scalar of SU(2)} \label{Dim13}
As shown in Appendix \ref{SubgroupsforSO3}, the vacuum of the 13-dimensional representation $\Phi_{(13)}$ of the $\text{SU(2)}_\text{D}$ group can be invariant under an $A_5$ subgroup. In this case, we premise the vacuum has an $A_5$ symmetry and then work out the concrete vacuum configuration using the method mentioned in Sec. ~\ref{VEVPatternSection}. 
The $A_5$ group contains fivefold axes, threefold axes, and twofold axes. We can rotate to the direction that one fivefold axis aligns to the $z$-axis, and one threefold axis and one twofold axis lies within the $x$-$z$ plane. With these configurations, the three generators of the $A_5$ group can be given by
\begin{eqnarray}
    S &=& e^{i J_3 \frac{2 \pi}{5}} \nonumber \\
    T &=& e^{i (J_1 \sin\theta_{53} + J_3 \cos\theta_{53}) \frac{2 \pi}{3}} \\
    K &=& e^{i (J_1 \sin\theta_{52} + J_3 \cos\theta_{52}) \pi},
\end{eqnarray}
where $\theta_{53}$ is the angle between the $Z_5$ and $Z_3$ rotation axes, while $\theta_{52}$ is the angle between the $Z_5$ and $Z_2$ rotation axes. These values are given by $\theta_{53}=-\arctan\left(3-\sqrt{5}\right)$, $\theta_{52}=-\arctan\left(\frac{\sqrt{5}+1}{2}\right)$. Therefore, the invariance of the VEV under the $V_{(13)}(S)$ transformation constrains its expression to be
\begin{equation}
	\vec{v}=a_5\vec{v}_5+a_0\vec{v}_0+a_{-5}\vec{v}_{-5},
\end{equation}
where $m$ in $a_m,\vec{v}_m$ is the eigenvalue of the basis.

Again, similar to the processes described in Sec. \ref{Dim9}, solving Eq.~\eqref{bm} by replacing $\vec{n} \cdot \vec{J}$ in Eq.~\eqref{OtherDirectionEigens} with $\theta=\theta_{53}$, one finally arrives at
%
\begin{equation}
	 \vec{v}=\left(0,a_5,0,0,0,0,\sqrt{\frac{11}{7}} a_5,0,0,0,0,a_5,0 \right)^\mathrm{T}, \label{VEV13Dim}
\end{equation}
where $a_5$ is an arbitrary real parameter.
%
%
%
%
%
With the aid of the character table displayed in Appendix \ref{charactertableandgenerators}, the 13-dimensional target space of $\text{SU(2)}_\text{D}$ can be decomposed into the direct sum of the $\textbf{13}=\textbf{1}+\textbf{3}+\textbf{4}+\textbf{5}$ subspaces, which are invariant under $A_5$. The $\textbf{1}$ subspace has been found in the form of Eq.~\eqref{VEV13Dim}. One can then in principle follow the algorithm described in Sec. ~\ref{ScalarClassificationSection} to solve all the detailed expressions of the $\textbf{3}$, $\textbf{4}$, $\textbf{5}$ representations. However, practical calculations show that the coefficients are extremely complicated, and it is very difficult for us to simplify the analytical expressions of the recursive radical signs. Therefore, in this subsection, we adopt some other flexible strategies in light of such situations.

The only one triplet of the $A_5$ group in this case is exactly the Goldstone degrees of freedom eaten by gauge bosons. By shifting $\Phi_{(13)}$ relative to the vacuum configuration in Eq.~\eqref{VEV13Dim}
\begin{eqnarray}
    &&\sqrt{2} \Phi_{(13)}=\nonumber\\
    &&\left(\Delta_6,a_5+\Delta_5^{\prime},\Delta_4,\Delta_3,\Delta_2,\Delta_1,\sqrt{\frac{11}{7}} a_5 + \Delta_0^{\prime}, \Delta_{-1}, \Delta_{-2}, \Delta_{-3}, \Delta_{-4},a_5+\Delta'_{-5},\Delta_{-6}\right)^\mathrm{T},\nonumber\\
\end{eqnarray}
and substituting this into $(D^{\mu} \Phi_{(13)})^{\dagger} D_{\mu} \Phi_{(13)}$ in Eq.~\eqref{Lagrangian}, one can acquire the expressions of Goldstone fields by collecting the bilinear mixing terms between the partial derivatives of the three gauge fields with $\Delta_m$. The results are given by
\begin{equation}
    G^+ = -\frac{\sqrt{3}}{5}\Delta_{6} - \frac{\sqrt{66}}{10} \Delta_{1} +\frac{\sqrt{22}}{10}\Delta_{-4},\qquad
    G^- = (G^+)^*,\qquad
    G^0 = -\sqrt{2} \text{Im} \Delta_5,
\end{equation}
where $G^{\pm},G^0$ are eaten by $X^{\pm},Z'$, respectively, as shown in Appendix~\ref{generatorsofSU2}. This is equivalent to the projection operator
\begin{equation}
    P_{\textbf{3}}=\left(\begin{array}{ccccccccccccc}
		\frac{-i\sqrt{3}}{5} & 0 & 0 & 0 & 0 & \frac{-i\sqrt{66}}{10} & 0 & 0 & 0 & 0 &  \frac{i\sqrt{22}}{10} & 0 & 0 \\
		0 & \frac{i}{\sqrt{2}} & 0 & 0 & 0 & 0 & 0 & 0 & 0 & 0 & 0 & \frac{-i}{\sqrt{2}} & 0 \\
		0 & 0 & \frac{-i\sqrt{22}}{10} & 0 & 0 & 0 & 0 & \frac{i\sqrt{66}}{10} & 0 & 0 & 0 & 0 & \frac{i\sqrt{3}}{5} \\
	\end{array}\right) \label{A5_13Dim_P3}.
\end{equation}
Similarly with Eq.~\eqref{eq:Real_Glodstones}, the real triplet formalism can also be written as
\begin{equation}
	\begin{pmatrix}
		G_1
		\\G_2
		\\G_3
	\end{pmatrix}
	=\begin{pmatrix}
		\frac{\sqrt{6}}{5}\text{Im}\Delta_6 + \frac{\sqrt{33}}{5}\text{Im}\Delta_1 + \frac{\sqrt{11}}{5}\text{Im}\Delta_4 \\
		\frac{\sqrt{6}}{5}\text{Re}\Delta_6 + \frac{\sqrt{33}}{5}\text{Re}\Delta_1 - \frac{\sqrt{11}}{5}\text{Re}\Delta_4 \\
		-\sqrt{2}\text{Im}\Delta_5
	\end{pmatrix}.
\end{equation}

Then let us turn to the $\textbf{5}$-subspace. Connecting the eigenvectors with the same eigenvalues of both the matrix form of $V_{13}(S)$ and the diagonalized $S_{\textbf{5}}$ defined in Eq.~\eqref{A5_S5} restricts the projection operator $P_\textbf{5}$, satisfying
\begin{equation}
	P_\textbf{5}	V_{(13)}(S)\Phi_{(13)}=S_\textbf{5}\Phi_{\textbf{5}},
\end{equation}
within the formalism of
\begin{equation}
     P_\textbf{5}=\left(
    \begin{array}{ccccccccccccc}
     0 & 0 & 0 & 0 & p_1 & 0 & 0 & 0 & 0 & p_2 & 0 & 0 & 0 \\
     p_3 & 0 & 0 & 0 & 0 & p_4 & 0 & 0 & 0 & 0 & p_5 & 0 & 0 \\
     0 & p_6 & 0 & 0 & 0 & 0 & p_7 & 0 & 0 & 0 & 0 & p_8 & 0 \\
     0 & 0 & p_9 & 0 & 0 & 0 & 0 & p_{10} & 0 & 0 & 0 & 0 & p_{11} \\
     0 & 0 & 0 & p_{12} & 0 & 0 & 0 & 0 & p_{13} & 0 & 0 & 0 & 0 \\
\end{array}
\right), \label{P5_13Dim}
\end{equation}
where $p_{1\text{-}13}$ are the undetermined parameters. Since all the subspaces corresponding to different irreducible representations should be orthogonal to each other, and each row of the projector operators can be regarded as the basis within the $\Phi_{(15)}$ space that can be expanded into the corresponding irreducible subspaces, each row of Eq.~\eqref{P5_13Dim} should be perpendicular to the rows in Eqs.~\eqref{A5_13Dim_P3} and \eqref{VEV13Dim}. Then it is easy to obtain that for the third row in Eq.~\eqref{P5_13Dim}, $p_{6,7,8}$ should satisfy
\begin{equation}\label{parametersofP53}
    p_7= -2 \sqrt{\frac{7}{11}}p_6,\quad p_8= p_6.
\end{equation}
As a result,
\begin{equation}
    P_{\textbf{5},3} \propto \left(
	0 , 1, 0 , 0 , 0  , 0 , -2 \sqrt{\frac{7}{11}} , 0 , 0 , 0 , 0 , 1 , 0 \\
	\right), \label{P5_3_ProportionalResults}
\end{equation}
where $P_{\textbf{5},k}$ means the $k$ th row of the projector operator $P_{\textbf{5}}$.

As we have illustrated after Eq.~\eqref{Normalized}, each column of $P_{\textbf{5}}^{\dagger}$ is a vector within the basis of the $\textbf{5}$ subspace and the condition that this subspace is closed under the transformation of any element of $A_5$ guarantees that
\begin{eqnarray}
    V_{(13)}(U) P_{\textbf{5},i}^{\dagger} =  (U_\textbf{5})_{ij} P_{\textbf{5},j}^{\dagger} \label{RotateP53General}
\end{eqnarray}
for any element $U \in A_5$, where $i,j$ run within $1$--$5$. Therefore, specifically, if we adopt $U=e^{iJ_{3} \pi}e^{i J_{2} \theta_{55}}$, where $\theta_{55}=2\theta_{52}$ is the angle between two next-to-nearest neighbor fivefold axes, one can verify by observing a regular dodecahedron that such a $U$ rotates a fivefold axis aligning to the $z$-direction toward its next-to-nearest neighbor by the $e^{i J_{2} \theta_{55}}$ operation, then match all the other threefold axes and twofold axes through the $e^{iJ_{3} \pi}$ operation to flip the direction of the pentagon at the top. Therefore, $U=e^{iJ_{3} \pi}e^{i J_{2} \theta_{55}} \in A_5$.
With Eq.~\eqref{P5_3_ProportionalResults}, one can compute that
\begin{eqnarray} 
    & & V_{(13)}(U) P_{\textbf{5},3}^\dagger = e^{i V_{(13)}(J_{3}) \pi}e^{i V_{(13)}(J_{2}) \theta_{55}} P_{\textbf{5},3}^\dagger \nonumber \\
    &\propto& \left(-\frac{\sqrt{66}}{25},-\frac{\sqrt{22}}{50},-\frac{9}{25},\frac{\sqrt{30}}{25},\frac{2 \sqrt{30}}{25},-\frac{\sqrt{3}}{25},\frac{\sqrt{14}}{25},-\frac{\sqrt{3}}{25},
    \frac{2 \sqrt{30}}{25}, \frac{\sqrt{30}}{25},-\frac{9}{25},
     \right. \nonumber \\
    & &
    \left.  -\frac{\sqrt{22}}{50},-\frac{\sqrt{66}}{25}\right)^\text{T} \in P_{\textbf{5}} \Phi_{(13)}. 
    \label{RotatedP53}
\end{eqnarray}
We can express that Eq.~\eqref{RotatedP53} is a member within the $\textbf{5}$ subspace, so it can be expanded in terms of the combination of the basis $P_{\textbf{5},1}$, $P_{\textbf{5},2}$, $P_{\textbf{5},3}$, $P_{\textbf{5},4}$, and $P_{\textbf{5},5}$ as denoted in Eq.~\eqref{RotateP53General}. Notice that the formalism of each $P_{\textbf{5},i}$ is highly restricted by Eq.~\eqref{P5_13Dim}; therefore, it is easy to compare the coefficients and find out the relations among $p_{1,2}$, $p_{3,4,5}$, $p_{9,10,11}$ and $p_{12,13}$. After doing the normalization, one acquires
\begin{equation}
    P_\textbf{5}=\left(
	\begin{array}{ccccccccccccc}
		0 & 0 & 0 & 0 & \frac{2}{\sqrt{5}} & 0 & 0 & 0 & 0 & \frac{1}{\sqrt{5}} & 0 & 0 & 0 \\
		\frac{\sqrt{11}}{5} & 0 & 0 & 0 & 0 & \frac{\sqrt{2}}{10} & 0 & 0 & 0 & 0 & \frac{3 \sqrt{6}}{10} & 0 & 0 \\
		0 & \frac{\sqrt{22}}{10} & 0 & 0 & 0 & 0 & -\frac{\sqrt{14}}{5} & 0 & 0 & 0 & 0 & \frac{\sqrt{22}}{10} & 0 \\
		0 & 0 & \frac{3 \sqrt{6}}{10} & 0 & 0 & 0 & 0 & \frac{\sqrt{2}}{10} & 0 & 0 & 0 & 0 & \frac{\sqrt{11}}{5} \\
		0 & 0 & 0 & \frac{1}{\sqrt{5}} & 0 & 0 & 0 & 0 & \frac{2}{\sqrt{5}} & 0 & 0 & 0 & 0 \\
	\end{array}
	\right),
\end{equation}
corresponding to a series of degenerate fields
\begin{equation}
    \Phi_{\textbf{5}}=\frac{1}{\sqrt{2}}\begin{pmatrix}
		D^{++}\\D^+\\D^0\\D^-\\D^{--}
	\end{pmatrix}=
    \left(
	\begin{array}{c}
		\sqrt{\frac{2}{5}} \Delta _2+\frac{1}{\sqrt{10}}\Delta _{-3} \\
		\frac{1}{10}\Delta _1+\frac{3 \sqrt{3} }{10}\Delta _{-4}+\frac{1}{5} \sqrt{\frac{11}{2}} \Delta _6 \\
		\frac{\sqrt{11}}{10}  \Delta _{-5}+\frac{\sqrt{11}}{10} \Delta _5-\frac{\sqrt{7} }{5}\Delta _0 \\
		\frac{1}{10}\Delta _{-1}+\frac{3 }{10}\sqrt{3} \Delta _4+\frac{1}{5} \sqrt{\frac{11}{2}} \Delta _{-6} \\
		\sqrt{\frac{2}{5}} \Delta _{-2}+\frac{1}{\sqrt{10}}\Delta _3 \\
	\end{array}
	\right).
\end{equation}
The corresponding real scalars can be defined by
\begin{align}
	&D_1=\frac{D^{++}+D^{--}}{\sqrt{2}}
    ,\qquad D_2=\frac{D^{++}-D^{--}}{i\sqrt{2}}, \nonumber
	\\
	&D_3=\frac{D^{+}+D^{-}}{\sqrt{2}}
	,\qquad
	D_4=\frac{D^{+}-D^{-}}{i\sqrt{2}},\qquad
	D_5=D^0,
\end{align}
so therefore
\begin{equation}
	\begin{pmatrix}
		D_1\\D_2\\D_3\\D_4\\D_5
	\end{pmatrix}
	=\left(
	\begin{array}{c}
		2 \sqrt{\frac{2}{5}} \text{Re$\Delta $}_2+\sqrt{\frac{2}{5}} \text{Re$\Delta $}_3 \\
		2 \sqrt{\frac{2}{5}} \text{Im$\Delta $}_2-\sqrt{\frac{2}{5}} \text{Im$\Delta $}_3 \\
		\frac{1}{5}\text{Re$\Delta $}_1+\frac{3 \sqrt{3} }{5}\text{Re$\Delta $}_4+\frac{\sqrt{22} }{5}\text{Re$\Delta $}_6 \\
		\frac{1}{5}\text{Im$\Delta $}_1-\frac{3 \sqrt{3} }{5}\text{Im$\Delta $}_4+\frac{\sqrt{22} }{5}\text{Im$\Delta $}_6 \\
		\frac{\sqrt{22} }{5}\text{Re$\Delta $}_5-\frac{\sqrt{14}}{5}  \Delta _0\\
	\end{array}
	\right).
\end{equation}

The remaining $\textbf{4}$ irreducible subspace is hence acquired following a similar algorithm, while in this subsection we neglect the detailed processes. The final results are displayed in Sec. \ref{SubspaceEnumeration}.

Besides, we also present the two-loop beta function of the gauge coupling coefficient $g_\text{D}$ given by~\cite{Jones:1981we} :
\begin{eqnarray}
    \beta_{g_\text{D}} &=& \frac{g_\text{D}^3}{16\pi^2}\left[-\frac{11}{3}\text{C}_2(A) + \frac{\eta_S}{6}\text{T}(R)\right]\nonumber\\
    &&+\frac{g_\text{D}^5}{(16\pi^2)^2}\left[-\frac{34}{3}(\text{C}_2(A))^2+\eta_S\left(\frac{1}{3}\text{C}_2(A)+2\text{C}_2(R)\right)\text{T}(R)\right]
\end{eqnarray}
where $\text{T}(R)=j(j+1)(2j+1)/3$, $\text{C}_2(A)=2$,~$\text{C}_2(R)=j(j+1)$ for $\text{SU}(2)_\text{D}$ and $\eta_S=1~(2)$ for the real (complex) scalar field. In this paper, when $j$ is an integer, we only consider the real scalar situation, and when $j$ is a half-integer, the scalar is complex. Therefore, $g_\text{D}$ is found to be asymptotically free only for the models with
\begin{equation}
g_D^2(v')\lesssim\left\{\begin{matrix}\infty&(j\leq1)\\ 24.63 & (j=3/2) \\11&(j=2)\\ 0.4 & (j=5/2) \\0.65&(j=3)\end{matrix}\right. .
\end{equation}
It means that for a model with $j>3$, there will be a Landau pole at some higher energy scale, so the model with high dimensional representation should be embedded in some UV completion models.

\section{Subspaces for odd dimensional representation of SU(2)} \label{SubspaceEnumeration}

In the previous sections, we have shown the algorithms adopted to calculate the vacuum and subspace decomposition with the discrete symmetry broken from the $\text{SU(2)}_\text{D}$ symmetry by a single $\Phi_{(n)}$ Higgs multiplet. In this section, we list the results of breaking the $\text{SU(2)}_\text{D}$ symmetry to the discrete groups for each odd $n=5$--$21$, while we neglect the detailed processes accomplishing them. As we have illustrated, for all the even $n$ selections up to $n=20$, only the simplest discrete $Z_N$ symmetries can remain, so we do not list the results here. For all the odd $n \leq 21$ values, all the possible nontrivial discrete subgroups that the $\text{SU(2)}_\text{D}$ group can break into have been calculated by Ref.~\cite{Etesi:1996urw} and are repeated in Table~\ref{SubgroupsforSO3table} of Appendix \ref{SubgroupsforSO3} for bookkeeping. For all these cases, we are going to show the results of the detailed VEV configurations as well as the irreducible subspaces of the Higgs multiplet, whereafter we will illustrate the orientations of the axes of the $A_4$, $A_5$ and $S_4$ groups in the Secs. \ref{Dim9} and \ref{Dim13}, respectively. From Table~\ref{SubgroupsforSO3table} we learn that when $n=13$, $17$, $19$ and $21$, the $\text{SU(2)}_\text{D}$ group can be broken into the $S_4$. In this case we rotate toward a direction so that a fourfold axis is along the $z$-axis, while one of the threefold axis is within the $x$-$z$ plane, similar to the instructions illustrated in Appendix \ref{S4_Characters_Generators}. Furthermore, when $n=13$, $19$, and $21$, the $\text{SU(2)}_\text{D}$ group can be broken into $A_4 \subset S_4$ as well, making it convenient to cast the $S_4$ results straightforwardly to the $A_4$ cases. Therefore, in the cases of $n=13$, $19$, and $21$, we adopt another orientation of the $A_4$ axes in accordance with the $S_4$ assignments, so that one of the twofold axes of the $A_4$ group is set aligning to the $z$-direction, while one of the threefold axes is within the $x$-$z$ plane.

In the following, the multiplets are listed corresponding to the irreducible subspace sequentially, which are presented in Table \ref{tab:VEV_Subspace_Results}  for $n<13$. For all these $n$-d representations of SU(2), we give their characters related to the discrete group in Appendix \ref{charactertableandgenerators}.
\begin{table}[h!]
	\begin{center}
		\resizebox{\textwidth}{!}{\begin{tabular}{|c|c|}
              \hline
			\textbf{5-d scalar} & \textbf{Vacuum with $Z_2\oplus Z_2$ symmetry}\\ $\Phi_{(5)}=\frac{1}{\sqrt{2}}\left(\Delta _2,\Delta _1,\Delta _0,\Delta _{-1},\Delta_{-2}\right)^\text{T}$ & $\frac{v^\prime}{\sqrt{2}}\left(\frac{c_\alpha}{\sqrt{2}},0,s_\alpha,0,\frac{c_\alpha}{\sqrt{2}}\right)^\text{T}$ \\
			\hline
			\multicolumn{2}{|c|}{\textbf{subspace}\qquad\textbf{5}= \textbf{1}+\textbf{1}+\textbf{$1_a$}+\textbf{$1_b$}+\textbf{$1_c$}}\\
			\multicolumn{2}{|c|}{$
					s_\alpha\Delta _0 +c_\alpha\sqrt{2} \text{Re$\Delta $}_2 ,\quad c_\alpha \Delta _0 - s_\alpha \sqrt{2} \text{Re$\Delta $}_2,\quad \sqrt{2}\text{Im}\Delta_1^\text{G},\quad  -\sqrt{2} \text{Re}\Delta_1^\text{G},\quad -\sqrt{2} \text{Im}\Delta_2^\text{G}
			$}\\
   \multicolumn{2}{|c|}{with the last three singlets to be Goldstone bosons corresponding to $G_1,G_2,G_3$ respectively. }\\
    \multicolumn{2}{|c|}{ \textbf{conditions}\qquad $0 = -\mu^2 + \frac{1}{2} \lambda_{H\Phi} v^2 + \lambda_1 v'^2 + 3\kappa' v'^2 s_{3\alpha},\quad 0 = \kappa' c_{3\alpha}$ }\\
   \multicolumn{2}{|c|}{\textbf{masses}\qquad $\mathcal{M}^2_{\rho}=2\lambda_1 v'^2,\quad m^2_{\textbf{1}}=0,\quad m_{X_1}^2=(2-c_{2\alpha}+\sqrt{3}s_{2\alpha}) g_{\text{D}}^2 v'^2,\quad m_{X_2}^2= (2-c_{2\alpha}-\sqrt{3}s_{2\alpha}) g_{\text{D}}^2 v'^2,\qquad m_{X_3}^2=4 c_\alpha^2 g_{\text{D}}^2 v'^2$}\\
            \hline
			\textbf{7-d scalar} & \textbf{Vacuum with $A_4$ symmetry}\\ $\Phi_{(7)}=\frac{1}{\sqrt{2}}\left(\Delta _3,\Delta _2,\Delta _1,\Delta _0,\Delta _{-1},\Delta _{-2},\Delta _{-3}\right)^\text{T}$ & $\frac{v^\prime}{\sqrt{2}}\left(0,\frac{\sqrt{2}}{2},0,0,0,\frac{\sqrt{2}}{2},0\right)^\text{T}$ \\
			\hline
			\multicolumn{2}{|c|}{\textbf{subspace}\qquad\textbf{7}= \textbf{1}+\textbf{3}+\textbf{3}}\\
			\multicolumn{2}{|c|}{$
					\sqrt{2}\text{Re}\Delta_2,\quad
					\left(
	           \begin{array}{c}
	           	\frac{\sqrt{5}}{2} \text{Im$\Delta $}_1+\frac{\sqrt{3} }{2} \text{Im$\Delta $}_3\\
	           	-\frac{\sqrt{5} }{2}\text{Re$\Delta $}_1+\frac{\sqrt{3}}{2} \text{Re$\Delta $}_3 \\
	           	-\sqrt{2} \text{Im$\Delta $}_2 \\
	           \end{array}
	        \right)^\text{G},\quad
				\begin{pmatrix}
						\frac{\sqrt{3}}{2}\text{Re} \Delta_1+\frac{\sqrt{5}}{2}\text{Re}\Delta_3
						\\\frac{\sqrt{3}}{2}\text{Im}\Delta_1-\frac{\sqrt{5}}{2}\text{Im}\Delta_3
						\\\Delta_0
					\end{pmatrix}
			$}\\
    \multicolumn{2}{|c|}{ \textbf{condition}\qquad $0 = -\mu^2 + \frac{1}{2} \lambda_{H\Phi} v^2 + (\lambda_1 + 48\lambda_2) v'^2$ }\\
    \multicolumn{2}{|c|}{\textbf{masses}\qquad $\mathcal{M}^2_{\rho}=2(\lambda_1+48\lambda_2) v'^2,\quad m_{\textbf{3}}^2= 60\lambda_2 v'^2,\quad m_{X_1}^2=m_{X_2}^2=m_{X_3}^2 = 4g_{\text{D}}^2 v'^2$}\\
			\hline
            \textbf{7-d scalar} & \textbf{Vacuum with $D_3$ symmetry}\\ $\Phi_{(7)}=\frac{1}{\sqrt{2}}\left(\Delta _3,\Delta _2,\Delta _1,\Delta _0,\Delta _{-1},\Delta _{-2},\Delta _{-3}\right)^\text{T}$ & $\frac{v^\prime}{\sqrt{2}}\left(\frac{\sqrt{2}}{2},0,0,0,0,0,\frac{\sqrt{2}}{2}\right)^\text{T}$ \\
			\hline
			\multicolumn{2}{|c|}{\textbf{subspace}\qquad\textbf{7}= \textbf{1}+\textbf{1'}+\textbf{1'}+\textbf{2}+\textbf{2}}\\
			\multicolumn{2}{|c|}{$
					\sqrt{2}\text{Re}\Delta_3,\quad -\sqrt{2}\text{Im}\Delta_{3}^\text{G},\quad \Delta_0,\quad
                    \begin{pmatrix}
                        \sqrt{2}\text{Im}\Delta_2\\
                   - \sqrt{2}\text{Re}\Delta_2
                     \end{pmatrix}^\text{G},\quad
                    \begin{pmatrix}
                        \sqrt{2}\text{Re}\Delta_1
                        \\
                        \sqrt{2}\text{Im}\Delta_1
                    \end{pmatrix}$}\\
            \multicolumn{2}{|c|}{
            with the second singlet and the first doublet to be Goldstone bosons corresponding to $G_3$, $(G_1,G_2)^T$.}
            \\
            \multicolumn{2}{|c|}{\textbf{condition}\qquad $0 = -\mu^2 + \frac{1}{2} \lambda_{H\Phi} v^2 + (\lambda_1 + \frac{171}{2}\lambda_2) v'^2$ }\\
            \multicolumn{2}{|c|}{\textbf{masses}\qquad $\mathcal{M}^2_{\rho}=2(\lambda_1+\frac{171}{2}\lambda_2) v'^2,\quad m_{\textbf{1'}}^2=-\frac{135}{2} \lambda_2 v'^2,\quad  m_{\textbf{2}}^2=-45 \lambda_2 v'^2,\quad m_{X_1}^2=m_{X_2}^2= \frac{3}{2} g_{\text{D}}^2 v'^2,\qquad m_{X_3}^2 = 9 g_{\text{D}}^2 v'^2$}\\
			\hline
				\textbf{9-d scalar} & \textbf{Vacuum with $S_4$ symmetry}\\ $\Phi_{(9)}=\frac{1}{\sqrt{2}}\left(\Delta_4,\Delta_3,\Delta_2,\Delta_1,\Delta_0,\Delta_{-1},\Delta_{-2},\Delta_{-3},\Delta_{-4}\right)^\text{T}$ & 
                $\frac{v^\prime}{\sqrt{2}}\left(\frac{1}{2}\sqrt{\frac{5}{6}},0,0,0,-\frac{1}{2}\sqrt{\frac{7}{3}},0,0,0,\frac{1}{2}\sqrt{\frac{5}{6}}\right)^\text{T}$\\
			\hline
			\multicolumn{2}{|c|}{\textbf{subspace}\qquad$\textbf{9}=\textbf{1}+\textbf{3}+\textbf{3'}+\textbf{2}$}\\
			\multicolumn{2}{|c|}{$
				\left(\sqrt{\frac{5}{6}} \text{Re$\Delta $}_4-\frac{1}{2} \sqrt{\frac{7}{3}} \Delta _0\right),\qquad
					\begin{pmatrix}
		-\frac{\sqrt{7}}{2}\text{Im}\Delta_1 + \frac{1}{2}\text{Im}\Delta_3 \\
		-\frac{\sqrt{7}}{2}\text{Re}\Delta_1 - \frac{1}{2}\text{Re}\Delta_3 \\
		-\sqrt{2}\text{Im}\Delta_4
	\end{pmatrix}^\text{G},\quad
        \begin{pmatrix}
				\frac{\sqrt{7}}{2} \text{Re$\Delta $}_3-\frac{1}{2}\text{Re$\Delta $}_1
				\\\frac{\sqrt{7}}{2} \text{Im$\Delta $}_3+\frac{1}{2}\text{Im$\Delta $}_1
				\\ \sqrt{2}\text{Re$\Delta $}_2
	\end{pmatrix},\quad
			\left(
        	\begin{array}{c}
	           \frac{1}{2} \sqrt{\frac{5}{3}}         \Delta _0+\sqrt{\frac{7}{6}} \text{Re$\Delta $}_4 \\
            	\sqrt{2} \text{Im$\Delta $}_2 \\
        	\end{array}
    	\right)$}\\
      \multicolumn{2}{|c|}{\textbf{condition}\qquad $0 = -\mu^2 + \frac{1}{2} \lambda_{H\Phi} v^2 + (\lambda_1 + \frac{400}{3}\lambda_2 - \sqrt{\frac{7}{3}}^3 \kappa') v'^2$ }\\
     \multicolumn{2}{|c|}{\textbf{masses}\qquad $\mathcal{M}^2_{\rho}=2(\lambda_1+\frac{400}{3}\lambda_2 - \frac{1}{2}\sqrt{\frac{7}{3}}^3 \kappa') v'^2,\quad m_{\textbf{3'}}^2 = (\frac{140}{3}\lambda_2 + \frac{20\sqrt{21}}{9}\kappa') v'^2,\quad m_{\textbf{2}}^2 = (\frac{560}{3}\lambda_2 + \frac{5\sqrt{21}}{9}\kappa') v'^2,\quad m_{X_1}^2=m_{X_2}^2=m_{X_3}^2= \frac{20}{3}g_{\text{D}}^2 v'^2$}\\
     \hline
		\end{tabular}}
	\end{center}
        \caption{Subspaces for $(2n+1)$-d representation of SU(2) with specific remnant symmetry and vacuum configuration. We also have to note that in the 5-d situation, the second minimal condition has two solutions. One is the angle $\alpha=(\pm\pi/2 + 2k\pi)/3$, and the Lagrangian actually satisfies $\text{U}(1)$ symmetry which is not discrete. Another one is $\kappa'=0$ and $\alpha\neq(\pm\pi/2 + 2k\pi)/3$, which correspond to the pure $Z_2\oplus Z_2$ symmetry. The latter solution is used in the mass spectra. Here, $\sin(\alpha)$ and $\cos(\alpha)$ are denoted as $s_\alpha$ and $c_\alpha$, respectively, for simplicity.}
        \label{tab:VEV_Subspace_Results}
\end{table}

For $n\geq 13$, sometimes there is more than one non-Abelian discrete symmetry group that can remain. Since usually the results are too lengthy to be placed inside a table, we list them in the text below. We also have to note that for all the multiplets both in Table \ref{tab:VEV_Subspace_Results} and below, the first triplet representation labeled as ``G" of the discrete group corresponds to the three Goldstone bosons.

Before listing all of our results, we would like to present some conventions. Once the VEV configuration is determined, there should still be some degrees of freedom indicating the multiplicity of the identical representations embedded in $\Phi_{(n)}$. We set the radial length of the $\text{SU}(2)_\text{D}$ VEV as $v'$ with
\begin{equation}
    \langle \Phi_{(n)} \rangle^\dagger \langle \Phi_{(n)} \rangle = \frac{v'^2}{2}.   
\end{equation}
And the VEV configuration $\langle \Phi_{(n)} \rangle$ is a combination of the identical representations with the multiplicity $p$, where the combination coefficients are the function of the $p$ dimension spherical coordinate angles.
The coupling constant conventions are according to Table~\ref{tab:odd_operators}. For the quartic couplings, the effective potential contains
\begin{eqnarray}
    V \supset \lambda_k (\Phi^{\dagger}_{(n)} \tau^{a_1} \tau^{a_2} \tau^{a_3} \dots \tau^{a_{2(k - 1)}} \Phi_{(n)})^2,
\end{eqnarray}
for all the possible $k \geq 1$ integers. For the trilinear terms $\kappa \xi_{(n)}^{p q r} \Delta_p \Delta_q \Delta_r$, only $\kappa$ is required; however, it is convenient to define a dimensionless $\kappa^{\prime} \equiv \frac{\kappa}{v'}$. It is also worth mentioning that in Table~\ref{tab:odd_operators}, $\xi^{0,0,0}_{(2j+1)}=(-1)^{(j/2)} \times 2\sqrt{2}$.

In addition, we also list the minimal conditions and the mass squared matrices of the dark sector. The minimal condition of the SM Higgs sector is given by
\begin{equation}
    0 = -m_{H}^2 + \lambda_{H} v^2 +\frac{1}{2}\lambda_{H\Phi} v'^2.
\end{equation}
Besides, the SM Higgs can also mix with the identical components of $\Phi_{(n)}$. Without loss of generality, we can always rotate to a basis that only one identical component receives nonzero VEV $v'$ while the remaining $p-1$ identical components vanish. In such a basis, the mass matrix of the SM Higgs and the $\text{SU}(2)_\text{D}$ discrete subgroup identical representation components appears in the form of
\begin{eqnarray}
    (\mathcal{M}^2_{h\rho})_{(p+1) \times (p+1)}=\left( \begin{array}{cc}
     2 \lambda_H v^2 & (\mathcal{M}^2_{h \rho})_{1 \times p}\\
    (\mathcal{M}^2_{h \rho})^{\text{T}}_{p \times 1} & (\mathcal{M}^2_{\rho})_{p \times p}
   \end{array} \right), \label{Mix_SM}\\ 
   \text{with} \qquad (\mathcal{M}^2_{h \rho})_{1 \times p} = \left(\begin{array}{cccc}
     \lambda_{H \Phi} v v' & 0 & \cdots & 0
   \end{array} \right).
\end{eqnarray}
Since all the components in Eq.~\eqref{Mix_SM} are universal in their forms in different $\text{SU}(2)_\text{D}$ representation cases except all the components in the submatrix $\mathcal{M}^2_{\rho}$, in the following of this paper, we only show $\mathcal{M}^2_{\rho}$ in each of the cases.


\begin{itemize}
    \item 
For the 13-d representation, the general scalar field is given by
\begin{equation}
    \Phi_{(13)}=\frac{1}{\sqrt{2}}\left(\Delta _6,\Delta _5,\Delta _4,\dots,\Delta _{-4},\Delta _{-5},\Delta _{-6}\right)^\text{T}.
\end{equation}
Specific vacuum configurations might break the $\text{SU(2)}_\text{D}$ symmetry into the $A_4$, $S_4$, and $A_5$ discrete groups. We are then going to discuss these cases separately.
\begin{itemize}
    \item[*]
Let us begin with the $A_5$ case. 
The vacuum configuration is calculated to be
\begin{equation}
    \vec{v}=\frac{v^\prime}{\sqrt{2}}\left(0,\frac{\sqrt{7}}{5},0,0,0,0,\frac{\sqrt{11}}{5},0,0,0,0,\frac{\sqrt{7}}{5},0\right)^\text{T}.
\end{equation}
The subspaces of $A_5$ corresponding to $\bf{13}=\bf{1}+\bf{3}+\bf{4}+\bf{5}$ are given by
\begin{equation}
	\left(\frac{\sqrt{11} }{5}\Delta _0+\frac{2 \sqrt{7} }{5}\text{Re$\Delta $}_5\right)
\end{equation},
\begin{equation}
    \left(
	\begin{array}{c}
		\frac{\sqrt{33}}{5} \text{Im$\Delta $}_1+\frac{\sqrt{11} }{5}\text{Im$\Delta $}_4+\frac{\sqrt{6} }{5} \text{Im$\Delta $}_6\\
		\frac{\sqrt{33} }{5}\text{Re$\Delta $}_1+\frac{\sqrt{6} }{5}\text{Re$\Delta $}_6-\frac{\sqrt{11} }{5}\text{Re$\Delta $}_4 \\
		-\sqrt{2} \text{Im$\Delta $}_5 \\
	\end{array}
	\right)^\text{G},
\end{equation}
\begin{equation}
    \left(
	\begin{array}{c}
		\frac{2\sqrt{2}}{\sqrt{5}} \text{Re$\Delta $}_3-\sqrt{\frac{2}{5}} \text{Re$\Delta $}_2 \\
		\sqrt{\frac{2}{5}} \text{Im$\Delta $}_2+	\frac{2\sqrt{2}}{\sqrt{5}} \text{Im$\Delta $}_3 \\
		-\frac{4 }{5}\text{Re$\Delta $}_1+\frac{\sqrt{22} }{5}\text{Re$\Delta $}_6-\frac{2 \sqrt{3} }{5} \text{Re$\Delta $}_4\\
		-\frac{4 }{5}\text{Im$\Delta $}_1+\frac{2 \sqrt{3} }{5}\text{Im$\Delta $}_4+\frac{\sqrt{22} }{5} \text{Im$\Delta $}_6\\
	\end{array}
	\right),
\end{equation}
\begin{equation}
   \left(
	\begin{array}{c}
		2 \sqrt{\frac{2}{5}} \text{Re$\Delta $}_2+\sqrt{\frac{2}{5}} \text{Re$\Delta $}_3 \\
		2 \sqrt{\frac{2}{5}} \text{Im$\Delta $}_2-\sqrt{\frac{2}{5}} \text{Im$\Delta $}_3 \\
		\frac{1}{5}\text{Re$\Delta $}_1+\frac{3 \sqrt{3} }{5}\text{Re$\Delta $}_4+\frac{\sqrt{22} }{5}\text{Re$\Delta $}_6 \\
		\frac{1}{5}\text{Im$\Delta $}_1-\frac{3 \sqrt{3} }{5}\text{Im$\Delta $}_4+\frac{\sqrt{22} }{5}\text{Im$\Delta $}_6 \\
		\frac{\sqrt{22} }{5}\text{Re$\Delta $}_5-\frac{\sqrt{14}}{5}  \Delta _0\\
	\end{array}
	\right).
\end{equation}
The minimal condition and squared mass (matrices) of the dark sector are given by
\begin{equation}
    0 = -\mu^2 + \frac{1}{2} \lambda_{H\Phi} v^2 + \left( \lambda_1 + 588 \lambda_2 + 612696 \lambda_3 - \frac{33 \sqrt{11}}{20} \kappa' \right)v'^2,
\end{equation}
\begin{equation}
    \mathcal{M}^2_{\rho} = 2(\lambda_1 + 588\lambda_2 + 612696\lambda_3 - \frac{33\sqrt{11}}{40}\kappa') v'^2,
\end{equation}
\begin{equation}
    m_{\textbf{4}}^2 = (124740\lambda_3 + \frac{63\sqrt{11}}{20}\kappa') v'^2,\quad m_{\textbf{5}}^2 = (462\lambda_2 + 815430\lambda_3 + \frac{21\sqrt{11}}{10}\kappa') v'^2,
\end{equation}
\begin{equation}
    m_{X_1}^2=m_{X_2}^2=m_{X_3}^2= \frac{7}{2} g_{\text{D}}^2 v'^2.
\end{equation}

\item[*] Now let us turn to the $S_4$ symmetry. The vacuum configuration is parametrized as
\begin{equation}\label{vacuumof13dS4}
	\vec{v}=\frac{v^\prime}{\sqrt{2}}\left(0,0,\frac{\sqrt{7}}{4},0,0,0,\frac{1}{2 \sqrt{2}} ,0,0,0,\frac{\sqrt{7}}{4},0,0\right)^\text{T}.
\end{equation}
The subspaces of $S_4$ corresponding to $\bf{13=1+1'+2+3+3'+3'}$ are given by
\begin{equation}
	\Delta_{S_4,13,1}=\left(\frac{1}{2 \sqrt{2}}\Delta _0+\frac{\sqrt{7} }{2}\text{Re$\Delta $}_4\right), \label{n13_Singlet1}
\end{equation}
\begin{equation}
    \Delta_{S_4,13,1'}=\left(-\frac{1}{2} \sqrt{\frac{11}{2}} \text{Im$\Delta $}_2-\frac{1}{2} \sqrt{\frac{5}{2}} \text{Im$\Delta $}_6
    \right), \label{n13_Singlet2}
\end{equation}
\begin{equation}
   \left(
        \begin{array}{c}
    	-\frac{1}{2} \sqrt{\frac{7}{2}} \Delta _0 + \frac{1}{2}\text{Re$\Delta $}_4\\
    	\frac{1}{2} \sqrt{\frac{5}{2}} \text{Im$\Delta $}_2-\frac{1}{2} \sqrt{\frac{11}{2}} \text{Im$\Delta $}_6 \\
        \end{array}
    \right), \label{n13_Doublet}
\end{equation}
\begin{equation}
        \Delta_{S_4,13,3}=\left(
	\begin{array}{c}
	\frac{1}{2} \sqrt{\frac{3}{2}} \text{Im$\Delta $}_1+\frac{\sqrt{15} }{4}\text{Im$\Delta $}_3+\frac{\sqrt{11} }{4}\text{Im$\Delta $}_5 \\
	\frac{1}{2} \sqrt{\frac{3}{2}} \text{Re$\Delta $}_1-\frac{\sqrt{15}}{4} \text{Re$\Delta $}_3+\frac{\sqrt{11}}{4} \text{Re$\Delta $}_5 \\
	-\sqrt{2} \text{Im$\Delta $}_4 \\
	\end{array}
	\right)^\text{G}, \label{n13_Triplet1}
\end{equation}
\begin{equation}
	\Delta_{S_4,13,3'_1}=\left(
    \begin{array}{c}
        \frac{1}{2} \sqrt{\frac{55}{14}} \text{Re$\Delta $}_1+\frac{3}{4} \sqrt{\frac{11}{7}} \text{Re$\Delta $}_3+\frac{1}{4} \sqrt{\frac{15}  {7}} \text{Re$\Delta $}_5 \\
        \frac{1}{2} \sqrt{\frac{55}{14}} \text{Im$\Delta $}_1-\frac{3}{4} \sqrt{\frac{11}{7}} \text{Im$\Delta $}_3+\frac{1}{4} \sqrt{\frac{15}{7}} \text{Im$\Delta $}_5 \\
        -\frac{1}{2} \sqrt{\frac{11}{7}} \text{Re$\Delta $}_2-\frac{3}{2} \sqrt{\frac{5}{7}} \text{Re$\Delta $}_6 \\
        \end{array}
	\right), \label{n13_Triplet2}
\end{equation}
\begin{equation}
	\left(
	\begin{array}{c}
	-\frac{3 }{\sqrt{14}}\text{Re$\Delta $}_1+\frac{1}{2} \sqrt{\frac{5}{7}} \text{Re$\Delta $}_3+\frac{1}{2} \sqrt{\frac{33}{7}} \text{Re$\Delta $}_5 \\
	-\frac{3}{\sqrt{14}} \text{Im$\Delta $}_1 - \frac{1}{2} \sqrt{\frac{5}{7}} \text{Im$\Delta $}_3 + \frac{1}{2} \sqrt{\frac{33}{7}} \text{Im$\Delta $}_5 \\
	-\frac{3}{2} \sqrt{\frac{5}{7}} \text{Re$\Delta $}_2 + \frac{1}{2} \sqrt{\frac{11}{7}} \text{Re$\Delta $}_6 \\
	\end{array}
	\right). \label{n13_S4_3Prime_2}
\end{equation}
The minimal condition and squared mass (matrices) are given by
\begin{equation}
    0 = -\mu^2 + \frac{1}{2} \lambda_{H\Phi} v^2 + \left( \lambda_1 + 588 \lambda_2 + 731766 \lambda_3 + \frac{3 \sqrt{2}}{10} \kappa' \right)v'^2,
\end{equation}
\begin{equation}
    \mathcal{M}^2_{\rho} = 2(\lambda_1 + 588\lambda_2 + 731766\lambda_3 + \frac{3\sqrt{2}}{20}\kappa') v'^2,
\end{equation}
\begin{equation}
    m_{\textbf{1'}}^2 = (-181440\lambda_3 + \frac{63\sqrt{2}}{10}\kappa') v'^2,\quad m_{\textbf{2}}^2 = (84\lambda_2 + 114240\lambda_3 + \frac{21\sqrt{2}}{10}\kappa') v'^2,
\end{equation}
\begin{equation}
    \mathcal{M}_{\textbf{3'}}^2 = v'^2 \left( \begin{array}{cc}
     330\lambda_2 + 436650\lambda_3 + 3\sqrt{2}\kappa' & 48\sqrt{55}\lambda_2 + 71760\sqrt{55}\lambda_3 + \frac{3\sqrt{110}}{20}\kappa'\\
      48\sqrt{55}\lambda_2 + 71760\sqrt{55}\lambda_3 + \frac{3\sqrt{110}}{20}\kappa' &  384\lambda_2 + 477690\lambda_3 - \frac{51\sqrt{2}}{10}\kappa'
   \end{array} \right),
\end{equation}
\begin{equation}
    m_{X_1}^2=m_{X_2}^2=m_{X_3}^2= \frac{7}{2} g_{\text{D}}^2 v'^2.
\end{equation}

\item[*] Finally, we consider the case of $A_4$ symmetry. In principle, it can be computed with the algorithm described in Sec. \ref{VEVPatternSection} and \ref{ScalarClassificationSection} independently. However, once the fact that $A_4 \subset S_4$ is noticed, we can straightforwardly utilize Eqs.~\eqref{vacuumof13dS4}--\eqref{n13_S4_3Prime_2} to write the $A_4$ results. By observing the character in Tables~\ref{tab:Character_S4} and \ref{tab:Character_A4}, one finds the relations between the representation of the two groups:
\begin{align}
	\bf{1},\bf{1'} \  \text{of}\  S_4 &\leftrightarrow \bf{1} \ \text{of}\  A_4, \nonumber
	\\ 	\bf{3},\bf{3'} \  \text{of}\  S_4 &\leftrightarrow \bf{3} \  \text{of}\  A_4, \nonumber
	\\ 	\bf{2} \  \text{of}\  S_4 &\leftrightarrow \bf{1'},\bf{1''} \ \text{of}\ A_4. \label{S4toA4}
\end{align}
One can therefore realize that Eq.~\eqref{vacuumof13dS4}, corresponding to Eq.~\eqref{n13_Singlet1} can also be part of the VEV configuration in the $A_4$ case. However, pure Eq.~\eqref{vacuumof13dS4} respects the $S_4$ symmetry, so we consider another possible part of the vacuum configuration as follows:

\begin{equation}
\begin{aligned}
   \vec{v}=\frac{v^{\prime}}{\sqrt{2}} &\left(-\frac{1}{4} i \sqrt{\frac{5}{2}} c_\alpha,0,\frac{\sqrt{7}}{4}  s_\alpha,0,-\frac{1}{4} i \sqrt{\frac{11}{2}} c_\alpha,0,\right.
   \\& \left.
   \frac{s_\alpha}{2 \sqrt{2}},0,\frac{1}{4} i \sqrt{\frac{11}{2}} c_\alpha,0,\frac{\sqrt{7}}{4}  s_\alpha,0,\frac{1}{4} i \sqrt{\frac{5}{2}} c_\alpha\right).
   \end{aligned}
\end{equation}\label{n13_A4_VEV}

Since the subspace of $A_4$ is decomposed into $\bf{13=1+1+1'+1''+3+3+3}$, from Eq.~\eqref{S4_3d_SK} one can immediately realize that the two $\bf{1}$, and the three $\bf{3}$ representations of the $A_4$ group can be expressed by Eqs.~\eqref{n13_Singlet1}, \eqref{n13_Singlet2}, \eqref{n13_Triplet1}, \eqref{n13_Triplet2}, and \eqref{n13_S4_3Prime_2} by linear combination. 
For Eq.\eqref{n13_Doublet}, it should be decomposed into the $\bf{1^{\prime}}$ and $\bf{1^{\prime \prime}}$ representations of the $A_4$ group. 
We show the results of the four singlets and the first two triplets as

\begin{equation}\label{n13_A4_singlet1}
	   s_\alpha \times \Delta_{S_4,13,1} + c_\alpha \times \Delta_{S_4,13,1'},
\end{equation}
 \begin{equation}\label{n13_A4_singlet2}
    c_\alpha \times\Delta_{S_4,13,1} - s_\alpha \times\Delta_{S_4,13,1'}  
    ,
\end{equation}
\begin{equation}\label{n13_A4_singlet3}
    \left(-\frac{\sqrt{7} }{4}\Delta _0+\frac{1}{2 \sqrt{2}}\text{Re$\Delta $}_4-i\frac{\sqrt{5}}{4}   \text{Im$\Delta $}_2+ i\frac{ \sqrt{11}}{4} \text{Im$\Delta $}_6\right),
\end{equation}
\begin{equation}\label{n13_A4_singlet4}
    \left(\frac{\sqrt{7} }{4}\Delta _0-\frac{1}{2 \sqrt{2}}\text{Re$\Delta $}_4-i\frac{ \sqrt{5}}{4}  \text{Im$\Delta $}_2+i\frac{\sqrt{11}}{4}   \text{Im$\Delta $}_6\right),
\end{equation}

\begin{equation}\label{n13_A4_triplet1}
    s_\alpha \times \Delta_{S_4,13,3} + c_\alpha \times \Delta_{S_4,13,3'_1}^\text{G},
\end{equation}
\begin{equation}\label{n13_A4_triplet2}
    c_\alpha \times\Delta_{S_4,13,3} - s_\alpha \times\Delta_{S_4,13,3'_1},
\end{equation}
The first two singlets Eqs.~\eqref{n13_A4_singlet1} and ~\eqref{n13_A4_singlet2} are the linear combinations of the Eqs.~\eqref{n13_Singlet1} and ~\eqref{n13_Singlet2}. The two triplets Eqs.~\eqref{n13_A4_triplet1} and ~\eqref{n13_A4_triplet2} are the linear combinations of the Eqs.~\eqref{n13_Triplet1} and ~\eqref{n13_Triplet2}. And the remaining one triplet is inherited from Eq.~\eqref{n13_S4_3Prime_2}.

The two minimal conditions are given by
\begin{align}
    0 =& -\mu^2 + \frac{1}{2} \lambda_{H\Phi} v^2 + [ \lambda_1 + 588 \lambda_2 + (654006 - 43200c_{2\alpha} + 34560 c_{4\alpha}) \lambda_3 \notag\\
    &+ (\frac{27\sqrt{2} s_\alpha}{10} + \frac{12\sqrt{2} s_{3\alpha}}{5})\kappa'] v'^2,\notag\\
    0 =& (3 c_\alpha + 8 c_{3\alpha})(\kappa' - 14400\sqrt{2} s_\alpha \lambda_3).
\end{align}
In the second condition, if $\alpha=\arctan \sqrt{11/21}$, or $\alpha=\pi/2$, the first bracket vanishes, corresponding to the restoration of the $A_5$ and $S_4$ symmetry respectively. If only the $A_4$ symmetry remains, the second bracket in the second condition needs to be zero. In this case, the (components of the) squared masses (matrices) are
\begin{equation}
    (\mathcal{M}^2_{\rho})_{11} = 2[\lambda_1 + 588\lambda_2 + (673446 - 45360 c_{2\alpha} + 17280 c_{4\alpha})\lambda_3] v'^2,
\end{equation}
\begin{equation}
    (\mathcal{M}^2_{\rho})_{22} = -(12960 + 47520 c_{2\alpha} + 34560 c_{4\alpha})\lambda_3 v'^2,
\end{equation}
\begin{equation}
     (\mathcal{M}^2_{\rho})_{12} = (21600 s_{2\alpha} - 34560 s_{4\alpha})\lambda_3 v'^2,
\end{equation}
\begin{equation}
    m_{\textbf{1'}}^2=m_{\textbf{1''}}^2 = [(372 + 288c_{2\alpha}) \lambda_2 + (570720 + 516960 c_{2\alpha})\lambda_3] v'^2,
\end{equation}
\begin{equation}
    \mathcal{M}_{\textbf{3}}^2 = v'^2 \left( \begin{array}{cc}
     330\lambda_2 + 523050\lambda_3 & -48 \sqrt{55} s_\alpha \lambda_2- 76080\sqrt{55} s_\alpha \lambda_3\\
      -48 \sqrt{55} s_\alpha \lambda_2- 76080\sqrt{55} s_\alpha \lambda_3 & 192(1-c_{2\alpha})\lambda_2 + (238170 - 92640 c_{2\alpha})\lambda_3
   \end{array} \right),
\end{equation}
\begin{equation}
    m_{X_1}^2=m_{X_2}^2=m_{X_3}^2= \frac{7}{2} g_{\text{D}}^2 v'^2.
\end{equation}

\end{itemize}

\item 
For the 15-d representation, the complete scalar field is given by
\begin{equation}
	\Phi_{(15)}=\frac{1}{\sqrt{2}}\left(\Delta _7,\Delta _6,\Delta _5,\Delta _4,\Delta _3,\Delta _2,\Delta _1,\Delta _0,\Delta _{-1},\Delta _{-2},\Delta _{-3},\Delta _{-4},\Delta _{-5},\Delta _{-6},\Delta _{-7}\right)^\text{T}.
\end{equation}
When the $\text{SU(2)}_\text{D}$ group is broken into the $A_4$ symmetry, the VEV configuration is given by
\begin{equation}
    \vec{v}=\frac{v^\prime}{\sqrt{2}}\left(0,\frac{4 \sqrt{11}}{27},0,0,\frac{2 \sqrt{13}}{27},0,0,-\frac{1}{9}\sqrt{\frac{91}{3}},0,0,\frac{2 \sqrt{13}}{27},0,0,\frac{4 \sqrt{11}}{27},0\right)^\text{T}.
\end{equation}
The subspaces of $A_4$ corresponding to the $\bf{15}=\bf{1}+\bf{1'}+\bf{1''}+\bf{3}+\bf{3}+\bf{3}+\bf{3}$ are given by
\begin{equation}
   \left(-\frac{1}{9} \sqrt{\frac{91}{3}} \Delta _0+\frac{4 \sqrt{13}}{27} \text{Re$\Delta $}_3+\frac{8 \sqrt{11} }{27}\text{Re$\Delta $}_6
   \right),
\end{equation}
\begin{equation}
   \left(-\frac{1}{9} \sqrt{\frac{22}{3}} \Delta _{-1}+\frac{4 \sqrt{22} }{27}\Delta _2+\frac{1}{27}\Delta _{-4}+\frac{8 \sqrt{2} }{27}\Delta _5+\frac{ \sqrt{182}}{27} \Delta _{-7}
   \right),
\end{equation}
\begin{equation}
    \left(-\frac{1}{9} \sqrt{\frac{22}{3}} \Delta _1+\frac{4\sqrt{22}}{27}  \Delta _{-2}+\frac{1}{27}\Delta _4+ \frac{8\sqrt{2}}{27}  \Delta _{-5}+\frac{\sqrt{182} }{27}	\Delta _7\right),
\end{equation}
\begin{equation}
    \left(
	\begin{array}{c}
	-\frac{\sqrt{91}}{9}  \text{Im$\Delta $}_1+\frac{5}{9} \sqrt{\frac{13}{21}} \text{Im$\Delta $}_2+\frac{1}{9} \sqrt{\frac{286}{21}} \text{Im$\Delta $}_4+\frac{2}{9} \sqrt{\frac{143}{21}} \text{Im$\Delta $}_5+\frac{2}{9} \sqrt{\frac{11}{3}} \text{Im$\Delta $}_7 \\
	-\frac{\sqrt{91}}{9}  \text{Re$\Delta $}_1-\frac{5}{9} \sqrt{\frac{13}{21}} \text{Re$\Delta $}_2+\frac{1}{9} \sqrt{\frac{286}{21}} \text{Re$\Delta $}_4-\frac{2}{9} \sqrt{\frac{143}{21}} \text{Re$\Delta $}_5+\frac{2}{9} \sqrt{\frac{11}{3}} \text{Re$\Delta $}_7 \\
	-\frac{1}{3} \sqrt{\frac{26}{21}} \text{Im$\Delta $}_3-\frac{4}{3} \sqrt{\frac{22}{21}} \text{Im$\Delta $}_6 \\
	\end{array}
	\right)^\text{G},
\end{equation}
\begin{equation}
   	\left(
	\begin{array}{c}
		\frac{2}{3} \sqrt{\frac{11}{21}} \text{Im$\Delta $}_2-\frac{11}{3} \sqrt{\frac{2}{21}} \text{Im$\Delta $}_4-\frac{1}{3 \sqrt{21}}\text{Im$\Delta $}_5+\frac{1}{3} \sqrt{\frac{13}{3}} \text{Im$\Delta $}_7 \\
		\frac{2}{3} \sqrt{\frac{11}{21}} \text{Re$\Delta $}_2+\frac{11}{3} \sqrt{\frac{2}{21}} \text{Re$\Delta $}_4-\frac{1}{3 \sqrt{21}}\text{Re$\Delta $}_5 -\frac{1}{3} \sqrt{\frac{13}{3}} \text{Re$\Delta $}_7\\
		-\frac{4}{3} \sqrt{\frac{22}{21}} \text{Im$\Delta $}_3+\frac{1}{3} \sqrt{\frac{26}{21}} \text{Im$\Delta $}_6 \\
	\end{array}
	\right),
\end{equation}
\begin{equation}
    	\left(
	\begin{array}{c}
		-\frac{1}{\sqrt{47}}\text{Re$\Delta $}_1-3 \sqrt{\frac{3}{47}} \text{Re$\Delta $}_2+\sqrt{\frac{22}{141}} \text{Re$\Delta $}_4+4 \sqrt{\frac{11}{141}} \text{Re$\Delta $}_5 \\
		-\frac{1}{\sqrt{47}}\text{Im$\Delta $}_1+3 \sqrt{\frac{3}{47}} \text{Im$\Delta $}_2+\sqrt{\frac{22}{141}} \text{Im$\Delta $}_4-4 \sqrt{\frac{11}{141}} \text{Im$\Delta $}_5 \\
		\frac{2}{3} \sqrt{\frac{14}{47}} \Delta _0-\frac{23}{3} \sqrt{\frac{2}{141}} \text{Re$\Delta $}_3+\frac{2}{3} \sqrt{\frac{286}{141}} \text{Re$\Delta $}_6 \\
	\end{array}
	\right),
\end{equation}
\begin{equation}
   	\left(
	\begin{array}{c}
		\frac{10}{9} \sqrt{\frac{77}{141}} \text{Re$\Delta $}_1-\frac{4}{27} \sqrt{\frac{77}{47}} \text{Re$\Delta $}_2+\frac{31}{27} \sqrt{\frac{14}{47}} \text{Re$\Delta $}_4-\frac{17}{27} \sqrt{\frac{7}{47}} \text{Re$\Delta $}_5+\frac{\sqrt{611} }{27}\text{Re$\Delta $}_7 \\
		\frac{10}{9} \sqrt{\frac{77}{141}} \text{Im$\Delta $}_1+\frac{4}{27} \sqrt{\frac{77}{47}} \text{Im$\Delta $}_2+\frac{31}{27} \sqrt{\frac{14}{47}} \text{Im$\Delta $}_4+\frac{17}{27} \sqrt{\frac{7}{47}} \text{Im$\Delta $}_5+\frac{\sqrt{611} }{27} \text{Im$\Delta $}_7\\
		\frac{16}{9} \sqrt{\frac{22}{141}} \Delta _0+\frac{14}{27} \sqrt{\frac{154}{47}} \text{Re$\Delta $}_3+\frac{5}{27} \sqrt{\frac{182}{47}} \text{Re$\Delta $}_6 \\
	\end{array}
	\right).
\end{equation}
The minimal condition and the squared mass (matrices) are
\begin{equation}
    0 = -\mu^2 + \frac{1}{2} \lambda_{H\Phi} v^2 + \left( \lambda_1 + \frac{3136}{3} \lambda_2 + \frac{5841440}{3} \lambda_3 \right)v'^2,
\end{equation}
\begin{equation}
    \mathcal{M}^2_{\rho} = 2(\lambda_1 + \frac{3136}{3}\lambda_2 + \frac{5841440}{3}\lambda_3) v'^2,
\end{equation}
\begin{equation}
    m_{\textbf{1'}}^2=m_{\textbf{1''}}^2 = (\frac{2288}{3}\lambda_2 + \frac{6883552}{3} \lambda_3)v'^2,
\end{equation}
\begin{equation}
    \mathcal{M}_{\textbf{3}}^2 = v'^2 \left( \begin{array}{ccc}
     687960\lambda_3 & 0 & 0\\
      0 & \frac{30056}{47}\lambda_2 + \frac{82996264}{47}\lambda_3 & \frac{3536\sqrt{231}}{141}\lambda_2 + \frac{11139232\sqrt{231}}{141}\lambda_3\\
      0 &  \frac{3536\sqrt{231}}{141}\lambda_2 + \frac{11139232\sqrt{231}}{141}\lambda_3 & \frac{32032}{141}\lambda_2 + \frac{129821432}{141}\lambda_3
   \end{array} \right),
\end{equation}
\begin{equation}
    m_{X_1}^2=m_{X_2}^2=m_{X_3}^2= \frac{56}{3} g_{\text{D}}^2 v'^2.
\end{equation}

\item
For the 17-d representation, the general scalar field is given by
\begin{equation}
    \Phi_{(17)}=\frac{1}{\sqrt{2}}\left(\Delta_8,\Delta _7,\Delta _6,\Delta _5,\dots,\Delta _{-5},\Delta _{-6},\Delta _{-7},\Delta_{-8}\right)^\text{T}.
\end{equation}
The vacuum configuration preserving the $S_4$ symmetry is given by
\begin{equation}\label{vacuumof17dS4}
   \vec{v}= \frac{v^\prime}{\sqrt{2}}\left(\frac{1}{8}\sqrt{\frac{65}{6}},0,0,0,-\frac{1}{4}\sqrt{\frac{7}{6}},0,0,0,\frac{\sqrt{33}}{8},0,0,0,-\frac{1}{4}\sqrt{\frac{7}{6}},0,0,0,\frac{1}{8}\sqrt{\frac{65}{6}}\right)^\text{T}.
\end{equation}
The subspaces of $S_4$ corresponding to $\bf{17=1+2+2+3+3+3'+3'}$ are given by
\begin{equation}
    \left(\frac{\sqrt{33} }{8}\Delta _0-\frac{1}{2} \sqrt{\frac{7}{6}} \text{Re$\Delta $}_4+\frac{1}{4} \sqrt{\frac{65}{6}} \text{Re$\Delta $}_8\right),
\end{equation}

\begin{equation}
	\left(
	\begin{array}{c}
		\sqrt{2} \text{Im$\Delta $}_6 \\
		\frac{1}{16} \sqrt{\frac{143}{2}} \Delta _0+\frac{\sqrt{91} }{8}\text{Re$\Delta $}_4-\frac{\sqrt{5}}{16} \text{Re$\Delta $}_8 \\
	\end{array}
	\right), 
\end{equation}
\begin{equation}
   \left(
	\begin{array}{c}
		\sqrt{2} \text{Im$\Delta $}_2 \\
		-\frac{1}{16} \sqrt{\frac{105}{2}} \Delta _0+\frac{1}{8} \sqrt{\frac{55}{3}} \text{Re$\Delta $}_4+\frac{1}{16} \sqrt{\frac{1001}{3}} \text{Re$\Delta $}_8 \\
	\end{array}
	\right),
\end{equation}
\begin{equation}
   \left(
	\begin{array}{c}
	\frac{3 \sqrt{11} }{8}\text{Im$\Delta $}_1-\frac{1}{8} \sqrt{\frac{35}{3}} \text{Im$\Delta $}_3-\frac{\sqrt{91} }{24} \text{Im$\Delta $}_5+\frac{\sqrt{65} }{24}\text{Im$\Delta $}_7\\
	\frac{3 \sqrt{11} }{8}\text{Re$\Delta $}_1+\frac{1}{8} \sqrt{\frac{35}{3}} \text{Re$\Delta $}_3-\frac{\sqrt{91} }{24}\text{Re$\Delta $}_5-\frac{\sqrt{65} }{24}\text{Re$\Delta $}_7 \\
	\frac{\sqrt{7} }{6}\text{Im$\Delta $}_4-\frac{\sqrt{65} }{6}\text{Im$\Delta $}_8 \\
	\end{array}
	\right)^\text{G},
\end{equation}
\begin{equation}
   \left(
	\begin{array}{c}
	\frac{1}{2} \sqrt{\frac{13}{3}} \text{Im$\Delta $}_3-\frac{\sqrt{5} }{6}\text{Im$\Delta $}_5+\frac{\sqrt{7} }{3}\text{Im$\Delta $}_7 \\
	\frac{1}{2} \sqrt{\frac{13}{3}} \text{Re$\Delta $}_3+\frac{\sqrt{5} }{6}\text{Re$\Delta $}_5+\frac{\sqrt{7} }{3}\text{Re$\Delta $}_7 \\
	\frac{\sqrt{65} }{6}\text{Im$\Delta $}_4+\frac{\sqrt{7} }{6} \text{Im$\Delta $}_8\\
	\end{array}
	\right),
\end{equation}
\begin{equation}
    \left(
	\begin{array}{c}
	-\frac{1}{2} \sqrt{\frac{13}{47}} \text{Re$\Delta $}_1-\sqrt{\frac{77}{47}} \text{Re$\Delta $}_5+\frac{1}{2} \sqrt{\frac{55}{47}} \text{Re$\Delta $}_7 \\
	\frac{1}{2} \sqrt{\frac{13}{47}} \text{Im$\Delta $}_1+\sqrt{\frac{77}{47}} \text{Im$\Delta $}_5+\frac{1}{2} \sqrt{\frac{55}{47}} \text{Im$\Delta $}_7 \\
	-\frac{1}{2} \sqrt{\frac{455}{94}} \text{Re$\Delta $}_2-\frac{3}{2} \sqrt{\frac{33}{94}} \text{Re$\Delta $}_6 \\
	\end{array}
	\right),
 \end{equation}
 \begin{equation}
        \left(
	\begin{array}{c}
	\frac{1}{8} \sqrt{\frac{1155}{47}} \text{Re$\Delta $}_1-\frac{\sqrt{47} }{8} \text{Re$\Delta $}_3+\frac{1}{8} \sqrt{\frac{195}{47}} \text{Re$\Delta $}_5+\frac{3}{8} \sqrt{\frac{273}{47}} \text{Re$\Delta $}_7\\
	-\frac{1}{8} \sqrt{\frac{1155}{47}} \text{Im$\Delta $}_1-\frac{\sqrt{47} }{8} \text{Im$\Delta $}_3-\frac{1}{8} \sqrt{\frac{195}{47}} \text{Im$\Delta $}_5+\frac{3}{8} \sqrt{\frac{273}{47}} \text{Im$\Delta $}_7\\
	\frac{1}{2} \sqrt{\frac{455}{94}} \text{Re$\Delta $}_6-\frac{3}{2} \sqrt{\frac{33}{94}} \text{Re$\Delta $}_2 \\
	\end{array}
	\right).
\end{equation}  
The minimal condition and the squared mass (matrices) are
\begin{equation}
    0 = -\mu^2 + \frac{1}{2} \lambda_{H\Phi} v^2 + \left( \lambda_1 + 1728 \lambda_2 + 6478176 \lambda_3 + \frac{3\sqrt{33}}{7}\kappa' \right)v'^2,
\end{equation}
\begin{equation}
    \mathcal{M}^2_{\rho} = 2(\lambda_1 + 1728\lambda_2 + 6478176\lambda_3 + \frac{3\sqrt{33}}{14}\kappa') v'^2,
\end{equation}
\begin{equation}
    \mathcal{M}_{\textbf{2}}^2 = v'^2 \left( \begin{array}{cc}
     \frac{102960}{41}\lambda_2 + \frac{445121760}{41}\lambda_3 - \frac{264\sqrt{33}}{1435} \kappa' & \frac{96\sqrt{15015}}{41}\lambda_2 + \frac{729600\sqrt{15015}}{41}\lambda_3 -\frac{186\sqrt{455}}{1435}\kappa' \\
      \frac{96\sqrt{15015}}{41}\lambda_2 + \frac{729600\sqrt{15015}}{41}\lambda_3 -\frac{186\sqrt{455}}{1435}\kappa' & \frac{1344}{41}\lambda_2 - \frac{5214720}{41}\lambda_3 - \frac{3\sqrt{33}}{41} \kappa'
   \end{array} \right),
\end{equation}
\begin{equation}
    m_{\textbf{3}}^2 = -(748440 \lambda_3 + \frac{54 \sqrt{33}}{35}\kappa') v'^2,
\end{equation}
\begin{equation}
    \mathcal{M}_{\textbf{3'}}^2 = v'^2 \left( \begin{array}{cc}
     \frac{13728}{47}\lambda_2 - \frac{23969640}{47}\lambda_3 + \frac{54\sqrt{33}}{329} \kappa' & \frac{112\sqrt{15015}}{47}\lambda_2 + \frac{747040\sqrt{15015}}{47}\lambda_3 -\frac{372\sqrt{455}}{1645}\kappa' \\
      \frac{112\sqrt{15015}}{47}\lambda_2 + \frac{747040\sqrt{15015}}{47}\lambda_3 -\frac{372\sqrt{455}}{1645}\kappa' & \frac{13720}{47}\lambda_2 + \frac{21040600}{47}\lambda_3 - \frac{48\sqrt{33}}{47} \kappa'
   \end{array} \right),
\end{equation}
\begin{equation}
    m_{X_1}^2=m_{X_2}^2=m_{X_3}^2= 24 g_{\text{D}}^2 v'^2.
\end{equation}

\item
For the 19-d representation, the general scalar field is given by
\begin{equation}
    \Phi_{(19)}=\frac{1}{\sqrt{2}}\left(\Delta_9,\Delta_8,\Delta _7,\Delta _6,\Delta _5,\dots,\Delta _{-5},\Delta _{-6},\Delta _{-7},\Delta_{-8},\Delta_{-9}\right)^\text{T}.
\end{equation}
In this case, the $\text{SU(2)}_\text{D}$ group might be broken into the $S_4$ or $A_4$ discrete group. We discuss them separately again.
\begin{itemize}
    \item[*]
The vacuum configuration preserving $S_4$ symmetry is given by
\begin{equation}\label{vacuumof19dS4}
    \vec{v} = i	\frac{v^\prime}{\sqrt{2}}\left(0,-\frac{1}{4}  \sqrt{\frac{7}{3}},0,0,0,-\frac{1}{4} \sqrt{\frac{17}{3}},0,0,0,0,0,0,0,\frac{1}{4}  \sqrt{\frac{17}{3}},0,0,0,\frac{1}{4} \sqrt{\frac{7}{3}},0\right)^\text{T}.
\end{equation}
The subspaces of $S_4$ corresponding to $\bf{19=1+1'+2+3+3+3+3'+3'}$ are given by
\begin{equation}
    \Delta_{S_4,19,1}=\left(-\frac{1}{2} \sqrt{\frac{17}{3}} \text{Im$\Delta $}_4 - \frac{1}{2} \sqrt{\frac{7}{3}} \text{Im$\Delta $}_8\right),  \label{n19_Singlet1}
\end{equation}
\begin{equation}
    \Delta_{S_4,19,1'}=\left(\frac{1}{2} \sqrt{\frac{3}{2}} \text{Re$\Delta $}_2+\frac{1}{2} \sqrt{\frac{13}{2}} \text{Re$\Delta $}_6\right), \label{n19_Singlet2}
\end{equation}
\begin{equation}
     \left(
	\begin{array}{c}\frac{1}{2} \sqrt{\frac{7}{3}} \text{Im$\Delta $}_4-\frac{1}{2} \sqrt{\frac{17}{3}} \text{Im$\Delta $}_8
		\\
		 - \frac{1}{2} \sqrt{\frac{13}{2}} \text{Re$\Delta $}_2 + \frac{1}{2} \sqrt{\frac{3}{2}} \text{Re$\Delta $}_6\\
		\end{array}\right), \label{n19_Doublet}
\end{equation}
\begin{equation}
    \Delta_{S_4,19,3_1}=\left(
	\begin{array}{c}
	\frac{1}{4} \sqrt{\frac{221}{15}} \text{Re$\Delta $}_3+\frac{\sqrt{119} }{12}\text{Re$\Delta $}_5+\frac{1}{12} \sqrt{\frac{119}{5}} \text{Re$\Delta $}_7+\frac{1}{4} \sqrt{\frac{7}{5}} \text{Re$\Delta $}_9 \\
	\frac{1}{4} \sqrt{\frac{221}{15}} \text{Im$\Delta $}_3 - \frac{\sqrt{119} }{12}\text{Im$\Delta $}_5 + \frac{1}{12} \sqrt{\frac{119}{5}} \text{Im$\Delta $}_7 - \frac{1}{4} \sqrt{\frac{7}{5}} \text{Im$\Delta $}_9\\
	-\frac{1}{3} \sqrt{\frac{34}{5}} \text{Re$\Delta $}_4-\frac{2}{3} \sqrt{\frac{14}{5}} \text{Re$\Delta $}_8 \\
	\end{array}
	\right)^\text{G}, \label{n19_Triplet1}
\end{equation}
\begin{equation}
    \left(
        \begin{array}{c}
    \frac{3}{64} \sqrt{\frac{715}{2}} \text{Re$\Delta $}_1 - \frac{13}{64} \sqrt{\frac{91}{15}} \text{Re$\Delta $}_3 - \frac{1}{192}\text{Re$\Delta $}_5 + \frac{763}{384 \sqrt{5}}\text{Re$\Delta $}_7 + \frac{29}{128} \sqrt{\frac{17}{5}} \text{Re$\Delta $}_9 \\
    \frac{3}{64} \sqrt{\frac{715}{2}} \text{Im$\Delta $}_1 + \frac{13}{64} \sqrt{\frac{91}{15}} \text{Im$\Delta $}_3 - \frac{1}{192}\text{Im$\Delta $}_5 - \frac{763}{384 \sqrt{5}}\text{Im$\Delta $}_7 + \frac{29}{128} \sqrt{\frac{17}{5}} \text{Im$\Delta $}_9 \\
    - \frac{2}{3} \sqrt{\frac{14}{5}} \text{Re$\Delta $}_4 + \frac{1}{3} \sqrt{\frac{34}{5}} \text{Re$\Delta $}_8 \\
    \end{array}
    \right), \label{n19_Triplet2}
\end{equation}
\begin{equation}
    \left(
	\begin{array}{c}
	\frac{21}{64} \sqrt{\frac{5}{2}} \text{Re$\Delta $}_1+\frac{\sqrt{1155} }{64}\text{Re$\Delta $}_3-\frac{3 \sqrt{143} }{64}\text{Re$\Delta $}_5-\frac{3 \sqrt{715} }{128}\text{Re$\Delta $}_7 +\frac{\sqrt{12155} }{128}\text{Re$\Delta $}_9\\
	\frac{21}{64} \sqrt{\frac{5}{2}} \text{Im$\Delta $}_1-\frac{\sqrt{1155} }{64}\text{Im$\Delta $}_3-\frac{3 \sqrt{143} }{64}\text{Im$\Delta $}_5+\frac{3 \sqrt{715}}{128} \text{Im$\Delta $}_7+\frac{\sqrt{12155} }{128}\text{Im$\Delta $}_9 \\
	\Delta _0 \\
	\end{array}
	\right), \label{n19_Triplet3}
\end{equation}
\begin{equation}
		\Delta_{S_4,19,3'_1}=\left(
	\begin{array}{c}
		\frac{1}{2} \sqrt{\frac{11}{10}} \text{Im$\Delta $}_1+\frac{1}{4} \sqrt{\frac{21}{5}} \text{Im$\Delta $}_3+\frac{\sqrt{13}}{4} \text{Im$\Delta $}_5+\frac{1}{2} \sqrt{\frac{13}{5}} \text{Im$\Delta $}_7 \\
		-\frac{1}{2} \sqrt{\frac{11}{10}} \text{Re$\Delta $}_1+\frac{1}{4} \sqrt{\frac{21}{5}} \text{Re$\Delta $}_3-\frac{\sqrt{13} }{4}\text{Re$\Delta $}_5+\frac{1}{2} \sqrt{\frac{13}{5}} \text{Re$\Delta $}_7 \\
		-\frac{1}{2 \sqrt{5}}\text{Im$\Delta $}_2-\frac{1}{2} \sqrt{\frac{39}{5}} \text{Im$\Delta $}_6 \\
	\end{array}
	\right) \label{n19_Triplet4}
\end{equation}
\begin{equation}
	\left(
	\begin{array}{c}
		\frac{1}{8} \sqrt{\frac{429}{10}} \text{Im$\Delta $}_1 - \frac{1}{8} \sqrt{\frac{91}{5}} \text{Im$\Delta $}_3 - \frac{\sqrt{3} }{8}\text{Im$\Delta $}_5 + \frac{1}{16} \sqrt{\frac{3}{5}} \text{Im$\Delta $}_7 - \frac{\sqrt{255} }{16}\text{Im$\Delta $}_9 \\
		-\frac{1}{8} \sqrt{\frac{429}{10}} \text{Re$\Delta $}_1 - \frac{1}{8} \sqrt{\frac{91}{5}} \text{Re$\Delta $}_3 + \frac{\sqrt{3} }{8}\text{Re$\Delta $}_5 + \frac{1}{16} \sqrt{\frac{3}{5}} \text{Re$\Delta $}_7 + \frac{\sqrt{255} }{16}\text{Re$\Delta $}_9 \\
		\frac{1}{2} \sqrt{\frac{39}{5}} \text{Im$\Delta $}_2-\frac{1}{2 \sqrt{5}}\text{Im$\Delta $}_6 \\
	\end{array}
	\right). \label{n19_Treplet5}
\end{equation}
The minimal condition  and the squared mass (matrices) are
\begin{equation}
    0 = -\mu^2 + \frac{1}{2} \lambda_{H\Phi} v^2 + \left( \lambda_1 + 2700 \lambda_2 + 13450710 \lambda_3 + 81456784830 \lambda_4 \right)v'^2,
\end{equation}
\begin{equation}
    \mathcal{M}^2_{\rho} = 2(\lambda_1 + 2700 \lambda_2 + 13450710 \lambda_3 + 81456784830 \lambda_4) v'^2,
\end{equation}
\begin{equation}
    m_{\textbf{1'}}^2 = (571200\lambda_3 + 8264692800\lambda_4)v'^2,
\end{equation}
\begin{equation}
    m_{\textbf{2}}^2 = (1428\lambda_2 + 9436224\lambda_3 + 63190410864\lambda_4)v'^2,
\end{equation}
\begin{equation}
    (\mathcal{M}_{\textbf{3}}^2)_{11}=0, \qquad (\mathcal{M}_{\textbf{3}}^2)_{12}=(-160650\sqrt{143}\lambda_3 - 1422234450\sqrt{143}\lambda_4)v'^2,
\end{equation}
\begin{equation}
    (\mathcal{M}_{\textbf{3}}^2)_{22}=(3462900\lambda_3 + 33864912900\lambda_4)v'^2,
\end{equation}
\begin{equation}
    (\mathcal{M}_{\textbf{3'}}^2)_{11} = (\frac{9282}{5}\lambda_2 + \frac{83130306}{5}\lambda_3 + \frac{621882054066}{5}\lambda_4)v'^2,
\end{equation}
\begin{equation}
    (\mathcal{M}_{\textbf{3'}}^2)_{12} = (-\frac{952\sqrt{39}}{5}\lambda_2 - \frac{5624416\sqrt{39}}{5}\lambda_3 - \frac{35792903776\sqrt{39}}{5}\lambda_4)v'^2,
\end{equation}
\begin{equation}
    (\mathcal{M}_{\textbf{3'}}^2)_{22} = (\frac{3808}{5}\lambda_2 + \frac{20444914}{5}\lambda_3 + \frac{121740637354}{5}\lambda_4)v'^2,
\end{equation}
\begin{equation}
    m_{X_1}^2=m_{X_2}^2=m_{X_3}^2= 30 g_{\text{D}}^2 v'^2.
\end{equation}

\item[*]
Similarly to the case of 13-d representation, from Eq.~\eqref{S4toA4}, we know that the $\bf{1'}$ representation of the $S_4$ group corresponding to Eq.~\eqref{n19_Singlet2} induces another part of the VEV configuration proportional to
\begin{equation}\label{n19_A4_VEV}
   \begin{aligned}
       \vec{v}=\frac{v^{\prime}}{\sqrt{2}} &\left(0,-\frac{1}{4} i \sqrt{\frac{7}{3}} \sin (\alpha ),0,\frac{1}{4} \sqrt{\frac{13}{2}} \cos (\alpha ),0,-\frac{1}{4} i \sqrt{\frac{17}{3}} \sin (\alpha ),0,\right.
       \\&\frac{1}{4} \sqrt{\frac{3}{2}} \cos (\alpha ),0,0,0,\frac{1}{4} \sqrt{\frac{3}{2}} \cos (\alpha ),0,\frac{1}{4} i \sqrt{\frac{17}{3}} \sin (\alpha ),0,
        \\&\left.
        \frac{1}{4} \sqrt{\frac{13}{2}} \cos (\alpha ),
       0,\frac{1}{4} i \sqrt{\frac{7}{3}} \sin (\alpha ),0\right)^\text{T},
       \end{aligned}
\end{equation}
to break the $S_4$ symmetry into the $A_4$ symmetry.
Since the subspaces of the $A_4$ group are calculated to be 
$\bf{19=1+1+1'+1''+3+3+3+3+3}$, subspaces given in order are 
\begin{equation}\label{n19_A4_Singlet1}
   \sin (\alpha )\times \Delta_{S_4,19,1}+\cos (\alpha )\times \Delta_{S_4,19,1'},
\end{equation}
\begin{equation}\label{n19_A4_Singlet2}
   \cos (\alpha )\times \Delta_{S_4,19,1}-\sin (\alpha )\times \Delta_{S_4,19,1'},
\end{equation}

\begin{equation}\label{n19_A4_Singlet3}
     \left(-\frac{\sqrt{13} }{4}\text{Re$\Delta $}_2 + \frac{\sqrt{3} }{4}\text{Re$\Delta $}_6 - i\frac{1}{2}  \sqrt{\frac{7}{6}} \text{Im$\Delta $}_4 + i\frac{1}{2}  \sqrt{\frac{17}{6}} \text{Im$\Delta $}_8
\right),
\end{equation}
\begin{equation}\label{n19_A4_Singlet4}
    \left(-\frac{\sqrt{13} }{4}\text{Re$\Delta $}_2 + \frac{\sqrt{3} }{4}\text{Re$\Delta $}_6 + i\frac{1}{2} \sqrt{\frac{7}{6}} \text{Im$\Delta $}_4 - i\frac{1}{2}  \sqrt{\frac{17}{6}} \text{Im$\Delta $}_8
    \right).
\end{equation}

\begin{equation}\label{n19_A4_Triplet1}
    \sin (\alpha )\times \Delta_{S_4,19,3_1}+\cos (\alpha )\times \Delta_{S_4,19,3'_1}^\text{G},
\end{equation}
\begin{equation}\label{n19_A4_Triplet2}
    \cos (\alpha )\times \Delta_{S_4,19,3_1}-\sin (\alpha )\times \Delta_{S_4,19,3'_1}.
\end{equation}
Two singlets Eqs.~\eqref{n19_A4_Singlet1} and ~\eqref{n19_A4_Singlet2} are the linear combinations of the Eqs.~\eqref{n19_Singlet1} and ~\eqref{n19_Singlet2}. Two triplets Eqs.~\eqref{n19_A4_Triplet1} and ~\eqref{n19_A4_Triplet2} are linear combinations of Eqs.~\eqref{n19_Triplet1} and \eqref{n19_Triplet4}. Equations~ \eqref{n19_Triplet2}, \eqref{n19_Triplet3}, and~\eqref{n19_Treplet5} can be inherited, while the doublet Eq.~\eqref{n19_Doublet} is decomposed into the $\bf{1'+1''}$ subspaces Eqs.~\eqref{n19_A4_Singlet3}
and \eqref{n19_A4_Singlet4}. Equation~\eqref{n19_Triplet1} is the Goldstone triplet corresponding to the VEV configuration Eq.~\eqref{vacuumof19dS4}, while Eq.~\eqref{n19_A4_Triplet1} is the Goldstone triplet corresponding to the VEV configuration Eq.~\eqref{n19_A4_VEV}. Two minimal conditions are given by
\begin{align}
    0 =& -\mu^2 + \frac{1}{2} \lambda_{H\Phi} v^2 + [ \lambda_1 + 2700 \lambda_2 + (14309910 + 955200 c_{2\alpha} + 96000 c_{4\alpha}) \lambda_3  \notag\\
    &+ (90193394430 + 8893915200 c_{2\alpha} + 157305600 c_{4\alpha}) \lambda_4 ]v'^2,\\
    0 =& [(199 + 80 c_{2\alpha})\lambda_3 + (1852899 + 131088c_{2\alpha}) \lambda_4]s_{2\alpha}.
\end{align}
The second minimal condition equation has more complicated solutions. When $\alpha=\pi/2$, the vacuum configuration restores the $S_4$ symmetry, and when $\alpha=0$, although only $A_4$ symmetry is respected, no coupling constant in the potential can be eliminated in the mass matrix forms. When $\alpha \neq 0$ and $\alpha \neq \pi/2$, the vacuum preserves $A_4$ symmetry while the second minimal condition eliminates a coupling constant in the mass matrix. Since we aim to do a more general discussion, we will neglect the second minimal condition, and only apply the first minimal condition to simplify the 19-d $A_4$ squared mass matrices
\begin{align}
    (\mathcal{M}^2_{\rho})_{11} =& 2[\lambda_1 + 2700 \lambda_2 + (14309910 + 955200 c_{2\alpha} + 96000 c_{4\alpha})\lambda_3 \notag\\
    & + (90193394430 + 8893915200 c_{2\alpha} + 157305600 \lambda_4 c_{4\alpha}) \lambda_4] v'^2,
\end{align}
\begin{equation}
     (\mathcal{M}^2_{\rho})_{22} = -[(955200 c_{2\alpha} + 384000 c_{4\alpha})\lambda_3 + (8893915200 c_{2\alpha} + 629222400 c_{4\alpha})\lambda_4]v'^2,
\end{equation}
\begin{equation}
     (\mathcal{M}^2_{\rho})_{12} = -7200 s_{2\alpha}[(199 + 80 c_{2\alpha})\lambda_3 + (1852899 + 131088c_{2\alpha}) \lambda_4]v'^2,
\end{equation}
\begin{align}
    m_{\textbf{1'}}^2 = m_{\textbf{1''}}^2 =& [(948 - 480 c_{2\alpha})\lambda_2 + (7411584 - 2120640 c_{2\alpha} - 96000 c_{4\alpha})\lambda_3 \notag\\
    &+ (52236216624 - 11111499840 c_{2\alpha} - 157305600 c_{4\alpha})\lambda_4]v'^2,
\end{align}
\begin{align}
    (\mathcal{M}_{\textbf{3}}^2)_{11} =& [\frac{9282}{5}\lambda_2 + (\frac{73578306}{5} - 955200 c_{2\alpha} - 192000 c_{4\alpha})\lambda_3 \notag\\
    &+ (\frac{578985534066}{5} - 8893915200 c_{2\alpha} - 314611200 c_{4\alpha})\lambda_4]v'^2,
\end{align}
\begin{align}
    (\mathcal{M}_{\textbf{3}}^2)_{22} =& [\frac{6591}{20}(1 + c_{2\alpha})\lambda_2 + (\frac{6643032}{5} + \frac{6163032}{5} c_{2\alpha} - 96000 c_{4\alpha})\lambda_3 \notag\\
    &+ (\frac{32029186377}{5} + \frac{31242658377}{5} c_{2\alpha} - 157305600 c_{4\alpha})\lambda_4]v'^2,
\end{align}
\begin{align}
    (\mathcal{M}_{\textbf{3}}^2)_{33} =& [\frac{1485}{4}(1 + c_{2\alpha})\lambda_2 + (2710620 - 848280 c_{2\alpha} - 96000 c_{4\alpha})\lambda_3 \notag\\
    &+ (19178215695 - 14844002805 c_{2\alpha} - 157305600 c_{4\alpha})\lambda_4]v'^2,
\end{align}
\begin{align}
    (\mathcal{M}_{\textbf{3}}^2)_{44} =& [\frac{1904}{5}(1 - c_{2\alpha})\lambda_2 - (\frac{991918}{5} + \frac{21916832}{5} c_{2\alpha} + 96000 c_{4\alpha})\lambda_3 \notag\\
    &- (\frac{49451410198}{5} + \frac{171978575552}{5} c_{2\alpha} + 157305600 c_{4\alpha})\lambda_4]v'^2,
\end{align}
\begin{equation}
    (\mathcal{M}_{\textbf{3}}^2)_{12} = \frac{3\sqrt{119}}{5} c_{\alpha}(169\lambda_2 + 1071202\lambda_3 + 7059122122\lambda_4)v'^2,
\end{equation}
\begin{equation}
    (\mathcal{M}_{\textbf{3}}^2)_{13} = -9\sqrt{17017} c_{\alpha}(\lambda_2 + 6418\lambda_3 + 42101818\lambda_4)v'^2,
\end{equation}
\begin{equation}
    (\mathcal{M}_{\textbf{3}}^2)_{14} = \frac{952\sqrt{39}}{5} s_{\alpha}(\lambda_2 + 5908\lambda_3 + 37597588\lambda_4)v'^2,
\end{equation}
\begin{align}
    (\mathcal{M}_{\textbf{3}}^2)_{23} =& -\frac{9\sqrt{143}}{4} [13(1 + c_{2\alpha})\lambda_2 + (157384 + 85984 c_{2\alpha})\lambda_3 \notag\\
    &+ (1235010484 + 602906284 c_{2\alpha})\lambda_4]v'^2,
\end{align}
\begin{equation}
    (\mathcal{M}_{\textbf{3}}^2)_{24} = \frac{2\sqrt{4641}}{5} s_{2\alpha}(13\lambda_2 + 89179\lambda_3 + 576606394\lambda_4)v'^2,
\end{equation}
\begin{equation}
    (\mathcal{M}_{\textbf{3}}^2)_{34} = -6\sqrt{3927} s_{2\alpha}(\lambda_2 + 6343\lambda_3 + 40685218\lambda_4)v'^2,
\end{equation}
\begin{equation}
    m_{X_1}^2=m_{X_2}^2=m_{X_3}^2 = 30 g_{\text{D}}^2 v'^2.
\end{equation}

\end{itemize}

\item 
For the 21-d representation, the scalar field is given by
\begin{equation}
    \Phi_{(21)}=\frac{1}{\sqrt{2}}\left( \Delta_{10},\Delta_9,\Delta_8,\Delta _7,\Delta _6,\Delta _5,\dots,\Delta _{-5},\Delta _{-6},\Delta _{-7},\Delta_{-8},\Delta_{-9},\Delta_{-10}\right)^\text{T}.
\end{equation}
In this case, the $\text{SU(2)}_\text{D}$ symmetry might be broken into the $A_4$, $S_4$, $A_5$ subgroups. We again discuss them separately.
\begin{itemize}
\item[*]
The VEV configuration preserving the $A_5$ symmetry is given by
\begin{equation}\label{vacuumof21dS4}
\begin{aligned}
    \vec{v}=\frac{v^\prime}{\sqrt{2}}&\left(\frac{1}{25}\sqrt{\frac{187}{3}},0,0,0,0,\frac{\sqrt{209}}{25},0,0,0,0,\frac{1}{25}\sqrt{\frac{247}{3}},0,0,0,0,\frac{\sqrt{209}}{25},\right.
    \\&\left.
    0,0,0,0,\frac{1}{25}\sqrt{\frac{187}{3}}\right)^\text{T}.
\end{aligned}
\end{equation}
The subspaces of $A_5$ corresponding to $\bf{21=1+3+3'+5+5+4}$ are given by
\begin{equation}
    \left(\frac{1}{25} \sqrt{\frac{247}{3}} \Delta _0+\frac{2 \sqrt{209} }{25}\text{Re$\Delta $}_5+\frac{2}{25} \sqrt{\frac{187}{3}} \text{Re$\Delta $}_{10}	\right),
\end{equation}
\begin{equation}
   \left(
	\begin{array}{c}
		\frac{\sqrt{247} }{25}\text{Im$\Delta $}_1+\frac{3 \sqrt{57}}{25} \text{Im$\Delta $}_4+\frac{2 \sqrt{114} }{25}\text{Im$\Delta $}_6+\frac{\sqrt{34} }{25}\text{Im$\Delta $}_9 \\
		\frac{\sqrt{247}}{25} \text{Re$\Delta $}_1-\frac{3 \sqrt{57} }{25}\text{Re$\Delta $}_4+\frac{2 \sqrt{114} }{25}\text{Re$\Delta $}_6-\frac{\sqrt{34} }{25}\text{Re$\Delta $}_9 \\
		-\frac{1}{5} \sqrt{\frac{114}{5}} \text{Im$\Delta $}_5-\frac{2}{5} \sqrt{\frac{34}{5}} \text{Im$\Delta $}_{10} \\
	\end{array}
	\right)^\text{G},
\end{equation}
\begin{equation}
    \left(
	\begin{array}{c}
	-\frac{\sqrt{221}}{25}  \text{Im$\Delta $}_2-\frac{4 \sqrt{34} }{25}\text{Im$\Delta $}_3-\frac{13 \sqrt{2} }{25}\text{Im$\Delta $}_7-\frac{7 \sqrt{3} }{25}\text{Im$\Delta $}_8 \\
	\frac{\sqrt{221} }{25}\text{Re$\Delta $}_2-\frac{4 \sqrt{34} }{25}\text{Re$\Delta $}_3+\frac{13 \sqrt{2} }{25}\text{Re$\Delta $}_7-\frac{7 \sqrt{3} }{25}\text{Re$\Delta $}_8 \\
	\frac{2}{5} \sqrt{\frac{34}{5}} \text{Im$\Delta $}_5-\frac{1}{5} \sqrt{\frac{114}{5}} \text{Im$\Delta $}_{10} \\
	\end{array}
	\right),
\end{equation}
\begin{equation}
    \left(
	\begin{array}{c}
		-\frac{3}{5} \sqrt{\frac{51}{115}} \text{Im$\Delta $}_2+\frac{14}{15} \sqrt{\frac{442}{345}} \text{Im$\Delta $}_3-\frac{37}{15} \sqrt{\frac{26}{345}} \text{Im$\Delta $}_7-\frac{1}{15} \sqrt{\frac{299}{5}} \text{Im$\Delta $}_8 \\
		\frac{3}{5} \sqrt{\frac{51}{115}} \text{Re$\Delta $}_2+\frac{14}{15} \sqrt{\frac{442}{345}} \text{Re$\Delta $}_3+\frac{37}{15} \sqrt{\frac{26}{345}} \text{Re$\Delta $}_7-\frac{1}{15} \sqrt{\frac{299}{5}} \text{Re$\Delta $}_8 \\
		-\frac{16}{15}  \sqrt{\frac{17}{115}} \text{Re$\Delta $}_1+\frac{8}{15} \sqrt{\frac{221}{345}} \text{Re$\Delta $}_4+\frac{13}{15} \sqrt{\frac{442}{345}} \text{Re$\Delta $}_6+\frac{2}{5} \sqrt{\frac{494}{115}} \text{Re$\Delta $}_9 \\
		-\frac{16}{15} \sqrt{\frac{17}{115}} \text{Im$\Delta $}_1-\frac{8}{15} \sqrt{\frac{221}{345}} \text{Im$\Delta $}_4+\frac{13}{15} \sqrt{\frac{442}{345}} \text{Im$\Delta $}_6-\frac{2}{5} \sqrt{\frac{494}{115}} \text{Im$\Delta $}_9 \\
		\frac{1}{3} \sqrt{\frac{494}{69}} \text{Re$\Delta $}_{10}-\frac{1}{3} \sqrt{\frac{374}{69}} \Delta _0 \\
	\end{array}
	\right),
\end{equation}
 \begin{equation}
     	\left(
	\begin{array}{c}
		\frac{6}{5} \sqrt{\frac{39}{115}} \text{Re$\Delta $}_2+\frac{19}{15} \sqrt{\frac{2}{345}} \text{Re$\Delta $}_3+\frac{16}{15} \sqrt{\frac{34}{345}} \text{Re$\Delta $}_7+\frac{2}{15} \sqrt{\frac{391}{5}} \text{Re$\Delta $}_8 \\
		-\frac{6}{5} \sqrt{\frac{39}{115}} \text{Im$\Delta $}_2 + \frac{19}{15} \sqrt{\frac{2}{345}} \text{Im$\Delta $}_3 - \frac{16}{15} \sqrt{\frac{34}{345}} \text{Im$\Delta $}_7 + \frac{2}{15} \sqrt{\frac{391}{5}} \text{Im$\Delta $}_8 \\
		\frac{37}{15}  \sqrt{\frac{13}{115}} \text{Im$\Delta $}_1 - \frac{277 }{15 \sqrt{345}}\text{Im$\Delta $}_4 + \frac{62}{15} \sqrt{\frac{2}{345}} \text{Im$\Delta $}_6 + \frac{1}{5} \sqrt{\frac{646}{115}} \text{Im$\Delta $}_9 \\
		\frac{37}{15} \sqrt{\frac{13}{115}} \text{Re$\Delta $}_1+\frac{277 }{15 \sqrt{345}}\text{Re$\Delta $}_4+\frac{62}{15} \sqrt{\frac{2}{345}} \text{Re$\Delta $}_6-\frac{1}{5} \sqrt{\frac{646}{115}} \text{Re$\Delta $}_9 \\
		\frac{19}{75} \sqrt{\frac{286}{69}} \Delta _0-\frac{3\sqrt{46}}{25}   \text{Re$\Delta $}_5+\frac{22}{75} \sqrt{\frac{646}{69}} \text{Re$\Delta $}_{10} \\
	\end{array}
	\right),
 \end{equation}
\begin{equation}
    \left(
	\begin{array}{c}
	-\frac{4\sqrt{39} }{25}  \text{Re$\Delta $}_2-\frac{2}{25} \sqrt{\frac{2}{3}} \text{Re$\Delta $}_3+\frac{7}{25} \sqrt{\frac{34}{3}} \text{Re$\Delta $}_7+\frac{2 \sqrt{17} }{25}\text{Re$\Delta $}_8 \\
	-\frac{4 \sqrt{39} }{25}  \text{Im$\Delta $}_2+\frac{2}{25} \sqrt{\frac{2}{3}} \text{Im$\Delta $}_3+\frac{7}{25} \sqrt{\frac{34}{3}} \text{Im$\Delta $}_7-\frac{2 \sqrt{17} }{25}\text{Im$\Delta $}_8 \\
	-\frac{6\sqrt{13}}{25}  \text{Im$\Delta $}_1-\frac{4 }{25 \sqrt{3}}\text{Im$\Delta $}_4+\frac{14}{25} \sqrt{\frac{2}{3}} \text{Im$\Delta $}_6+\frac{\sqrt{646} }{25}\text{Im$\Delta $}_9 \\
	\frac{6 \sqrt{13} }{25}\text{Re$\Delta $}_1-\frac{4 }{25 \sqrt{3}}\text{Re$\Delta $}_4-\frac{14}{25} \sqrt{\frac{2}{3}} \text{Re$\Delta $}_6+\frac{\sqrt{646}}{25} \text{Re$\Delta $}_9 \\
	\end{array}
	\right).
\end{equation}
The minimal condition and the squared mass (matrices) are given by
\begin{align}
    0 =& -\mu^2 + \frac{1}{2} \lambda_{H\Phi} v^2 + (\lambda_1 + \frac{12100}{3} \lambda_2 + \frac{87318440}{3} \lambda_3 + 263763839240 \lambda_4 \notag\\
    &- \frac{475\sqrt{741}}{2646} \kappa')v'^2,
\end{align}
\begin{equation}
    \mathcal{M}^2_{\rho} = 2(\lambda_1 + \frac{12100}{3} \lambda_2 + \frac{87318440}{3} \lambda_3 + 263763839240 \lambda_4 - \frac{475\sqrt{741}}{5292} \kappa') v'^2,
\end{equation}
\begin{equation}
    m_{\textbf{3'}}^2 = (-9243234000\lambda_4 + \frac{55 \sqrt{741}}{126}\kappa') v'^2,
\end{equation}
\begin{equation}
    m_{\textbf{4}}^2 = (4702500\lambda_3 + 66430336500\lambda_4 + \frac{55 \sqrt{741}}{882}\kappa') v'^2,
\end{equation}
\begin{equation}
    (\mathcal{M}_{\textbf{5}}^2)_{11} = (\frac{1478048}{621}\lambda_2 + \frac{15281907784}{621}\lambda_3 + \frac{21758682400216}{9}\lambda_4 + \frac{63305\sqrt{741}}{547722}\kappa')v'^2,
\end{equation}
\begin{equation}
    (\mathcal{M}_{\textbf{5}}^2)_{22} = (\frac{512050}{621}\lambda_2 + \frac{10547747210}{621}\lambda_3 + \frac{1845087724370}{9}\lambda_4 + \frac{99275\sqrt{741}}{273861}\kappa')v'^2,
\end{equation}
\begin{equation}
    (\mathcal{M}_{\textbf{5}}^2)_{12} = (\frac{58520\sqrt{221}}{621}\lambda_2 + \frac{663457960\sqrt{221}}{621}\lambda_3 + \frac{102211274440\sqrt{221}}{9}\lambda_4 + \frac{12826\sqrt{969}}{273861}\kappa')v'^2,
\end{equation}
\begin{equation}
    m_{X_1}^2=m_{X_2}^2=m_{X_3}^2= \frac{110}{3} g_{\text{D}}^2 v'^2.
\end{equation}

\item[*]
The VEV configuration preserving $S_4$ symmetry is given by
\begin{equation}
\begin{aligned}
	\vec{v}=\frac{v^\prime}{\sqrt{2}}&\left(0,0,\frac{1}{16}\sqrt{\frac{187}{3}},0,0,0,-\frac{\sqrt{11}}{8},0,0,0,-\frac{1}{8}\sqrt{\frac{65}{6}},0,0,0,-\frac{\sqrt{11}}{8},\right.
    \\&\left.
    0,0,0,\frac{1}{16}\sqrt{\frac{187}{3}},0,0\right)^\text{T}. \label{n21_VEV}
\end{aligned}
\end{equation}
The subspaces of $S_4$ corresponding to $\bf{21=1+1'+2+2+3+3+3'+3'+3'}$ are given by
\begin{equation}
    \Delta_{S_4,21,1}=\left(-\frac{1}{8} \sqrt{\frac{65}{6}} \Delta _0-\frac{\sqrt{11}}{4}  \text{Re$\Delta $}_4+\frac{1}{8} \sqrt{\frac{187}{3}} \text{Re$\Delta $}_8
    \right), \label{n21_Singlet1}
\end{equation}
\begin{equation}
    \Delta_{S_4,21,1'}=\left(
    -\frac{1}{8} \sqrt{\frac{247}{3}} \text{Im$\Delta $}_2 + \frac{1}{8} \sqrt{\frac{19}{6}} \text{Im$\Delta $}_6 + \frac{1}{8} \sqrt{\frac{85}{2}} \text{Im$\Delta $}_{10}
    \right), \label{n21_Singlet2}
\end{equation}
\begin{equation}
	\left(
	\begin{array}{c}
		\frac{1}{3} \sqrt{\frac{2}{3}} \text{Im$\Delta $}_2+\frac{2}{3} \sqrt{\frac{13}{3}} \text{Im$\Delta $}_6 \\
		-\frac{1}{12} \sqrt{\frac{55}{3}} \Delta _0-\frac{1}{3} \sqrt{\frac{13}{2}} \text{Re$\Delta $}_4-\frac{1}{6} \sqrt{\frac{221}{6}} \text{Re$\Delta $}_8 \\
	\end{array}
	\right), \label{n21_Doublet1}
\end{equation}
\begin{equation}
	\left(
	\begin{array}{c}
		-\frac{1}{24} \sqrt{\frac{1105}{3}} \text{Im$\Delta $}_2+\frac{1}{24} \sqrt{\frac{85}{6}} \text{Im$\Delta $}_6-\frac{3}{8} \sqrt{\frac{19}{2}} \text{Im$\Delta $}_{10} \\
		-\frac{1}{24} \sqrt{\frac{2431}{6}} \Delta _0+\frac{\sqrt{85} }{12}\text{Re$\Delta $}_4-\frac{1}{24} \sqrt{\frac{5}{3}} \text{Re$\Delta $}_8 \\
	\end{array}
	\right), \label{n21_doublet2}
\end{equation}
\begin{equation}
   \Delta_{S_4,21,3_1}= \left(
	\begin{array}{c}
	-\frac{1}{8} \sqrt{\frac{65}{2}} \text{Im$\Delta $}_1-\frac{7}{8} \sqrt{\frac{3}{5}} \text{Im$\Delta $}_3-\frac{3 \sqrt{3} }{8}\text{Im$\Delta $}_5+\frac{3}{16} \sqrt{\frac{51}{5}} \text{Im$\Delta $}_7+\frac{1}{16} \sqrt{\frac{323}{5}} \text{Im$\Delta $}_9 \\
	-\frac{1}{8} \sqrt{\frac{65}{2}} \text{Re$\Delta $}_1+\frac{7}{8} \sqrt{\frac{3}{5}} \text{Re$\Delta $}_3-\frac{3 \sqrt{3} }{8} \text{Re$\Delta $}_5-\frac{3}{16} \sqrt{\frac{51}{5}} \text{Re$\Delta $}_7+\frac{1}{16} \sqrt{\frac{323}{5}} \text{Re$\Delta $}_9\\
	\sqrt{\frac{3}{10}} \text{Im$\Delta $}_4-\sqrt{\frac{17}{10}} \text{Im$\Delta $}_8 \\
	\end{array}
	\right)^\text{G}, \label{n21_triplet1}
\end{equation}
\begin{equation}
    \left(
	\begin{array}{c}
	\frac{1}{4} \sqrt{\frac{17}{5}} \text{Re$\Delta $}_3 + \frac{\sqrt{17} }{4}\text{Re$\Delta $}_5 + \frac{1}{4 \sqrt{5}} \text{Re$\Delta $}_7 + \frac{1}{4} \sqrt{\frac{57}{5}} \text{Re$\Delta $}_9 \\
	-\frac{1}{4} \sqrt{\frac{17}{5}} \text{Im$\Delta $}_3+\frac{\sqrt{17} }{4}\text{Im$\Delta $}_5-\frac{1}{4 \sqrt{5}}\text{Im$\Delta $}_7+\frac{1}{4} \sqrt{\frac{57}{5}} \text{Im$\Delta $}_9\\
	\sqrt{\frac{17}{10}} \text{Im$\Delta $}_4+\sqrt{\frac{3}{10}} \text{Im$\Delta $}_8 \\
	\end{array}
	\right), \label{n21_triplet2}
\end{equation}
\begin{equation}
		\Delta_{S_4,21,3'_1}=\left(
\begin{array}{c}
 \frac{3}{8} \sqrt{\frac{741}{110}} \text{Re$\Delta $}_1+\frac{13}{8} \sqrt{\frac{19}{55}} \text{Re$\Delta $}_3-\frac{1}{8} \sqrt{\frac{19}{11}} \text{Re$\Delta $}_5-\frac{1}{16} \sqrt{\frac{323}{55}} \text{Re$\Delta $}_7-\frac{1}{16} \sqrt{\frac{255}{11}} \text{Re$\Delta $}_9 \\
 \frac{3}{8} \sqrt{\frac{741}{110}} \text{Im$\Delta $}_1-\frac{13}{8} \sqrt{\frac{19}{55}} \text{Im$\Delta $}_3-\frac{1}{8} \sqrt{\frac{19}{11}} \text{Im$\Delta $}_5+\frac{1}{16} \sqrt{\frac{323}{55}} \text{Im$\Delta $}_7-\frac{1}{16} \sqrt{\frac{255}{11}} \text{Im$\Delta $}_9 \\
 -\frac{1}{4} \sqrt{\frac{247}{110}} \text{Re$\Delta $}_2+\frac{3}{8} \sqrt{\frac{19}{55}} \text{Re$\Delta $}_6+\frac{5}{8} \sqrt{\frac{51}{11}} \text{Re$\Delta $}_{10} \\
\end{array}
\right), \label{n21_triplet3}
\end{equation}
\begin{equation}
	\left(
	\begin{array}{c}
		\frac{1}{8} \sqrt{\frac{255}{61}} \text{Re$\Delta $}_1-\frac{1}{2} \sqrt{\frac{221}{122}} \text{Re$\Delta $}_5+\frac{11}{8} \sqrt{\frac{65}{122}} \text{Re$\Delta $}_7+\frac{1}{8} \sqrt{\frac{3705}{122}} \text{Re$\Delta $}_9 \\
		\frac{1}{8} \sqrt{\frac{255}{61}} \text{Im$\Delta $}_1 - \frac{1}{2} \sqrt{\frac{221}{122}} \text{Im$\Delta $}_5 - \frac{11}{8} \sqrt{\frac{65}{122}} \text{Im$\Delta $}_7 + \frac{1}{8} \sqrt{\frac{3705}{122}} \text{Im$\Delta $}_9 \\
		\frac{9}{8} \sqrt{\frac{85}{61}} \text{Re$\Delta $}_2+\frac{1}{8} \sqrt{\frac{1105}{122}} \text{Re$\Delta $}_6+\frac{1}{8} \sqrt{\frac{741}{122}} \text{Re$\Delta $}_{10} \\
	\end{array}
	\right), \label{n21_triplet4}
\end{equation}
\begin{equation}
	\left(
	\begin{array}{c}
		-\frac{89}{8} \sqrt{\frac{13}{3355}} \text{Re$\Delta $}_1+\frac{1}{2} \sqrt{\frac{183}{110}} \text{Re$\Delta $}_3+2 \sqrt{\frac{6}{671}} \text{Re$\Delta $}_5+\frac{71}{8} \sqrt{\frac{51}{6710}} \text{Re$\Delta $}_7-\frac{5}{8} \sqrt{\frac{1615}{1342}} \text{Re$\Delta $}_9 \\
		-\frac{89}{8} \sqrt{\frac{13}{3355}} \text{Im$\Delta $}_1 - \frac{1}{2} \sqrt{\frac{183}{110}} \text{Im$\Delta $}_3 + 2 \sqrt{\frac{6}{671}} \text{Im$\Delta $}_5 - \frac{71}{8} \sqrt{\frac{51}{6710}} \text{Im$\Delta $}_7 - \frac{5}{8} \sqrt{\frac{1615}{1342}} \text{Im$\Delta $}_9 \\
		-\frac{23}{8}  \sqrt{\frac{39}{3355}} \text{Re$\Delta $}_2+\frac{509}{8} \sqrt{\frac{3}{6710}} \text{Re$\Delta $}_6-\frac{5}{8} \sqrt{\frac{323}{1342}} \text{Re$\Delta $}_{10} \\
	\end{array}
	\right). \label{n21_triplet5}
\end{equation}
The minimal condition and the squared mass (matrices) are
\begin{align}
    0 =& -\mu^2 + \frac{1}{2} \lambda_{H\Phi} v^2 + ( \lambda_1 + \frac{12100}{3} \lambda_2 + \frac{89768690}{3} \lambda_3 + 275301739790 \lambda_4 \notag\\
    &- \frac{314\sqrt{390}}{1323} \kappa' )v'^2,
\end{align}
\begin{equation}
    \mathcal{M}^2_{\rho} = 2(\lambda_1 + \frac{12100}{3} \lambda_2 + \frac{89768690}{3} \lambda_3 + 275301739790 \lambda_4 - \frac{157\sqrt{390}}{1323} \kappa') v'^2,
\end{equation}
\begin{equation}
    m_{\textbf{1'}}^2 = -(3168000\lambda_3 + 44753068800\lambda_4 + \frac{22 \sqrt{390}}{441}\kappa') v'^2,
\end{equation}
\begin{equation}
    m_{\textbf{3}}^2 = (6756750\lambda_3 + 91168827750\lambda_4 + \frac{55 \sqrt{390}}{126}\kappa') v'^2,
\end{equation}
\begin{equation}
    (\mathcal{M}_{\textbf{2}}^2)_{11} = (\frac{56056}{27}\lambda_2 + \frac{743120048}{27}\lambda_3 + \frac{2733532387696}{9}\lambda_4 + \frac{3938\sqrt{390}}{11907}\kappa')v'^2,
\end{equation}
\begin{equation}
    (\mathcal{M}_{\textbf{2}}^2)_{22} = (\frac{3740}{27}\lambda_2 + \frac{63345040}{27}\lambda_3 + \frac{177473130560}{9}\lambda_4 + \frac{3982\sqrt{390}}{11907}\kappa')v'^2,
\end{equation}
\begin{equation}
    (\mathcal{M}_{\textbf{2}}^2)_{12} = (\frac{308\sqrt{2210}}{27}\lambda_2 + \frac{5755024\sqrt{2210}}{27}\lambda_3 + \frac{23862632288\sqrt{2210}}{9}\lambda_4 + \frac{6886\sqrt{51}}{11907}\kappa')v'^2,
\end{equation}
\begin{equation}
    (\mathcal{M}_{\textbf{3'}}^2)_{11} = (\frac{3078}{5}\lambda_2 + \frac{25506774}{5}\lambda_3 + \frac{206812607094}{5}\lambda_4 + \frac{124\sqrt{390}}{1323}\kappa')v'^2,
\end{equation}
\begin{equation}
    (\mathcal{M}_{\textbf{3'}}^2)_{22} = (\frac{194480}{61}\lambda_2 + \frac{2303984650}{61}\lambda_3 + \frac{24391570595370}{61}\lambda_4 + \frac{91355\sqrt{390}}{312812}\kappa')v'^2,
\end{equation}
\begin{equation}
    (\mathcal{M}_{\textbf{3'}}^2)_{33} = (\frac{55696}{915}\lambda_2 + \frac{774028118}{915}\lambda_3 + \frac{728728684786}{305}\lambda_4 + \frac{161616\sqrt{390}}{312812}\kappa')v'^2,
\end{equation}
\begin{equation}
    (\mathcal{M}_{\textbf{3'}}^2)_{12} = 36\sqrt{\frac{92378}{61}}(-\lambda_2 - 8308\lambda_3 - 73703848\lambda_4 + \frac{163\sqrt{390}}{6191640}\kappa')v'^2,
\end{equation}
\begin{equation}
    (\mathcal{M}_{\textbf{3'}}^2)_{13} = \frac{708}{5}\sqrt{\frac{114}{61}}(\lambda_2 + \frac{978172}{59}\lambda_3 + \frac{12278380632}{59}\lambda_4 - \frac{1325\sqrt{390}}{1873368}\kappa')v'^2,
\end{equation}
\begin{equation}
    (\mathcal{M}_{\textbf{3'}}^2)_{23} = -\frac{944}{61}\sqrt{\frac{2431}{3}}(\lambda_2 + \frac{1210213}{236}\lambda_3 + \frac{6813613203}{236}\lambda_4 + \frac{34357\sqrt{390}}{108239040}\kappa')v'^2.
\end{equation}
\begin{equation}
    m_{X_1}^2=m_{X_2}^2=m_{X_3}^2= \frac{110}{3} g_{\text{D}}^2 v'^2.
\end{equation}

\item[*]
Again, besides Eq.~\eqref{n21_VEV}, the $\bf{1'}$ representation of $S_4$ corresponding to Eq.~\eqref{n21_Singlet2} induces
\begin{equation}\label{n21_A4_VEV}
    \begin{aligned}
    \vec{v}=\frac{v^{\prime}}{\sqrt{2}}&\left(\frac{1}{16} i \sqrt{\frac{85}{2}} \cos (\alpha ),0,\frac{1}{16} \sqrt{\frac{187}{3}} \sin (\alpha ),0,\frac{1}{16} i \sqrt{\frac{19}{6}} \cos (\alpha ),0,-\frac{1}{8} \sqrt{11} \sin (\alpha ),\right.
    \\&
    0,-\frac{1}{16} i \sqrt{\frac{247}{3}} \cos (\alpha ),0,
    -\frac{1}{8} \sqrt{\frac{65}{6}} \sin (\alpha ),0,\frac{1}{16} i \sqrt{\frac{247}{3}} \cos (\alpha ),0,-\frac{1}{8} \sqrt{11} \sin (\alpha ),\\& \left. 0,-\frac{1}{16} i \sqrt{\frac{19}{6}} \cos (\alpha ),0,
   \frac{1}{16} \sqrt{\frac{187}{3}} \sin (\alpha ),0,-\frac{1}{16} i \sqrt{\frac{85}{2}} \cos (\alpha )\right)^\text{T} ,
    \end{aligned}
\end{equation}
to break the $S_4$ symmetry into the $A_4$ symmetry.
Since $\bf{21}$=$\bf{1}$+$\bf{1}$+$\bf{1'}$+$\bf{1'}$+$\bf{1''}$+$\bf{1''}$+ $\bf{3}$+$\bf{3}$+$\bf{3}$+$\bf{3}$+$\bf{3}$, the subspaces are given in order as

\begin{equation}\label{n21_A4_singlet1}
\sin(\alpha)\times\Delta_{S_4,21,1}+\cos(\alpha)\times\Delta_{S_4,21,1'},
\end{equation}
\begin{equation}\label{n21_A4_singlet2}
    \cos(\alpha)\times\Delta_{S_4,21,1}-\sin(\alpha)\times\Delta_{S_4,21,1'},
\end{equation}

\begin{equation}\label{n21_A4_singlet3}
	\left(\sqrt{\frac{66}{197}} \Delta _0-\sqrt{\frac{65}{197}} \text{Re$\Delta $}_4+i\frac{25}{4}  \sqrt{\frac{5}{591}} \text{Im$\Delta $}_2-i\frac{1}{4}  \sqrt{\frac{65}{1182}} \text{Im$\Delta $}_6+i\frac{1}{4}  \sqrt{\frac{4199}{394}} \text{Im$\Delta $}_{10}
        \right),
\end{equation}
\begin{equation}\label{n21_A4_singlet4}
\begin{aligned}
        &\left(\frac{1}{16} \sqrt{\frac{12155}{591}} \Delta _0+\frac{11}{4} \sqrt{\frac{17}{394}} \text{Re$\Delta $}_4+\frac{1}{8} \sqrt{\frac{197}{6}} \text{Re$\Delta $}_8-i\frac{1}{8}  \sqrt{\frac{663}{394}} \text{Im$\Delta $}_2\right.
        \\&\left.-i\frac{31}{16}  \sqrt{\frac{51}{197}} \text{Im$\Delta $}_6+i\frac{1}{16}  \sqrt{\frac{95}{197}} \text{Im$\Delta $}_{10}
        \right),
\end{aligned}
\end{equation}
\begin{equation}\label{n21_A4_singlet5}
	\left(-\sqrt{\frac{66}{197}} \Delta _0+\sqrt{\frac{65}{197}} \text{Re$\Delta $}_4+i\frac{25}{4}  \sqrt{\frac{5}{591}} \text{Im$\Delta $}_2-i\frac{1}{4}  \sqrt{\frac{65}{1182}} \text{Im$\Delta $}_6+i\frac{1}{4}  \sqrt{\frac{4199}{394}} \text{Im$\Delta $}_{10}
        \right),
\end{equation}
\begin{equation}\label{n21_A4_singlet6}
\begin{aligned}
	&\left(-\frac{1}{16} \sqrt{\frac{12155}{591}} \Delta _0-\frac{11}{4} \sqrt{\frac{17}{394}} \text{Re$\Delta $}_4-\frac{1}{8} \sqrt{\frac{197}{6}} \text{Re$\Delta $}_8- i\frac{1}{8} \sqrt{\frac{663}{394}} \text{Im$\Delta $}_2\right.
 \\&\left.-i\frac{31}{16}  \sqrt{\frac{51}{197}} \text{Im$\Delta $}_6+i\frac{1}{16}  \sqrt{\frac{95}{197}} \text{Im$\Delta $}_{10}
        \right),
\end{aligned}
\end{equation}

\begin{equation}\label{n21_A4_triplet1}    \sin(\alpha)\times\Delta_{S_4,21,3_1}+\cos(\alpha)\times\Delta_{S_4,21,3'_1}^\text{G},
\end{equation}
\begin{equation}\label{n21_A4_triplet2}
    \cos(\alpha)\times\Delta_{S_4,21,3_1}-\sin(\alpha)\times\Delta_{S_4,21,3'_1},
\end{equation}
while the $\bf{1'}$, $\bf{1'}$, $\bf{1''}$, $\bf{1''}$ subspaces are, respectively, decomposed from the $S_4$-$\bf{2}$ representations denoted by Eqs.~\eqref{n21_Doublet1} and \eqref{n21_doublet2} considering Eq.(\ref{S4toA4}).
The three triplets  Eqs.~\eqref{n21_triplet2}, \eqref{n21_triplet4}, and \eqref{n21_triplet5} are inherited.
Of note, Eq.~\eqref{n21_A4_triplet1} is the Goldstone triplet corresponding to the VEV configuration Eq.~\eqref{n21_A4_VEV}.

The two minimal conditions are given by
\begin{align}
    0 =& -\mu^2 + \frac{1}{2} \lambda_{H\Phi} v^2 + [\lambda_1 + \frac{12100}{3} \lambda_2 + (\frac{94088690}{3} + 2976000 c_{2\alpha} + 1536000 c_{4\alpha}) \lambda_3 \notag\\
    &+ (295644043790 +  42040761600 c_{2\alpha} + 21698457600 c_{4\alpha}) \lambda_4 \notag\\
    &- \frac{2\sqrt{390}}{1323}(189 s_\alpha + 32 s_{3\alpha})\kappa']v'^2,\\
    0 =& (63 c_\alpha + 32 c_{3\alpha})[\kappa' + \frac{12700800 s_\alpha}{\sqrt{390}}(5\lambda_3 + 70633\lambda_4)].
\end{align}
When $\alpha = \arctan(\sqrt{95/33})$ or $\alpha = \pi/2$, $A_5$ or $S_4$ symmetry, respectively, is restored. The $A_4$ symmetry requires the disappearance of the square bracket of the second condition. The squared mass matrix components are
\begin{align}
    (\mathcal{M}^2_{\rho})_{11} =& 2[\lambda_1 + \frac{12100}{3} \lambda_2 + (\frac{107696690}{3} - 792000 c_{2\alpha} + 768000 c_{4\alpha})\lambda_3 \notag\\
    &+ (359722301390 -  11188267200 c_{2\alpha} + 10849228800 c_{4\alpha})\lambda_4] v'^2,
\end{align}
\begin{align}
    (\mathcal{M}^2_{\rho})_{22} =& -[(3024000 + 4560000 c_{2\alpha} + 1536000 c_{4\alpha})\lambda_3 \notag\\
    &+ (42718838400 + 64417296000 c_{2\alpha} + 21698457600 c_{4\alpha})\lambda_4] v'^2,
\end{align}
\begin{align}
    (\mathcal{M}^2_{\rho})_{12} =& -9600(31 s_{2\alpha} + 32 s_{4\alpha})(5\lambda_3 + 70633\lambda_4) v'^2,
\end{align}
\begin{align}
    (\mathcal{M}^2_{\textbf{1,1''}})_{11} =& \frac{40}{591}[(45227 + 43940c_{2\alpha})\lambda_2 + (356630230 + 629580160 c_{2\alpha})\lambda_3 \notag\\
    &+ (3152540596218 + 7183215367440 c_{2\alpha})\lambda_4] v'^2,
\end{align}
\begin{align}
    (\mathcal{M}^2_{\textbf{1,1''}})_{22} =& \frac{4}{591}[(158627 - 155720 c_{2\alpha})\lambda_2 - (261483884 + 1163179360 c_{2\alpha})\lambda_3 \notag\\
    &- (11514454088304 + 11446663522560 c_{2\alpha})\lambda_4] v'^2,
\end{align}
\begin{align}
    (\mathcal{M}^2_{\textbf{1,1''}})_{12} =& \frac{4}{197}\sqrt{\frac{170}{13}}[(585 + 5278 c_{2\alpha})\lambda_2 - (26655300 + 35850616 c_{2\alpha})\lambda_3 \notag\\
    &- (354615938640 + 705606428256 c_{2\alpha})\lambda_4]v'^2 \notag\\
    &+ i\frac{8}{3}\sqrt{\frac{35530}{13}}s_{2\alpha}(13\lambda_2 + 142364\lambda_3 + 1472784384\lambda_4)v'^2 ,
\end{align}
\begin{align}
    (\mathcal{M}_{\textbf{3}}^2)_{11} =& \frac{114}{5}(27 \lambda_2 - 37309 \lambda_3 - 1873640429\lambda_4)v'^2,
\end{align}
\begin{align}
    (\mathcal{M}_{\textbf{3}}^2)_{22} =& [\frac{5168}{15}(1 + c_{2\alpha})\lambda_2 + (-\frac{156737006}{15} + \frac{157711744}{15} c_{2\alpha})\lambda_3 \notag\\
    &+ (-\frac{820570139362}{5} + \frac{681532481888}{5}c_{2\alpha})\lambda_4]v'^2,
\end{align}
\begin{align}
    (\mathcal{M}_{\textbf{3}}^2)_{33} =& \frac{110}{61}[884(1 - c_{2\alpha})\lambda_2 + (3612163 - 7367152 c_{2\alpha})\lambda_3 \notag\\
    &+ (12838718235 - 68117137032 c_{2\alpha})\lambda_4]v'^2,
\end{align}
\begin{align}
    (\mathcal{M}_{\textbf{3}}^2)_{44} =& \frac{2}{915}[13924(1 - c_{2\alpha})\lambda_2 + (-7939784533 + 6212791408 c_{2\alpha})\lambda_3 \notag\\
    &+ (-115586648024973 + 88715231041848 c_{2\alpha})\lambda_4]v'^2,
\end{align}
\begin{align}
    (\mathcal{M}_{\textbf{3}}^2)_{12} =& \frac{456\sqrt{51}}{5}c_\alpha(\lambda_2 + 10308\lambda_3 + 108678248\lambda_4)v'^2,
\end{align}
\begin{equation}
    (\mathcal{M}_{\textbf{3}}^2)_{13} = 12\sqrt{\frac{7106}{793}} s_\alpha(39\lambda_2 + 389212\lambda_3 + 3795504392\lambda_4)v'^2,
\end{equation}
\begin{equation}
    (\mathcal{M}_{\textbf{3}}^2)_{14} = -\frac{12}{5}\sqrt{\frac{114}{61}} s_\alpha(59\lambda_2 + 3628172\lambda_3 + 49713870632\lambda_4)v'^2,
\end{equation}
\begin{equation}
    (\mathcal{M}_{\textbf{3}}^2)_{23} = 136\sqrt{\frac{418}{2379}} s_{2\alpha}(13\lambda_2 + 27254\lambda_3 - 105520236\lambda_4)v'^2,
\end{equation}
\begin{equation}
    (\mathcal{M}_{\textbf{3}}^2)_{24} = -\frac{8}{15}\sqrt{\frac{646}{61}} s_{2\alpha}(59\lambda_2 + 6479422\lambda_3 + 91181965632\lambda_4)v'^2,
\end{equation}
\begin{align}
    (\mathcal{M}_{\textbf{3}}^2)_{34} =& -\frac{4}{61}\sqrt{\frac{187}{39}}[1534(1 - c_{2\alpha})\lambda_2 + (481097 + 46590928 c_{2\alpha})\lambda_3 \notag\\
    &+ (-62050919973 + 722997781548 c_{2\alpha})\lambda_4]v'^2.
\end{align}
\begin{equation}
    m_{X_1}^2=m_{X_2}^2=m_{X_3}^2= \frac{110}{3} g_{\text{D}}^2 v'^2.
\end{equation}

\end{itemize}

\end{itemize}

\section{Summary}

In this paper, we dissect the models of vector dark matter stemming from a hidden $\text{SU(2)}_\text{D}$ gauge group, which is broken into some remnant discrete subgroups by the VEV of a single Higgs multiplet $\Phi_{(n)}$. We provide the general algorithms for writing all possible renormalizable couplings, and work out all the remnant discrete subgroups as well as the corresponding VEV configurations and the detailed forms of the Higgs multiplet subspaces. Up to $j=10$ or $n=21$, we find that the nontrivial discrete subgroups (beyond $Z_n$) can only exist in some cases with odd $n$. We list the results for each $n$ and find that the vector bosons of $\text{SU(2)}_\text{D}$ form a $\textbf{3}$ representation and therefore become degenerate in most of the cases. The remnant discrete groups can stabilize the vector bosons and thus allow them to play the role of the dark matter candidate. Since most of the non-Abelian Lie groups include at least one SU(2) as their subgroup, our algorithms can potentially be generalized to models with other Lie groups, especially to the models with higher-dimensional compact Lie groups. In those cases, the SU(2) subgroups play important roles in enumerating all their possible finite dimensional irreducible representations. We leave these as the direction of our future study.

\begin{acknowledgements}
    We thank to Zhao-Huan Yu and Pyungwon Ko for helpful discussions. C. C. is supported by the National Natural Science Foundation of China (NSFC) under Grant No.~11905300 and the Guangzhou Science and Technology Planning Project under Grant No.~2023A04J0008. 
Y.-L.T. is supported by NSFC under Grant No.~12005312.
H.-H.Z. is supported by NSFC under Grant No.~12275367.
This work is also supported by the Sun Yat-Sen University Science Foundation.  
\end{acknowledgements}
\newpage
\appendix

\section{Generators of SU(2)}
\label{generatorsofSU2}
In our convention, the ladder operators $J_+$ and $J_-$ in $(2j+1)$-d representation are defined as
\begin{equation}
	J_+\left| j,m\right\rangle=\left\{
	\begin{aligned}
		&-\sqrt{(j-m)(j+m+1)} \left| j,m+1\right\rangle \quad m\ge 0
		\\& \sqrt{(j-m)(j+m+1)} \left| j,m+1\right\rangle \quad m<0
	\end{aligned}
	\right. .\label{JpJm}
\end{equation}
We denote the matrix forms of ladder operators $J_+$ and $J_-$ as $\tau^+$ and $\tau^-$, respectively.
Then the generator matrices of SU(2) are given by
\begin{equation}
	\tau_{(n)}^1=\frac{\tau_{(n)}^+ + \tau_{(n)}^-}{2},\qquad \tau_{(n)}^2=\frac{\tau_{(n)}^+ - \tau_{(n)}^-}{2i},\qquad
	\tau_{(n)}^3=\text{Diag}\{j,j-1,...,-j+1,-j\}. \label{J123}
\end{equation}

The scalar multiplet $\Phi_{(n)}$ in $\text{SU(2)}_\text{D}$ $n$-d representation is shown in Eq.~\eqref{eq:Phi},
\begin{equation}
	\Phi_{(n)}=\xi_{(n)}\left(\Delta_j,\Delta_{j-1},\dots,\Delta_{-j+1},\Delta_{-j}\right)^\mathrm{T}.
\end{equation}
Its gauge covariant derivative term is given by
\begin{align}
	\mathcal{L}_\text{kin} =& (D^\mu \Phi_{(n)})^\dagger D_\mu \Phi_{(n)} \notag\\ 
	=& (\partial^\mu\Phi_{(n)}^\dagger) \partial_\mu\Phi_{(n)} + i g_\text{D} X^{a \mu} [\Phi_{(n)}^\dagger \tau_{(n)}^a \partial_\mu \Phi_{(n)} - (\partial_\mu \Phi_{(n)}^\dagger) \tau_{(n)}^a \Phi_{(n)}] + g_\text{D}^2 X_\mu^a X^{b\mu} \Phi_{(n)}^\dagger \tau_{(n)}^a \tau_{(n)}^b \Phi_{(n)}
\end{align}
with $D_{\mu}\Phi=\partial_{\mu}\Phi_{(n)} - i g_\text{D} X_{\mu}^a \tau_{(n)}^a \Phi_{(n)}$.
Here, $X^a(a=1,2,3)$ are three gauge bosons which can also be written as
\begin{equation}
	X^1= \frac{X^-+X^+}{\sqrt{2}},\qquad X^2= \frac{X^--X^+}{i \sqrt{2}},\qquad X^3=Z'.
\end{equation}

\section{Discrete symmetry groups of SO(3)}
\label{SubgroupsforSO3}
The following table shows the  discrete subgroups of the group SO(3) in an
arbitrary $2j+1$ dimensional irreducible representation.

The representation of weight $j$ contains the following stabilizer subgroups
systematically: 
\begin{itemize}
	\item If $j$ is even and $j\geq 4$, then
	$${\bf 1}, Z_2\oplus Z_2, Z_3,...,Z_j, D_3,...D_j.$$

	\item If $j$ is odd and $j\geq 5$, then
	$${\bf 1}, Z_2\oplus Z_2, Z_3,...,Z_j, D_3,...,D_j.$$
	
\end{itemize}
Beyond these, the nonsystematical groups are shown in Table \ref{SubgroupsforSO3table}.

\begin{table}
	\centering
          \setlength\tabcolsep{0.5em}
          \renewcommand{\arraystretch}{1}
       \scriptsize
	\begin{tabular}{|c|c|c|c|c|c|c|c|c|c|c|c|}\hline
    $\dim n$  & 5 & 7 & 9 & 11 & 13 & 15 & 17 & 19 &21 \\\hline
    subgroup  & ${\bf Z}_2\oplus{\bf Z}_2$ & ${\bf 1}$ & $S_4$ & & $A_4$ & $A_4$ & $S_4$ & $A_4$ & $A_4$ \\
		   &   & $A_4$ &  & & $S_4$  & & & $S_4$ & $S_4$\\
	   	 &   & ${\bf Z}_3$ & & & $A_5$ & & & &$A_5$\\
		   &  & $D_3$ &  & & & & & & \\
        \hline
	\end{tabular}
 \caption{Discrete subgroups that the $\text{SU(2)}_{D}$ groups can break into by a $\Phi_{(n)}$ Higgs multiplet.}\label{SubgroupsforSO3table}
\end{table}

\section{The number of the minimal conditions}\label{app:MinimalConditions}

As stated in Sec.~\ref{SubspaceEnumeration}, if there exist $p$ independent identity representations of the discrete symmetry, we can rotate them into the form that one identical field as a radical component receives the nonzero VEV $v'$, while the remainder $p-1$ identical VEVs vanish. Thus, the scalar fields can be written as
\begin{equation}
    \Phi'=\left(v' + \rho_1, \rho_2, ...,\rho_p,..., \rho_{2j+1}\right), \label{PhiP}
\end{equation}
where $\rho_{1,2,\dots,2j+1}$ are the fluctuations according to the VEV configurations. The scalar potential of the dark sector in Eq.~\eqref{EffectivePotential_MassSimplified} can formally be written in the form
\begin{align}
     V (\Phi') =&  B_a^1 \left[ \mu^2 + \frac{\lambda_{H \Phi}}{2} (v + h)^2
  \right] v' \rho_a + B_{a_1 a_2}^2 \left[ \mu^2 + \frac{\lambda_{H \Phi}}{2}
  (v + h)^2 \right] \rho_{a_1} \rho_{a_2} \nonumber\\
  &  + C^1_a v'^2 \rho_a + C^2_{a_1 a_2} v' \rho_{a_1}
  \rho_{a_2} + C^3_{a_1 a_2 a_3} \rho_{a_1} \rho_{a_2} \rho_{a_3}
  \nonumber\\
  &  + D^1_a v'^3 \rho_a + D^2_{a_1 a_2} v'^2
  \rho_{a_1} \rho_{a_2} + D^3_{a_1 a_2 a_3} v' \rho_{a_1} \rho_{a_2}
  \rho_{a_3} + D^4_{a_1 a_2 a_3 a_4} \rho_{a_1} \rho_{a_2} \rho_{a_3}
  \rho_{a_4},\label{eq:V_MinimalProof}
\end{align}
where the three rows are contributed by the quadratic and Higgs portal terms, trilinear terms and quartic terms respectively. $B,C,D$ are some tensors and are only the functions of the coupling constants of the coupling terms defined in Eq.~\eqref{EffectivePotential} and Table~\ref{tab:odd_operators}. For any $a$, the minimal condition of the dark sector is given by
\begin{align}
0 = \left. \frac{\partial V}{\partial \rho_a} \right|_{\vec{\rho} = 0, h =
  0} & =  B_a^1 \left( \mu^2 + \frac{\lambda_{H \Phi}}{2} v^2 \right) v' +
  C^1_a v'^2 + D^1_a v'^3 \notag\\
  & = \left[ B_a^1 \left( \mu^2 + \frac{\lambda_{H \Phi}}{2} v^2 \right) +
  C^1_a v' + D^1_a v'^2 \right] v'. \label{eq:MinimalProof}
\end{align}

Notice that only $B^1_a$, $C^1_a$, $D^1_a$ with a single group index remain, while others disappear due to the null assignments of $h$ and $\rho_{1,2,\dots,2j+1}$. Since the system respects the remaining symmetry, therefore Eq.~\eqref{eq:MinimalProof} should also be invariant under the transformation of the remained group. The left-hand side of Eq.~\eqref{eq:MinimalProof}, 0, is definitely the identical representation, so the right-hand side should also be identical. Here, $v^{\prime}$, $\mu^2$, $\lambda_{H \Phi}$, $\kappa$, $\lambda$ are also identical, and therefore, $B_{a}^1$, $C_{a}^1$, $D_{a}^1$ have no other choice but to be identical. As a result, when $a>p$, since $B_{a}^1$, $C_{a}^1$, $D_{a}^1$ belong to various nonidentical representations, they must disappear. Therefore the only nontrivial equations of the minimal conditions are when $a \leq p$, and in this case $B_{a}^1$, $C_{a}^1$, $D_{a}^1$ themselves belong to the identical representations and can become nonzero.

We should mention that when $p>1$, no rule, as we know, guarantees the independence of the $p$ nontrivial minimal condition equations, so the number of independent equations of the minimal conditions should be no larger than the multiplicity $p$ of the identical representations.

\section{Eigenvectors for 4-, 9-, and 13-d representation}\label{sec:eigenvector}

We list the eigenvectors $\vec{v}'_{m'}$ of the generator $J_1 \sin(\theta) + J_3\cos(\theta)$, in which the rotation axis lies within the $x$-$z$ plane with an angle $\theta$ from the $z$-axis.
All of them can be expressed in the functions of $\tan(\theta/2)$, and we will denote $\tan(\theta/2)$ as $t_{\theta/2}$ for simplicity.
In this appendix we do not normalize $\vec{v}'_{m'}$ since in our analysis it is not necessary, e.g., in Eq.~\eqref{bm}.
The following vectors are proportional to the normalized eigenvectors $\vec{v}'_{m'}$ with the last components equal to one.
Then they satisfy
\begin{equation}
   v_{-m',m}(t_{\theta/2}) \propto (-1)^{j+m}\times  v_{m',m}(t_{(\pi-\theta)/2}),
\end{equation}
where $v_{m',m} \equiv \vec{v'}^{\dagger}_{m'}(t_{\theta/2}) \vec{v}_{m}$ is defined in Eq.~\eqref{eq:vmm}.
Here, the $m$ component $(\vec{v}'_{m'})_m$ of the eigenvector $\vec{v}'_{m'}$ is just $v_{m',m}$.

The eigenvectors for 4-d representation are as follows:
\begin{equation}
    \vec{v}'_{\frac{3}{2}} \propto \left( \begin{array}{c}
        - t^{-3}_{\theta/2}\\
        \sqrt{3} t^{-2}_{\theta/2}\\
        \sqrt{3} t^{-1}_{\theta/2}\\
        1
    \end{array} \right), \qquad
    \vec{v}'_{\frac{1}{2}} \propto \left( \begin{array}{c}
        t^{-1}_{\theta/2}\\
        - \frac{1}{\sqrt{3}} (2 - t^{-2}_{\theta/2})\\
        - \frac{1}{\sqrt{3}} (t_{\theta/2} - 2 t^{-1}_{\theta/2})\\
        1
        \end{array} \right),
\end{equation}
\begin{equation}
    \vec{v}'_{-\frac{1}{2}} \propto \left( \begin{array}{c}
        - t_{\theta / 2}\\
        - \frac{1}{\sqrt{3}} (2 - t^2_{\theta / 2})\\
        \frac{1}{\sqrt{3}} (t^{- 1}_{\theta / 2} - 2 t_{\theta / 2})\\
        1
   \end{array} \right), \qquad
   \vec{v}'_{-\frac{3}{2}} \propto \left( \begin{array}{c}
        t^3_{\theta / 2}\\
        \sqrt{3} t^2_{\theta / 2}\\
        - \sqrt{3} t_{\theta / 2}\\
        1
   \end{array} \right).
\end{equation}

The eigenvectors for 9-d representation are as follows:
\begin{footnotesize}
\begin{equation}\label{JTheta_1}
  \vec{v}'_4 \propto \left( \begin{array}{c}
    t^{- 8}_{\theta / 2}\\
    - 2 \sqrt{2} t^{- 7}_{\theta / 2}\\
    2 \sqrt{7} t^{- 6}_{\theta / 2}\\
    - 2 \sqrt{14} t^{- 5}_{\theta / 2}\\
    \sqrt{70} t^{- 4}_{\theta / 2}\\
    2 \sqrt{14} t^{- 3}_{\theta / 2}\\
    2 \sqrt{7} t^{- 2}_{\theta / 2}\\
    2 \sqrt{2} t^{- 1}_{\theta / 2}\\
    1
  \end{array} \right), \quad
  \vec{v}'_3 \propto \left( \begin{array}{c}
    - t^{- 6}_{\theta / 2}\\
    \frac{1}{2 \sqrt{2}} (7 t^{- 5}_{\theta / 2} - t^{- 7}_{\theta / 2})\\
    \frac{- \sqrt{7}}{2} (3 t^{- 4}_{\theta / 2} - t^{- 6}_{\theta / 2})\\
    \frac{1}{2} \sqrt{\frac{7}{2}} (5 t^{- 3}_{\theta / 2} - 3 t^{- 5}_{\theta
    / 2})\\
    - \sqrt{\frac{35}{2}} (t^{- 2}_{\theta / 2} - t^{- 4}_{\theta / 2})\\
    \frac{1}{2} \sqrt{\frac{7}{2}} (5 t^{- 3}_{\theta / 2} - 3 t^{- 1}_{\theta
    / 2})\\
    \frac{\sqrt{7}}{2} (3 t^{- 2}_{\theta / 2} - 1)\\
    \frac{1}{2 \sqrt{2}} (7 t^{- 1}_{\theta / 2} - t_{\theta / 2})\\
    1
  \end{array} \right), \quad
  \vec{v}'_2 \propto \left( \begin{array}{c}
    t^{- 4}_{\theta / 2}\\
    \frac{- 1}{\sqrt{2}} (3 t^{- 3}_{\theta / 2} - t^{- 5}_{\theta / 2})\\
    \frac{1}{2 \sqrt{7}} (15 t^{- 2}_{\theta / 2} - 12 t^{- 4}_{\theta / 2} +
    t^{- 6}_{\theta / 2})\\
    \frac{- 1}{\sqrt{14}} (10 t^{- 1}_{\theta / 2} - 15 t^{- 3}_{\theta / 2} +
    3 t^{- 5}_{\theta / 2})\\
    \sqrt{\frac{5}{14}} (3 - 8 t^{- 2}_{\theta / 2} + 3 t^{- 4}_{\theta /
    2})\\
    \frac{1}{\sqrt{14}} (10 t^{- 3}_{\theta / 2} - 15 t^{- 1}_{\theta / 2} + 3
    t_{\theta / 2})\\
    \frac{1}{2 \sqrt{7}} (15 t^{- 2}_{\theta / 2} - 12 + t^2_{\theta / 2})\\
    \frac{1}{\sqrt{2}} (3 t^{- 1}_{\theta / 2} - t_{\theta / 2})\\
    1
  \end{array} \right),
\end{equation}
\begin{equation}\label{JTheta_2}
  \vec{v}'_1 \propto \left( \begin{array}{c}
    - t^{- 2}_{\theta / 2}\\
    \frac{1}{2 \sqrt{2}} (5 t^{- 1}_{\theta / 2} - 3 t^{- 3}_{\theta / 2})\\
    \frac{- 1}{2 \sqrt{7}} (10 - 15 t^{- 2}_{\theta / 2} + 3 t^{- 4}_{\theta /
    2})\\
    \frac{1}{2 \sqrt{14}} (10 t_{\theta / 2} - 30 t^{- 1}_{\theta / 2} + 15
    t^{- 3}_{\theta / 2} - t^{- 5}_{\theta / 2})\\
    - \sqrt{\frac{5}{14}} (t^2_{\theta / 2} - 6 + 6 t^{- 2}_{\theta / 2} -
    t^{- 4}_{\theta / 2})\\
    \frac{1}{2 \sqrt{14}} (10 t^{- 3}_{\theta / 2} - 30 t^{- 1}_{\theta / 2} +
    15 t_{\theta / 2} - t^3_{\theta / 2})\\
    \frac{1}{2 \sqrt{7}} (10 t^{- 2}_{\theta / 2} - 15 + 3 t^2_{\theta / 2})\\
    \frac{1}{2 \sqrt{2}} (5 t^{- 1}_{\theta / 2} - 3 t_{\theta / 2})\\
    1
  \end{array} \right), \qquad
  \vec{v}'_0 \propto \left( \begin{array}{c}
    1\\
    2 \sqrt{2} t^{- 1}_{\theta}\\
    \frac{- 2}{\sqrt{7}} (1 - 6 t^{- 2}_{\theta})\\
    - 2 \sqrt{\frac{2}{7}} (3 t^{- 1}_{\theta} - 4 t^{- 3}_{\theta})\\
    \sqrt{\frac{2}{35}} (35 - 40 t^{- 2}_{\theta} + 8 t^{- 4}_{\theta})\\
    - 2 \sqrt{\frac{2}{7}} (3 t^{- 1}_{\theta} - 4 t^{- 3}_{\theta})\\
    \frac{- 2}{\sqrt{7}} (1 - 6 t^{- 2}_{\theta})\\
    2 \sqrt{2} t^{- 1}_{\theta}\\
    1
  \end{array} \right),
\end{equation}
\begin{equation}\label{JTheta_3}
  \vec{v}'_{- 1} \propto \left( \begin{array}{c}
    - t^2_{\theta / 2}\\
    \frac{- 1}{2 \sqrt{2}} (5 t_{\theta / 2} - 3 t^3_{\theta / 2})\\
    \frac{- 1}{2 \sqrt{7}} (10 - 15 t^2_{\theta / 2} + 3 t^4_{\theta / 2})\\
    \frac{- 1}{2 \sqrt{14}} (10 t^{- 1}_{\theta / 2} - 30 t_{\theta / 2} + 15
    t^3_{\theta / 2} - t^5_{\theta / 2})\\
    - \sqrt{\frac{5}{14}} (t^{- 2}_{\theta / 2} - 6 + 6 t^2_{\theta / 2} -
    t^4_{\theta / 2})\\
    \frac{- 1}{2 \sqrt{14}} (10 t^3_{\theta / 2} - 30 t_{\theta / 2} + 15 t^{-
    1}_{\theta / 2} - t^{- 3}_{\theta / 2})\\
    \frac{1}{2 \sqrt{7}} (10 t^2_{\theta / 2} - 15 + 3 t^{- 2}_{\theta / 2})\\
    \frac{- 1}{2 \sqrt{2}} (5 t_{\theta / 2} - 3 t_{\theta / 2}^{- 1})\\
    1
  \end{array} \right), \qquad
  \vec{v}'_{- 2} \propto \left( \begin{array}{c}
    t^4_{\theta / 2}\\
    \frac{1}{\sqrt{2}} (3 t^3_{\theta / 2} - t^5_{\theta / 2})\\
    \frac{1}{2 \sqrt{7}} (15 t^2_{\theta / 2} - 12 t^4_{\theta / 2} +
    t^6_{\theta / 2})\\
    \frac{1}{\sqrt{14}} (10 t_{\theta / 2} - 15 t^3_{\theta / 2} + 3
    t^5_{\theta / 2})\\
    \sqrt{\frac{5}{14}} (3 - 8 t^2_{\theta / 2} + 3 t^4_{\theta / 2})\\
    \frac{- 1}{\sqrt{14}} (10 t^3_{\theta / 2} - 15 t_{\theta / 2} + 3 t^{-
    1}_{\theta / 2})\\
    \frac{1}{2 \sqrt{7}} (15 t^2_{\theta / 2} - 12 + t^{- 2}_{\theta / 2})\\
    \frac{- 1}{\sqrt{2}} (3 t_{\theta / 2} - t^{- 1}_{\theta / 2})\\
    1
  \end{array} \right),
\end{equation}
\begin{equation}\label{JTheta_4}
  \vec{v}'_{- 3} \propto \left( \begin{array}{c}
    - t^6_{\theta / 2}\\
    \frac{- 1}{2 \sqrt{2}} (7 t^5_{\theta / 2} - t^7_{\theta / 2})\\
    \frac{- \sqrt{7}}{2} (3 t^4_{\theta / 2} - t^6_{\theta / 2})\\
    \frac{- 1}{2} \sqrt{\frac{7}{2}} (5 t^3_{\theta / 2} - 3 t^5_{\theta /
    2})\\
    - \sqrt{\frac{35}{2}} (t^2_{\theta / 2} - t^4_{\theta / 2})\\
    \frac{- 1}{2} \sqrt{\frac{7}{2}} (5 t^3_{\theta / 2} - 3 t_{\theta / 2})\\
    \frac{\sqrt{7}}{2} (3 t^2_{\theta / 2} - 1)\\
    \frac{- 1}{2 \sqrt{2}} (7 t_{\theta / 2} - t^{- 1}_{\theta / 2})\\
    1
  \end{array} \right), \qquad
  \vec{v}'_{- 4} \propto \left( \begin{array}{c}
    t^8_{\theta / 2}\\
    2 \sqrt{2} t^7_{\theta / 2}\\
    2 \sqrt{7} t^6_{\theta / 2}\\
    2 \sqrt{14} t^5_{\theta / 2}\\
    \sqrt{70} t^4_{\theta / 2}\\
    - 2 \sqrt{14} t^3_{\theta / 2}\\
    2 \sqrt{7} t^2_{\theta / 2}\\
    - 2 \sqrt{2} t_{\theta / 2}\\
    1
  \end{array} \right).
\end{equation}
\end{footnotesize}

The eigenvectors of 13-d representation are as follows:
\begin{footnotesize}
\begin{equation}
  \vec{v}'_6 \propto \left( \begin{array}{c}
    t^{- 12}_{\theta / 2}\\
    - 2 \sqrt{3} t^{- 11}_{\theta / 2}\\
    \sqrt{66} t^{- 10}_{\theta / 2}\\
    - 2 \sqrt{55} t^{- 9}_{\theta / 2}\\
    3 \sqrt{55} t^{- 8}_{\theta / 2}\\
    - 6 \sqrt{22} t^{- 7}_{\theta / 2}\\
    2 \sqrt{231} t^{- 6}_{\theta / 2}\\
    6 \sqrt{22} t^{- 5}_{\theta / 2}\\
    3 \sqrt{55} t^{- 4}_{\theta / 2}\\
    2 \sqrt{55} t^{- 3}_{\theta / 2}\\
    \sqrt{66} t^{- 2}_{\theta / 2}\\
    2 \sqrt{3} t^{- 1}_{\theta / 2}\\
    1
  \end{array} \right), \quad
  \vec{v}'_5 \propto  \left( \begin{array}{c}
    - t^{- 10}_{\theta / 2}\\
    \frac{1}{2 \sqrt{3}} (11 t^{- 9}_{\theta / 2} - t^{- 11}_{\theta / 2})\\
    - \sqrt{\frac{11}{6}} (5 t^{- 8}_{\theta / 2} - t^{- 10}_{\theta / 2})\\
    \frac{1}{2} \sqrt{55} (3 t^{- 7}_{\theta / 2} - t^{- 9}_{\theta / 2})\\
    - \sqrt{55} (2 t^{- 6}_{\theta / 2} - t^{- 8}_{\theta / 2})\\
    \sqrt{\frac{11}{2}} (7 t^{- 5}_{\theta / 2} - 5 t^{- 7}_{\theta / 2})\\
    - \sqrt{231} (t^{- 4}_{\theta / 2} - t^{- 6}_{\theta / 2})\\
    \sqrt{\frac{11}{2}} (7 t^{- 5}_{\theta / 2} - 5 t^{- 3}_{\theta / 2})\\
    \sqrt{55} (2 t^{- 4}_{\theta / 2} - t^{- 2}_{\theta / 2})\\
    \frac{1}{2} \sqrt{55} (3 t^{- 3}_{\theta / 2} - t^{- 1}_{\theta / 2})\\
    \sqrt{\frac{11}{6}} (5 t^{- 2}_{\theta / 2} - 1)\\
    \frac{1}{2 \sqrt{3}} (11 t^{- 1}_{\theta / 2} - t_{\theta / 2})\\
    1
  \end{array} \right),\quad
  \vec{v}'_4 \propto  \left( \begin{array}{c}
     t^{- 8}_{\theta / 2}\\
     \frac{- 1}{\sqrt{3}} (5 t^{- 7}_{\theta / 2} - t^{- 9}_{\theta / 2})\\
     \frac{1}{\sqrt{66}} (45 t^{- 6}_{\theta / 2} - 20 t^{- 8}_{\theta / 2} +
     t^{- 10}_{\theta / 2})\\
     - \sqrt{\frac{5}{11}} (12 t^{- 5}_{\theta / 2} - 9 t^{- 7}_{\theta / 2} +
     t^{- 9}_{\theta / 2})\\
     \sqrt{\frac{5}{11}} (14 t^{- 4}_{\theta / 2} - 16 t^{- 6}_{\theta / 2} +
     3 t^{- 8}_{\theta / 2})\\
     - \sqrt{\frac{2}{11}} (21 t^{- 3}_{\theta / 2} - 35 t^{- 5}_{\theta / 2}
     + 10 t^{- 7}_{\theta / 2})\\
     \sqrt{\frac{21}{11}} (5 t^{- 2}_{\theta / 2} - 12 t^{- 4}_{\theta / 2} +
     5 t^{- 6}_{\theta / 2})\\
     \sqrt{\frac{2}{11}} (21 t^{- 5}_{\theta / 2} - 35 t^{- 3}_{\theta / 2} +
     10 t^{- 1}_{\theta / 2})\\
     \sqrt{\frac{5}{11}} (14 t^{- 4}_{\theta / 2} - 16 t^{- 2}_{\theta / 2} +
     3)\\
     \sqrt{\frac{5}{11}} (12 t^{- 3}_{\theta / 2} - 9 t^{- 1}_{\theta / 2} +
     t_{\theta / 2})\\
     \frac{1}{\sqrt{66}} (45 t^{- 2}_{\theta / 2} - 20 + t^2_{\theta / 2})\\
     \frac{1}{\sqrt{3}} (5 t^{- 1}_{\theta / 2} - t_{\theta / 2})\\
     1
   \end{array} \right),
\end{equation}
\begin{equation}
\vec{v}'_3 \propto \left( \begin{array}{c}
     - t^{- 6}_{\theta / 2}\\
     \frac{\sqrt{3}}{2} (3 t^{- 5}_{\theta / 2} - t^{- 7}_{\theta / 2})\\
     - \sqrt{\frac{3}{22}} (12 t^{- 4}_{\theta / 2} - 9 t^{- 6}_{\theta / 2} +
     t^{- 8}_{\theta / 2})\\
     \frac{1}{2 \sqrt{55}} (84 t^{- 3}_{\theta / 2} - 108 t^{- 5}_{\theta / 2}
     + 27 t^{- 7}_{\theta / 2} - t^{- 9}_{\theta / 2})\\
     \frac{- 3}{\sqrt{55}} (14 t^{- 2}_{\theta / 2} - 28 t^{- 4}_{\theta / 2}
     + 12 t^{- 6}_{\theta / 2} - t^{- 8}_{\theta / 2})\\
     \frac{3}{\sqrt{22}} (7 t^{- 1}_{\theta / 2} - 21 t^{- 3}_{\theta / 2} +
     14 t^{- 5}_{\theta / 2} - 2 t^{- 7}_{\theta / 2})\\
     - \sqrt{\frac{21}{11}} (2 - 9 t^{- 2}_{\theta / 2} + 9 t^{- 4}_{\theta /
     2} - 2 t^{- 6}_{\theta / 2})\\
     \frac{- 3}{\sqrt{22}} (2 t_{\theta / 2} - 14 t^{- 1}_{\theta / 2} + 21
     t^{- 3}_{\theta / 2} - 7 t^{- 5}_{\theta / 2})\\
     \frac{- 3}{\sqrt{55}} (t^2_{\theta / 2} - 12 + 28 t^{- 2}_{\theta / 2} -
     14 t^{- 4}_{\theta / 2})\\
     \frac{- 1}{2 \sqrt{55}} (t^3_{\theta / 2} - 27 t_{\theta / 2} + 108 t^{-
     1}_{\theta / 2} - 84 t^{- 3}_{\theta / 2})\\
     \sqrt{\frac{3}{22}} (t^2_{\theta / 2} - 9 + 12 t^{- 2}_{\theta / 2})\\
     - \frac{\sqrt{3}}{2} (t_{\theta / 2} - 3 t^{- 1}_{\theta / 2})\\
     1
   \end{array} \right), 
   \vec{v}'_2 \propto \left( \begin{array}{c}
     t^{- 4}_{\theta / 2}\\
     \frac{- 2}{\sqrt{3}} (2 t^{- 3}_{\theta / 2} - t^{- 5}_{\theta / 2})\\
     \sqrt{\frac{2}{33}} (14 t^{- 2}_{\theta / 2} - 16 t^{- 4}_{\theta / 2} +
     3 t^{- 6}_{\theta / 2})\\
     \frac{- 2}{\sqrt{55}} (14 t^{- 1}_{\theta / 2} - 28 t^{- 3}_{\theta / 2}
     + 12 t^{- 5}_{\theta / 2} - t^{- 7}_{\theta / 2})\\
     \frac{1}{3 \sqrt{55}} (70 - 224 t^{- 2}_{\theta / 2} + 168 t^{-
     4}_{\theta / 2} - 32 t^{- 6}_{\theta / 2} + t^{- 8}_{\theta / 2})\\
     \frac{- 2}{3} \sqrt{\frac{2}{11}} (7 t_{\theta / 2} - 35 t^{- 1}_{\theta
     / 2} + 42 t^{- 3}_{\theta / 2} - 14 t^{- 5}_{\theta / 2} + t^{-
     7}_{\theta / 2})\\
     2 \sqrt{\frac{7}{33}} (t_{\theta / 2}^2 - 8 + 15 t^{- 2}_{\theta / 2} - 8
     t^{- 4}_{\theta / 2} + t^{- 6}_{\theta / 2})\\
     \frac{2}{3} \sqrt{\frac{2}{11}} (t^3_{\theta / 2} - 14 t_{\theta / 2} +
     42 t^{- 1}_{\theta / 2} - 35 t^{- 3}_{\theta / 2} + 7 t^{- 5}_{\theta /
     2})\\
     \frac{1}{3 \sqrt{55}} (t^4_{\theta / 2} - 32 t^2_{\theta / 2} + 168 - 224
     t^{- 2}_{\theta / 2} + 70 t^{- 4}_{\theta / 2})\\
     \frac{- 2}{\sqrt{55}} (t^3_{\theta / 2} - 12 t_{\theta / 2} + 28 t^{-
     1}_{\theta / 2} - 14 t^{- 3}_{\theta / 2})\\
     \sqrt{\frac{2}{33}} (3 t^2_{\theta / 2} - 16 + 14 t^{- 2}_{\theta / 2})\\
     \frac{- 2}{\sqrt{3}} (t_{\theta / 2} - 2 t^{- 1}_{\theta / 2})\\
     1
   \end{array} \right),
\end{equation}
\begin{equation}
\vec{v}'_1 \propto \left( \begin{array}{c}
     - t^{- 2}_{\theta / 2}\\
     \frac{1}{2 \sqrt{3}} (7 t^{- 1}_{\theta / 2} - 5 t^{- 3}_{\theta / 2})\\
     \frac{- 1}{\sqrt{66}} (21 - 35 t^{- 2}_{\theta / 2} + 10 t^{- 4}_{\theta
     / 2})\\
     \frac{1}{2} \sqrt{\frac{5}{11}} (14 t_{\theta / 2} - 21 t^{- 1}_{\theta /
     2} + 14 t^{- 3}_{\theta / 2} - 2 t^{- 5}_{\theta / 2})\\
     \frac{- 1}{3} \sqrt{\frac{5}{11}} (7 t^2_{\theta / 2} - 35 + 42 t^{-
     2}_{\theta / 2} - 14 t^{- 4}_{\theta / 2} + t^{- 6}_{\theta / 2})\\
     \frac{1}{6 \sqrt{22}} (21 t^3_{\theta / 2} - 175 t_{\theta / 2} + 350
     t^{- 1}_{\theta / 2} - 210 t^{- 3}_{\theta / 2} + 35 t^{- 5}_{\theta / 2}
     - t^{- 7}_{\theta / 2})\\
     \frac{- 1}{2} \sqrt{\frac{7}{33}} (t^4_{\theta / 2} - 15 t^2_{\theta / 2}
     + 50 - 50 t^{- 2}_{\theta / 2} + 15 t^{- 4}_{\theta / 2} - t^{-
     6}_{\theta / 2})\\
     \frac{- 1}{6 \sqrt{22}} (t^5_{\theta / 2} - 35 t^3_{\theta / 2} + 210
     t_{\theta / 2} - 350 t^{- 1}_{\theta / 2} + 175 t^{- 3}_{\theta / 2} - 21
     t^{- 5}_{\theta / 2})\\
     \frac{1}{3} \sqrt{\frac{5}{11}} (t^4_{\theta / 2} - 14 t^2_{\theta / 2} +
     42 - 35 t^{- 2}_{\theta / 2} + 7 t^{- 4}_{\theta / 2})\\
     \frac{- 1}{2} \sqrt{\frac{5}{11}} (2 t^3_{\theta / 2} - 14 t_{\theta / 2}
     + 21 t^{- 1}_{\theta / 2} - 14 t^{- 3}_{\theta / 2})\\
     \frac{1}{\sqrt{66}} (10 t^2_{\theta / 2} - 35 + 21 t^{- 2}_{\theta /
     2})\\
     \frac{- 1}{2 \sqrt{3}} (5 t_{\theta / 2} - 7 t^{- 1}_{\theta / 2})\\
     1
   \end{array} \right),
   \vec{v}'_0 \propto \left( \begin{array}{c}
     1\\
     2 \sqrt{3} t^{- 1}_{\theta}\\
     - \sqrt{\frac{6}{11}} (1 - 10 t^{- 2}_{\theta})\\
     - 2 \sqrt{\frac{5}{11}} (3 t^{- 1}_{\theta} - 8 t^{- 3}_{\theta})\\
     \sqrt{\frac{5}{11}} (1 - 16 t^{- 2}_{\theta} + 16 t^{- 4}_{\theta})\\
     2 \sqrt{\frac{2}{11}} (5 t^{- 1}_{\theta} - 20 t^{- 3}_{\theta} + 8 t^{-
     5}_{\theta})\\
     \frac{- 2}{\sqrt{231}} (5 - 90 t^{- 2}_{\theta} + 120 t^{- 4}_{\theta} -
     16 t^{- 6}_{\theta})\\
     2 \sqrt{\frac{2}{11}} (5 t_{\theta} - 20 t^3_{\theta} + 8 t^5_{\theta})\\
     \sqrt{\frac{5}{11}} (1 - 16 t^{- 2}_{\theta} + 16 t^{- 4}_{\theta})\\
     - 2 \sqrt{\frac{5}{11}} (3 t^{- 1}_{\theta} - 8 t^{- 3}_{\theta})\\
     - \sqrt{\frac{6}{11}} (1 - 10 t^{- 2}_{\theta})\\
     2 \sqrt{3} t^{- 1}_{\theta}\\
     1
   \end{array} \right),
\end{equation}
\begin{equation}
 \vec{v}'_{- 1} \propto \left( \begin{array}{c}
     - t^2_{\theta / 2}\\
     \frac{- 1}{2 \sqrt{3}} (7 t^1_{\theta / 2} - 5 t^3_{\theta / 2})\\
     \frac{- 1}{\sqrt{66}} (21 - 35 t^2_{\theta / 2} + 10 t^4_{\theta / 2})\\
     \frac{- 1}{2} \sqrt{\frac{5}{11}} (14 t^{- 1}_{\theta / 2} - 21 t_{\theta
     / 2} + 14 t^3_{\theta / 2} - 2 t^5_{\theta / 2})\\
     \frac{- 1}{3} \sqrt{\frac{5}{11}} (7 t^{- 2}_{\theta / 2} - 35 + 42
     t^2_{\theta / 2} - 14 t^4_{\theta / 2} + t^6_{\theta / 2})\\
     \frac{- 1}{6 \sqrt{22}} (21 t^{- 3}_{\theta / 2} - 175 t^{- 1}_{\theta /
     2} + 350 t_{\theta / 2} - 210 t^3_{\theta / 2} + 35 t^5_{\theta / 2} -
     t^7_{\theta / 2})\\
     \frac{- 1}{2} \sqrt{\frac{7}{33}} (t^{- 4}_{\theta / 2} - 15 t^{-
     2}_{\theta / 2} + 50 - 50 t^2_{\theta / 2} + 15 t^4_{\theta / 2} -
     t^6_{\theta / 2})\\
     \frac{1}{6 \sqrt{22}} (t^{- 5}_{\theta / 2} - 35 t^{- 3}_{\theta / 2} +
     210 t^{- 1}_{\theta / 2} - 350 t_{\theta / 2} + 175 t^3_{\theta / 2} - 21
     t^5_{\theta / 2})\\
     \frac{1}{3} \sqrt{\frac{5}{11}} (t^{- 4}_{\theta / 2} - 14 t^{-
     2}_{\theta / 2} + 42 - 35 t^2_{\theta / 2} + 7 t^4_{\theta / 2})\\
     \frac{1}{2} \sqrt{\frac{5}{11}} (2 t^{- 3}_{\theta / 2} - 14 t^{-
     1}_{\theta / 2} + 21 t_{\theta / 2} - 14 t^3_{\theta / 2})\\
     \frac{1}{\sqrt{66}} (10 t^{- 2}_{\theta / 2} - 35 + 21 t^2_{\theta /
     2})\\
     \frac{1}{2 \sqrt{3}} (5 t^{- 1}_{\theta / 2} - 7 t_{\theta / 2})\\
     1
   \end{array} \right),
\end{equation}
\begin{equation}
 \vec{v}'_{- 2} \propto \left( \begin{array}{c}
     t^4_{\theta / 2}\\
     \frac{2}{\sqrt{3}} (2 t^3_{\theta / 2} - t^5_{\theta / 2})\\
     \sqrt{\frac{2}{33}} (14 t^2_{\theta / 2} - 16 t^4_{\theta / 2} + 3
     t^6_{\theta / 2})\\
     \frac{2}{\sqrt{55}} (14 t_{\theta / 2} - 28 t^3_{\theta / 2} + 12
     t^5_{\theta / 2} - t^7_{\theta / 2})\\
     \frac{1}{3 \sqrt{55}} (70 - 224 t^2_{\theta / 2} + 168 t^4_{\theta / 2} -
     32 t^6_{\theta / 2} + t^8_{\theta / 2})\\
     \frac{2}{3} \sqrt{\frac{2}{11}} (7 t^{- 1}_{\theta / 2} - 35 t_{\theta /
     2} + 42 t^3_{\theta / 2} - 14 t^5_{\theta / 2} + t^7_{\theta / 2})\\
     2 \sqrt{\frac{7}{33}} (t_{\theta / 2}^{- 2} - 8 + 15 t^2_{\theta / 2} - 8
     t^4_{\theta / 2} + t^6_{\theta / 2})\\
     \frac{- 2}{3} \sqrt{\frac{2}{11}} (t^{- 3}_{\theta / 2} - 14 t^{-
     1}_{\theta / 2} + 42 t_{\theta / 2} - 35 t^3_{\theta / 2} + 7 t^5_{\theta
     / 2})\\
     \frac{1}{3 \sqrt{55}} (t^{- 4}_{\theta / 2} - 32 t^{- 2}_{\theta / 2} +
     168 - 224 t^2_{\theta / 2} + 70 t^4_{\theta / 2})\\
     \frac{2}{\sqrt{55}} (t^{- 3}_{\theta / 2} - 12 t^{- 1}_{\theta / 2} + 28
     t_{\theta / 2} - 14 t^3_{\theta / 2})\\
     \sqrt{\frac{2}{33}} (3 t^{- 2}_{\theta / 2} - 16 + 14 t^2_{\theta / 2})\\
     \frac{2}{\sqrt{3}} (t^{- 1}_{\theta / 2} - 2 t_{\theta / 2})\\
     1
   \end{array} \right),
 \vec{v}'_{- 3} \propto \left( \begin{array}{c}
     - t^6_{\theta / 2}\\
     - \frac{\sqrt{3}}{2} (3 t^5_{\theta / 2} - t^7_{\theta / 2})\\
     \sqrt{\frac{3}{22}} (12 t^4_{\theta / 2} - 9 t^6_{\theta / 2} +
     t^8_{\theta / 2})\\
     \frac{- 1}{2 \sqrt{55}} (84 t^3_{\theta / 2} - 108 t^5_{\theta / 2} + 27
     t^7_{\theta / 2} - t^9_{\theta / 2})\\
     \frac{- 3}{\sqrt{55}} (14 t^2_{\theta / 2} - 28 t^4_{\theta / 2} + 12
     t^6_{\theta / 2} - t^8_{\theta / 2})\\
     \frac{- 3}{\sqrt{22}} (7 t_{\theta / 2} - 21 t^3_{\theta / 2} + 14
     t^5_{\theta / 2} - 2 t^7_{\theta / 2})\\
     - \sqrt{\frac{21}{11}} (2 - 9 t^2_{\theta / 2} + 9 t^4_{\theta / 2} - 2
     t^6_{\theta / 2})\\
     \frac{3}{\sqrt{22}} (2 t^{- 1}_{\theta / 2} - 14 t_{\theta / 2} + 21
     t^3_{\theta / 2} - 7 t^5_{\theta / 2})\\
     \frac{- 3}{\sqrt{55}} (t^{- 2}_{\theta / 2} - 12 + 28 t^2_{\theta / 2} -
     14 t^4_{\theta / 2})\\
     \frac{1}{2 \sqrt{55}} (t^{- 3}_{\theta / 2} - 27 t^{- 1}_{\theta / 2} +
     108 t_{\theta / 2} - 84 t^3_{\theta / 2})\\
     \sqrt{\frac{3}{22}} (t^{- 2}_{\theta / 2} - 9 + 12 t^2_{\theta / 2})\\
     \frac{\sqrt{3}}{2} (t^{- 1}_{\theta / 2} - 3 t_{\theta / 2})\\
     1
   \end{array} \right),
\end{equation}
\begin{equation}
 \vec{v}'_{- 4} \propto \left( \begin{array}{c}
     t^8_{\theta / 2}\\
     \frac{1}{\sqrt{3}} (5 t^7_{\theta / 2} - t^9_{\theta / 2})\\
     \frac{1}{\sqrt{66}} (45 t^6_{\theta / 2} - 20 t^8_{\theta / 2} +
     t^{10}_{\theta / 2})\\
     \sqrt{\frac{5}{11}} (12 t^5_{\theta / 2} - 9 t^7_{\theta / 2} +
     t^9_{\theta / 2})\\
     \sqrt{\frac{5}{11}} (14 t^4_{\theta / 2} - 16 t^6_{\theta / 2} + 3
     t^8_{\theta / 2})\\
     \sqrt{\frac{2}{11}} (21 t^3_{\theta / 2} - 35 t^5_{\theta / 2} + 10
     t^7_{\theta / 2})\\
     \sqrt{\frac{21}{11}} (5 t^2_{\theta / 2} - 12 t^4_{\theta / 2} + 5
     t^6_{\theta / 2})\\
     - \sqrt{\frac{2}{11}} (21 t^5_{\theta / 2} - 35 t^3_{\theta / 2} + 10
     t_{\theta / 2})\\
     \sqrt{\frac{5}{11}} (14 t^4_{\theta / 2} - 16 t^2_{\theta / 2} + 3)\\
     - \sqrt{\frac{5}{11}} (12 t^3_{\theta / 2} - 9 t_{\theta / 2} + t^{-
     1}_{\theta / 2})\\
     \frac{1}{\sqrt{66}} (45 t^2_{\theta / 2} - 20 + t^{- 2}_{\theta / 2})\\
     \frac{- 1}{\sqrt{3}} (5 t_{\theta / 2} - t_{\theta / 2}^{- 1})\\
     1
   \end{array} \right), \quad 
  \vec{v}'_{- 5} \propto \left( \begin{array}{c}
    - t^{10}_{\theta / 2}\\
    \frac{- 1}{2 \sqrt{3}} (11 t^9_{\theta / 2} - t^{11}_{\theta / 2})\\
    - \sqrt{\frac{11}{6}} (5 t^8_{\theta / 2} - t^{10}_{\theta / 2})\\
    \frac{- 1}{2} \sqrt{55} (3 t^7_{\theta / 2} - t^9_{\theta / 2})\\
    - \sqrt{55} (2 t^6_{\theta / 2} - t^8_{\theta / 2})\\
    - \sqrt{\frac{11}{2}} (7 t^5_{\theta / 2} - 5 t^7_{\theta / 2})\\
    - \sqrt{231} (t^4_{\theta / 2} - t^6_{\theta / 2})\\
    - \sqrt{\frac{11}{2}} (7 t^5_{\theta / 2} - 5 t^3_{\theta / 2})\\
    \sqrt{55} (2 t^4_{\theta / 2} - t^2_{\theta / 2})\\
    \frac{- 1}{2} \sqrt{55} (3 t^3_{\theta / 2} - t_{\theta / 2})\\
    \sqrt{\frac{11}{6}} (5 t^2_{\theta / 2} - 1)\\
    \frac{- 1}{2 \sqrt{3}} (11 t_{\theta / 2} - t^{- 1}_{\theta / 2})\\
    1
  \end{array} \right), \quad 
     \vec{v}'_{- 6} \propto \left( \begin{array}{c}
    t^{12}_{\theta / 2}\\
    2 \sqrt{3} t^{11}_{\theta / 2}\\
    \sqrt{66} t^{10}_{\theta / 2}\\
    2 \sqrt{55} t^9_{\theta / 2}\\
    3 \sqrt{55} t^8_{\theta / 2}\\
    6 \sqrt{22} t^7_{\theta / 2}\\
    2 \sqrt{231} t^6_{\theta / 2}\\
    - 6 \sqrt{22} t^5_{\theta / 2}\\
    3 \sqrt{55} t^4_{\theta / 2}\\
    - 2 \sqrt{55} t^3_{\theta / 2}\\
    \sqrt{66} t^2_{\theta / 2}\\
    - 2 \sqrt{3} t_{\theta / 2}\\
    1
  \end{array} \right). 
\end{equation}
\end{footnotesize}

\section{Character tables and generators of irreducible group}
\label{charactertableandgenerators}
\subsection{$S_4$} \label{S4_Characters_Generators}
The character table of the $S_4$ group irreducible representations is given in Table~\ref{tab:Character_S4}. 

\begin{table}
   \centering
    \setlength\tabcolsep{0.5em}
    \renewcommand{\arraystretch}{1}
    \small
		\begin{tabular}{|c|ccccc|}
			\hline
			$S_4$ & $C_{(e)}$ & $C_{(12)}$ & $C_{(12)(34)}$ & $C_{(123)}$ & $C_{(1234)}$\\
			\hline
			$|C|$ & 1 & 6 & 3 & 8 & 6 \\
			\hline
			$\bf{1}$ & 1 & 1 & 1 & 1 & 1 \\
			$\bf{1'}$ & 1 & --1 & 1 & 1 & --1 \\
			$\bf{2}$ & 2 & 0 & 2 & --1 & 0 \\
			$\bf{3'}$ & 3 & 1 & --1 &0 & --1 \\
			$\bf{3}$ & 3 & --1 & --1 & 0 & 1 \\
			\hline
		\end{tabular}
    \caption{Character table of $S_4$}
    \label{tab:Character_S4}
\end{table}

The three generators of the $S_4$ group are denoted as $S$, $T$ and $K$ corresponding to $Z_4$, $Z_3$ and $Z_2$ rotation along three unparallel axes, respectively, are set lying on the same plane. The angle between the $Z_4$ and $Z_3$ rotation axes is $\arctan\left(\sqrt{2}\right)$, while the angle between the $Z_4$ and $Z_2$ rotation axes is $\pi/2$.

For the $S_4$ 9-d reducible representation, the matrix of $S, K$, and $T$ is written as
\begin{equation}
	V_{(9)}(S)=e^{i V_{(9)}(J_3)\frac{\pi}{2}}=\text{Diag}\{1,-i,-1,i,1,-i,-1,i,1\},
\end{equation}
\begin{footnotesize}
\begin{align}\label{S4_n9_T}
		&V_{(9)}(T)=e^{i V_{(9)}(J_1\sin\theta+J_3\cos\theta)\frac{2\pi}{3}}
		\\&=\left(
		\begin{array}{ccccccccc}
			\frac{1}{2} & -1-i & i \sqrt{7} & (1-i) \sqrt{7} & -\sqrt{\frac{35}{2}} & (-1-i) \sqrt{7} & -i \sqrt{7} & 1-i & \frac{1}{2} \\
			-1-i & 3 i & (1-i) \sqrt{7} & -\sqrt{7} & 0 & -i \sqrt{7} & (1-i) \sqrt{7} & 3 & 1+i \\
			i \sqrt{7} & (1-i) \sqrt{7} & -2 & -1-i & i \sqrt{10} & -1+i & 2 & (1+i) \sqrt{7} & i \sqrt{7} \\
			(1-i) \sqrt{7} & -\sqrt{7} & -1-i & 3 i & 0 & -3 & -1-i & i \sqrt{7} & (-1+i) \sqrt{7} \\
			-\sqrt{\frac{35}{2}} & 0 & i \sqrt{10} & 0 & 3 & 0 & -i \sqrt{10} & 0 & -\sqrt{\frac{35}{2}} \\
			(-1-i) \sqrt{7} & -i \sqrt{7} & -1+i & -3 & 0 & -3 i & -1+i & -\sqrt{7} & (1+i) \sqrt{7} \\
			-i \sqrt{7} & (1-i) \sqrt{7} & 2 & -1-i & -i \sqrt{10} & -1+i & -2 & (1+i) \sqrt{7} & -i \sqrt{7} \\
			1-i & 3 & (1+i) \sqrt{7} & i \sqrt{7} & 0 & -\sqrt{7} & (1+i) \sqrt{7} & -3 i & -1+i \\
			\frac{1}{2} & 1+i & i \sqrt{7} & (-1+i) \sqrt{7} & -\sqrt{\frac{35}{2}} & (1+i) \sqrt{7} & -i \sqrt{7} & -1+i & \frac{1}{2} \notag\\
		\end{array}
		\right), 
\end{align}
\end{footnotesize}
\begin{equation}
		V_{(9)}(K)=e^{i V_{(9)}(J_1)\pi}=\text{Antidiag}\{1,-1,1,-1,1,-1,1,-1,1 \},
\end{equation}
\begin{eqnarray}
    T^\prime_{(9)}=R^\dagger V_{(9)}(T)R=\text{Diag}\{e^{-i\frac{2\pi}{3}}, e^{i\frac{2\pi}{3}},1,1 ,e^{i\frac{2\pi}{3}},e^{-i\frac{2\pi}{3}},e^{-i\frac{2\pi}{3}}, e^{i\frac{2\pi}{3}},1\},
\end{eqnarray}
where $R$ consists of the eigenvectors of $V_{(9)}(J_1\sin\theta+J_3\cos\theta)$ corresponding to eigenvalues $-4,4,-3,3,-2,2,-1,1,0$.\\
For the \textbf{3} irreducible representation of $S_4$,
\begin{equation}
	S_{\textbf{3}}=e^{i V_{(3)}(J_3) \frac{\pi}{2}}=\text{Diag}\{i,1,-i\},
 \quad
         K_{\textbf{3}}=e^{i V_{(3)}(J_1) \pi}=
        \text{Antidiag}\{1,-1,1\}, \label{S4_3d_SK}
\end{equation}
\begin{equation}
	T_{\textbf{3}}=e^{i V_{(3)}(J_1\sin\theta+J_3\cos\theta) \frac{2\pi}{3}}=\left(
	\begin{array}{ccc}
		\frac{i}{2} & \frac{1}{2}-\frac{i}{2} & \frac{1}{2} \\
		\frac{1}{2}-\frac{i}{2} & 0 & \frac{1}{2}+\frac{i}{2} \\
		\frac{1}{2} & \frac{1}{2}+\frac{i}{2} & -\frac{i}{2} \\
	\end{array}
	\right), \label{S4_3d_T}
\end{equation}
\begin{eqnarray}
    T^\prime_{\textbf{3}}=Q^\dagger T_{\textbf{3}}Q=\text{Diag}\{e^{-i\frac{2\pi}{3}},e^{i\frac{2\pi}{3}},1\},
\end{eqnarray}
where $Q$ consists of the eigenvectors of $V_{(3)}(J_1\sin\theta+J_3\cos\theta)$ corresponding to eigenvalues $-1,1,0$.\\
For the $\textbf{3}'$ irreducible representation,
\begin{align}
	S_{\textbf{3}'}=-e^{i V_{(3)}(J_3) \frac{\pi}{2}}=
        \text{Diag}\{-i,-1,i\},
 \quad
        K_{\textbf{3}'}=-e^{i V_{(3)}(J_1) \pi}=
 \text{Antidiag}\{-1,0,-1\},
\end{align}
\begin{equation}
	T_{\textbf{3}'}=e^{i V_{(3)}(J_1\sin\theta+J_3\cos\theta)\frac{2\pi}{3}}=\left(
	\begin{array}{ccc}
		\frac{i}{2} & \frac{1}{2}-\frac{i}{2} & \frac{1}{2} \\
		\frac{1}{2}-\frac{i}{2} & 0 & \frac{1}{2}+\frac{i}{2} \\
		\frac{1}{2} & \frac{1}{2}+\frac{i}{2} & -\frac{i}{2} \\
	\end{array}
	\right).
\end{equation}
\begin{eqnarray}
    T^\prime_{\textbf{3'}}=Q^\dagger T_{\textbf{3'}}Q=\text{Diag}\{e^{-i\frac{2\pi}{3}},e^{i\frac{2\pi}{3}},1\}
\end{eqnarray}
For \textbf{2} irreducible representation,
\begin{equation}
		S_{\bf{2}}=\left(
	\begin{array}{cc}
		-1 & 0 \\
		0 & 1 \\
	\end{array}
	\right),\quad
	T_{\bf{2}}=\left(
	\begin{array}{cc}
		-\frac{1}{2} & \frac{3}{2 \sqrt{5}} \\
		-\frac{\sqrt{5}}{2} & -\frac{1}{2} \\
	\end{array}
	\right),\quad
	K_{\bf{2}}=\left(
	\begin{array}{cc}
		1 & 0 \\
		0 & 1 \\
	\end{array}
	\right),
\end{equation}
\begin{equation}
    T'_{\textbf{2}}=Q^\dagger T_{\bf{2}}Q=\left(
	\begin{array}{cc}
		e^{\frac{2 i \pi }{3}} & 0 \\
		0 & e^{-\frac{1}{3} (2 i \pi )} \\
	\end{array}
	\right)\quad \text{with} \quad 
    Q=\left(
\begin{array}{cc}
 -\frac{1}{2} i \sqrt{\frac{3}{2}} & \frac{1}{2} i \sqrt{\frac{3}{2}} \\
 \frac{1}{2}\sqrt{\frac{5}{2}} & \frac{1}{2} \sqrt{\frac{5}{2}}\\
\end{array}
\right).
\end{equation}
\\
For the \textbf{1} irreducible representation,
\begin{equation}
	S_{\textbf{1}}=1,\qquad T_{\textbf{1}}=1,\qquad K_{\textbf{1}}=1.
\end{equation}
For the $\textbf{1}'$ irreducible representation,
\begin{equation}
     S_{\textbf{1}'}=-1, \qquad T_{\textbf{1}'}=1, \qquad 
     K_{\textbf{1}'}=-1.
\end{equation}
\subsection{$A_4$}
\begin{table}
     \centering
    \setlength
    \tabcolsep{0.5em}
    \renewcommand{\arraystretch}{1}
    \small
		\begin{tabular}{|c|cccc|}
				\hline
				$A_4$ & $C_{(e)}$ & $C_{(12)(34)}$ & $C_{(123)}$ & $C_{(132)}$\\
				\hline
				$|C|$ & 1 & 3 & 4 & 4\\
				\hline
				$\bf{1}$ & 1 & 1 & $1$ & $1$\\
				$\bf{1'}$ & 1 & 1 & $e^{i4\pi/3}$ & $e^{i2\pi/3}$\\
				$\bf{1''}$ & 1 & 1 & $e^{i2\pi/3}$ & $e^{i4\pi/3}$\\
				$\bf{3}$ & 3 & --1 &0 & 0\\
				\hline
		\end{tabular}
    \caption{Character table of $A_4$}
    \label{tab:Character_A4}
\end{table}
The character table of the $A_4$ irreducible representations is given in Table~\ref{tab:Character_A4}. 

The two generators of the $A_4$ group are denoted as $S$ and $T$ corresponding to the $Z_3$ and $Z_2$ rotations, respectively. The angle between two rotation axes is $\theta=\arctan(\sqrt{2})$.

For the \textbf{3} irreducible representation of $A_4$,
\begin{equation}
		S_{\textbf{3}}=e^{i V_{(3)}(J_3)\frac{2\pi}{3}}=\text{Diag}\{e^{\frac{2i\pi}{3}},1, e^{-\frac{1}{3}(2i\pi)}\}
,
\end{equation}
\begin{equation}
		T_{\textbf{3}}=e^{i V_{(3)}(J_1\sin\theta+J_3\cos\theta)\frac{2\pi}{3}}=\left(
	\begin{array}{ccc}
		\frac{i}{2} & \frac{1}{2}-\frac{i}{2} & \frac{1}{2} \\
		\frac{1}{2}-\frac{i}{2} & 0 & \frac{1}{2}+\frac{i}{2} \\
		\frac{1}{2} & \frac{1}{2}+\frac{i}{2} & -\frac{i}{2} \\
	\end{array}
	\right),
\end{equation}
\begin{equation}
	T'_{\textbf{3}}=Q^\dagger T_{\textbf{3}}Q=\text{Diag}\{-1,-1,-1\}
\end{equation}
where $Q$ consists of the eigenvectors of $V_{(3)}(J_1\sin\theta+J_3\cos\theta)$ corresponding to eigenvalues $-1,1,0$.\\
For 1-d representation, $\textbf{1}, \textbf{1}'$ and $\textbf{1}''$:
\begin{equation}
	S_{\textbf{1}}=1,T_{\textbf{1}}=1, \qquad
	S_{\textbf{1}'}=e^{\frac{4i\pi }{3}},T_{\textbf{1}'}=1, \qquad
	S_{\textbf{1}''}=e^{i\frac{2\pi}{3}},T_{\textbf{1}''}=1.
\end{equation}

\subsection{$A_5$}
\begin{table}
   \centering
    \setlength\tabcolsep{0.5em}
    \renewcommand{\arraystretch}{1}
    \small
		\begin{tabular}{|c|ccccc|}
			\hline
			$A_5$ & $C_{(e)}$ & $C_{(123)}$ & $C_{(12)(34)}$ & $C_{(12345)}$ & $C_{(13452)}$\\
			\hline
			$|C|$ & 1 & 20 & 15 & 12 & 12 \\
			\hline
			$\bf{1}$ & 1 & 1 & 1 & 1 & 1 \\
			$\bf{4}$ & 4 & 1 & 0 & --1 & --1 \\
			$\bf{5}$ & 5 & --1 & 1 & 0 & 0 \\
			$\bf{3}$ & 3 & 0 & --1 & $\frac{1+\sqrt{5}}{2}$ & $\frac{1-\sqrt{5}}{2}$ \\
			$\bf{3'}$ & 3 & 0 & --1 & $\frac{1-\sqrt{5}}{2}$ & $\frac{1+\sqrt{5}}{2}$ \\
			\hline
		\end{tabular}
    \caption{Character table of $A_5$.}
    \label{tab:Character_A5}
\end{table}
The character table of the $A_5$ irreducible representations is given in Table~\ref{tab:Character_A5}. 

The three generators of the $A_5$ group are denoted $S$, $T$, and $K$ corresponding to the $Z_5$, $Z_3$, and $Z_2$ rotations respectively. The angle between the $Z_5$ and $Z_3$ rotation axes is $\theta_{53}=-\arctan\left(3-\sqrt{5}\right)$, and the angle between the $Z_5$ and $Z_2$ rotation axes is $\theta_{52}=-\arctan\left(\frac{\sqrt{5}+1}{2}\right)$.

For 13-d representation, 
\begin{equation}
    V_{(13)}(S)=\text{Diag}\{e^{\frac{2i\pi}{5}},1,e^{-\frac{2i\pi}{5}},e^{-\frac{4 i\pi}{5}},e^{\frac{4 i\pi}{5}},e^{\frac{2i\pi}{5}},1,e^{-\frac{2i\pi}{5}},e^{-\frac{4 i\pi}{5}},e^{\frac{4 i\pi}{5}},e^{\frac{2i\pi}{5}},1,e^{-\frac{2i\pi}{5}}\},
\end{equation}
\begin{equation}
    V_{(13)}(T)=e^{i V_{(13)}(J_1\sin\theta_{53}+J_3\cos\theta_{53})\frac{2\pi}{3}},
\end{equation}
\begin{equation}
    V_{(13)}(K)=e^{i V_{(13)}(J_1\sin\theta_{52}+J_3\cos\theta_{52})\pi}.
\end{equation}
For \textbf{5} irreducible representation of $A_5$,
\begin{equation}
    S_{\textbf{5}}=e^{i V_{(5)}(J_3)\frac{2\pi}{5}}=\text{Diag}\{e^{\frac{4 i\pi}{5}},e^{\frac{2i\pi}{5}},1,e^{-\frac{2i\pi}{5}},e^{-\frac{4 i\pi}{5}}\},\label{A5_S5}
\end{equation}
\begin{equation}
    T_{\textbf{5}}=e^{i V_{(5)}(J_1\sin\theta_{53}+J_3\cos\theta_{53})\frac{2\pi}{3}},
\end{equation}
\begin{equation}
    K_{\textbf{5}}=e^{i V_{(5)}(J_1\sin\theta_{52}+J_3\cos\theta_{52})\pi}.
\end{equation}
For \textbf{3} representation,
\begin{equation}
	S_{\textbf{3}}=e^{i V_{(3)}(J_3)\frac{2\pi}{5}}=\text{Diag}\{e^{\frac{2i\pi}{5}},1,e^{-\frac{2i\pi }{5}}\} ,\qquad
	T_{\textbf{3}}
	=e^{i V_{(3)}(J_1\sin\theta_{53}+J_3\cos\theta_{53})\frac{2\pi}{3}}.
\end{equation}
\begin{equation}
    K_{\textbf{3}}
	=e^{i V_{(3)}(J_1\sin\theta_{52}+J_3\cos\theta_{52})\pi}.
\end{equation}
For $\textbf{3}'$ representation,
\begin{equation}
	S_{\textbf{3}'}=e^{i V_{(3)}(J_3)\frac{4\pi}{5}}=\text{Diag}\{e^{\frac{4i\pi}{5}},1,e^{-\frac{4i\pi }{5}}\} ,\qquad
	T_{\textbf{3}'}
	=e^{i V_{(3)}(J_1\sin\theta_{53}+J_3\cos\theta_{53})\frac{4\pi}{3}},
\end{equation}
\begin{equation}
    K_{\textbf{3}'}
	=e^{i V_{(3)}(J_1\sin\theta_{52}+J_3\cos\theta_{52})\pi}.
\end{equation}
For \textbf{4} representation,
\begin{equation}
	S_{\textbf{4}}=e^{i V_{(4)}(J_3)\frac{4\pi}{5}}=\text{Diag}\{e^{-\frac{4i\pi}{5}},e^{\frac{2i\pi}{5}},e^{-\frac{2i\pi }{5}},e^{\frac{4i\pi}{5}}\} ,\qquad
	T_{\textbf{4}}
	=e^{i V_{(4)}(J_1\sin\theta_{53}+J_3\cos\theta_{53})\frac{4\pi}{3}},
\end{equation}
\begin{equation}
    K_{\textbf{4}}
	=e^{i V_{(4)}(J_1\sin\theta_{52}+J_3\cos\theta_{52})\pi}.
\end{equation}
For \textbf{1} representation,
\begin{equation}
    S_{\textbf{1}}=1,\qquad T_{\textbf{1}}=1,\qquad K_{\textbf{1}}=1.
\end{equation}

We diagonalize $V_{(13)}(T)$, $T_\textbf{5}$, and $T_\textbf{3}$ as
\begin{align}
     T'_{(13)}&=R^\dagger V_{(13)}(T)R
        \\&=\text{Diag}\{1,1,e^{\frac{2 i \pi }{3}},e^{-\frac{2 i \pi }{3}},e^{-\frac{2 i \pi }{3}},e^{\frac{2 i \pi }{3}},1,1,e^{\frac{2 i \pi }{3}},e^{-\frac{2 i \pi }{3}},e^{-\frac{2 i \pi }{3}},e^{\frac{2 i \pi }{3}},1\}
     \\T'_{\textbf{5}}&=V^\dagger T_{\textbf{3}}V=
     \text{Diag}\{e^{\frac{2 i \pi }{3}},e^{-\frac{2 i \pi }{3}},e^{-\frac{2 i \pi }{3}},e^{\frac{2 i \pi }{3}},1 \}
     \\T'_{\textbf{3}}&=Q^\dagger T_{\textbf{3}}Q=
     \text{Diag}\{e^{-\frac{2 i \pi }{3}},e^{\frac{2 i \pi }{3}},1 \}
       \\T'_{\textbf{3'}}&=Q^\dagger T_{\textbf{3'}}Q=
     \text{Diag}\{e^{\frac{2 i \pi }{3}},-e^{\frac{2 i \pi }{3}},1 \},
\end{align}
where $R,V,Q$ consist of the eigenvectors of $V_{(n)}(J_1\sin\theta_{53} + J_3\cos\theta_{53})$ corresponding to eigenvalues $-\frac{n-1}{2},\frac{n-1}{2},-\frac{n-3}{2},\frac{n-3}{2}\dots,0$ in 13-d, 5-d, and 3-d representations respectively.

\subsection{$K_4(Z_2\oplus Z_2)$}
\begin{table}
     \centering
    \setlength
    \tabcolsep{0.5em}
    \renewcommand{\arraystretch}{1}
    \small
		\begin{tabular}{|c|cccc|}
				\hline
				$K_4$ & $C_{(e)}$ & $C_{(a)}$ & $C_{(b)}$ & $C_{(c)}$\\
				\hline
				$|C|$ & 1 & 1 & 1 & 1\\
				\hline
				$\bf{1}$ & 1 & 1 & $1$ & $1$\\
				$\chi_a$ & 1 & 1 & --1 & --1\\
				$\chi_b$ & 1 & --1 & 1 & --1\\
				$\chi_c$ & 1 & --1 & --1 & 1\\
				\hline
		\end{tabular}
   \caption{Character table of $K_4$}
   \label{tab:Character_K4}
\end{table}
The character table of the $K_4(Z_2\oplus Z_2)$ irreducible representations is given in Table~\ref{tab:Character_K4}, where $a,b,c$ there are the $Z_2$ rotation symmetry along $x-,y-,$ and $z-$axes respectively.

\subsection{The characters of $n$-d representation of SU(2) related to the discrete symmetry group }

The characters of $(2j+1)$-d representation of SU(2) are given by  
\begin{equation}
    \chi_j(\varphi)=\frac{\sin[(j+1/2)\varphi]}{\sin(\varphi/2)}
\end{equation}
For 5-d representation of SU(2), the characters related to the $Z_2\oplus Z_2$ group are
\begin{equation}
	\chi_2(e)=5,\qquad \chi_2 \left(\pi\right)=\frac{\sin\left(\frac{5\pi}{2}\right)}{\sin\left(\frac{\pi}{2}\right)}=1.
\end{equation}
For 7-d representation of SU(2), the characters related to the $A_4$ group are
\begin{align}
	\chi_3(e)=7,\quad
	\chi_3 (\pi)=\frac{\sin\left(\frac{7}{2}\pi\right)}{\sin\left(\frac{\pi}{2}\right)}=-1,\quad 
	\chi_3 \left(\frac{2\pi}{3}\right)=\frac{\sin\left(\frac{7}{2}\frac{2\pi}{3}\right)}{\sin\left(\frac{1}{2}\frac{2\pi}{3}\right)}=1,\quad 
	\chi_3 \left(\frac{4\pi}{3}\right)=\frac{\sin\left(\frac{7}{2}\frac{4\pi}{3}\right)}{\sin\left(\frac{1}{2}\frac{4\pi}{3}\right)}=1.
\end{align}
For 9-d representation of SU(2), the characters related to the $S_4$ group are
\begin{align}
    \chi_4(e)=9,\quad
    \chi_4(\pi)=\frac{\sin\left(\frac{9}{2}\pi\right)}{\sin\left(\frac{\pi}{2}\right)}=1,\quad 		\chi_4\left(\frac{\pi}{2}\right)=\frac{\sin\left(\frac{9}{2}\frac{\pi}{2}\right)}{\sin\left(\frac{1}{2}\frac{\pi}{2}\right)}=1,\quad
    \chi_4\left(\frac{2\pi}{3}\right)=\frac{\sin\left(\frac{9}{2}\frac{2\pi}{3}\right)}{\sin\left(\frac{1}{2}\frac{2\pi}{3}\right)}=0.
\end{align}
For 13-d representation of SU(2), the characters related to the $A_5$ group are
\begin{align}
	&\chi_6(e)=13,\quad
        \chi_6(\pi)=\frac{\sin\left(\frac{13}{2}\pi\right)}{\sin\left(\frac{1}{2}\pi\right)}=1,\quad
        \chi_6\left(\frac{2\pi}{3}\right)=\frac{\sin\left(\frac{13}{2}\frac{2\pi}{3}\right)}{\sin\left(\frac{1}{2}\frac{2\pi}{3}\right)}=1,
	\\&
	\chi_6\left(\frac{2\pi}{5}\right)=\frac{\sin\left(\frac{13}{2}\frac{2\pi}{5}\right)}{\sin\left(\frac{1}{2}\frac{2\pi}{5}\right)}=\frac{1+\sqrt{5}}{2},\quad \chi_6\left(\frac{4\pi}{5}\right)=\frac{\sin\left(\frac{13}{2}\frac{4\pi}{5}\right)}{\sin\left(\frac{1}{2}\frac{4\pi}{5}\right)}=\frac{1-\sqrt{5}}{2}.
\end{align}
For 15-d representation of SU(2), the characters related to the $A_4$ group are
\begin{align}
	\chi_7(e)=15,\quad
	\chi_7 (\pi)=\frac{\sin\left(\frac{15}{2}\pi\right)}{\sin\left(\frac{\pi}{2}\right)}=-1,\quad 
	\chi_7 \left(\frac{2\pi}{3}\right)=\frac{\sin\left(\frac{15}{2}\frac{2\pi}{3}\right)}{\sin\left(\frac{1}{2}\frac{2\pi}{3}\right)}=0,\quad
	\chi_7 \left(\frac{4\pi}{3}\right)=\frac{\sin\left(\frac{15}{2}\frac{4\pi}{3}\right)}{\sin\left(\frac{1}{2}\frac{4\pi}{3}\right)}=0.
\end{align}
For 17-d representation of SU(2), the characters related to the $S_4$ group are
\begin{align}
	\chi_8(e)=17,\quad
	\chi_8 (\pi)=\frac{\sin\left(\frac{17}{2}\pi\right)}{\sin\left(\frac{\pi}{2}\right)}=1,\quad 
	\chi_8 \left(\frac{2\pi}{3}\right)=\frac{\sin\left(\frac{17}{2}\frac{2\pi}{3}\right)}{\sin\left(\frac{1}{2}\frac{2\pi}{3}\right)}=-1,\quad
	\chi_8 \left(\frac{\pi}{2}\right)=\frac{\sin\left(\frac{17}{2}\frac{\pi}{2}\right)}{\sin\left(\frac{1}{2}\frac{\pi}{2}\right)}=1.
\end{align}
For 19-d representation of SU(2), the characters related to the $S_4$ group are
\begin{align}
	\chi_9(e)=19,\quad
	\chi_9 (\pi)=\frac{\sin\left(\frac{19}{2}\pi\right)}{\sin\left(\frac{\pi}{2}\right)}=-1,\quad 
	\chi_9 \left(\frac{2\pi}{3}\right)=\frac{\sin\left(\frac{19}{2}\frac{2\pi}{3}\right)}{\sin\left(\frac{1}{2}\frac{2\pi}{3}\right)}=1,\quad 
	\chi_9 \left(\frac{\pi}{2}\right)=\frac{\sin\left(\frac{19}{2}\frac{\pi}{2}\right)}{\sin\left(\frac{1}{2}\frac{\pi}{2}\right)}=1.\quad 
\end{align}
For 21-d representation of SU(2), the characters related to the $A_5$ group are
\begin{align}
	&\chi_{10}(e)=21,\quad
	\chi_{10} (\pi)=\frac{\sin\left(\frac{21}{2}\pi\right)}{\sin\left(\frac{\pi}{2}\right)}=1,\quad 
	\chi_{10} \left(\frac{2\pi}{3}\right)=\frac{\sin\left(\frac{21}{2}\frac{2\pi}{3}\right)}{\sin\left(\frac{1}{2}\frac{2\pi}{3}\right)}=0,
 \\&
	\chi_{10} \left(\frac{2\pi}{5}\right)=\frac{\sin\left(\frac{21}{2}\frac{2\pi}{5}\right)}{\sin\left(\frac{1}{2}\frac{2\pi}{5}\right)}=1,\quad \chi_{10}\left(\frac{4\pi}{5}\right)=\frac{\sin\left(\frac{21}{2}\frac{4\pi}{5}\right)}{\sin\left(\frac{1}{2}\frac{4\pi}{5}\right)}=1.
\end{align}

\bibliography{ref.bib}
\bibliographystyle{utphys}
\end{document}